\documentclass[12pt,fleqn,reqno]{article}
\usepackage{amsmath, amssymb,amsthm,cite}
\usepackage[dvips]{graphics,color}
\usepackage{fullpage}
\usepackage{url}

\newtheorem{theorem}{Theorem}

\newtheorem{lemma}[theorem]{Lemma}
\newtheorem{proposition}{Proposition}

\newtheorem{corollary}[theorem]{Corollary}

\newtheorem{definition}{Definition}

\numberwithin{equation}{section}

\def\const{\hbox{\rm const}}

\def\d{\mathop{\hbox{\rm d}}}

\def\th{\mathop{\hbox{\rm th}}}

\def\div{\mathop{\hbox{\rm div}}}
\def\curl{\mathop{\hbox{\rm curl}}}
\def\grad{\mathop{\hbox{\rm grad}}}

\def\mbs{\boldsymbol}

\def\V{{\mathcal V}}
\def\S{{\mathcal S}}
\def\C{{\mathcal C}}
\def\A{{\mathcal A}}

\def\dV{dV}
\def\dS{d\vec{S}}
\def\ds{d\vec{\ell}}

\def\volT{T}
\def\volX{\vec{\Psi}}

\def\surfT{\vec{T}}
\def\surfX{\vec{\Psi}}

\def\curvT{\vec{T}}
\def\curvX{\Psi}

\def\Esp{{\mathcal E}}
\def\triv{{\rm triv}}
\def\u{\vec{u}}
\def\E{\vec{E}}
\def\B{\vec{B}}
\def\J{\vec{J}}
\def\A{\vec{A}}
\def\vort{\vec{\omega}}
\def\Eop{{\rm E}}
\def\depvar{v}
\def\depvars{\boldsymbol{\depvar}}

\def\pde{G}
\def\pdesys{\boldsymbol{\pde}}
\def\pdedom{\Rnum^3}

\def\Div{\mathop{\hbox{\rm Div}}}
\def\Curl{\mathop{\hbox{\rm Curl}}}
\def\Grad{\mathop{\hbox{\rm Grad}}}
\def\flux{\mathcal{F}}

\def\hook{\rfloor}
\def\d{\mathbf{d}}
\def\D{\mathbf{D}}

\def\x{\vec{x}}

\def\Rnum{\mathbb{R}}

\def\t{{\rm t}}

\def\hook{\rfloor}
\def\nor{\hat\nu}

\def\vol{\epsilon}
\def\volform{{\boldsymbol{\vol}}}
\def\volV{\varepsilon}
\def\volformV{{\boldsymbol{\volV}}}

\def\p{\partial}

\def\bs{\boldsymbol}
\def\th{\text{th}}

\def\parderop#1{\partial/\partial{#1}}

\def\tbox#1#2{$\genfrac{}{}{0pt}{}{\hbox{#1}}{\hbox{#2}}$}

\begin{document}

\title{On the different types of global and local\\ conservation laws for partial differential equations\\ in three spatial dimensions: review and recent developments}

\author{
Stephen C. Anco$^{\rm a}$\\
{\small Department of Mathematics and Statistics}\\
{\small Brock University, St. Catharines, ON, L2S 3A1, Canada}\\\\
Alexei F. Cheviakov$^{\rm b}$\\
{\small Department of Mathematics and Statistics}\\
{\small University of Saskatchewan, Saskatoon, SK, S7N 5E6 Canada}
}

\footnotetext[1]{Corresponding author. Electronic mail: sanco@brocku.ca}
\footnotetext[2]{Electronic mail: chevaikov@math.usask.ca}

\maketitle

\begin{abstract}
For systems of partial differential equations in three spatial dimensions,
dynamical conservation laws holding on volumes, surfaces, and curves, 
as well as topological conservation laws holding on surfaces and curves, 
are studied in a unified framework.
Both global and local formulations of these different conservation laws are discussed,
including the forms of global constants of motion. 
The main results consist of providing an explicit characterization
for when two conservation laws are locally or globally equivalent,
and for when a conservation law is locally or globally trivial,
as well as deriving relationships among the different types of conservation laws.
In particular, the notion of a ``trivial'' conservation law is clarified
for all of the types of conservation laws.
Moreover, as further new results,
conditions under which a trivial local conservation law on a domain
can yield a non-trivial global conservation law on the domain boundary
are determined and shown to be related to differential identities that hold 
for PDE systems containing both evolution equations and spatial constraint equations. 
Numerous physical examples from
fluid flow, gas dynamics, electromagnetism, and magnetohydrodynamics
are used as illustrations.
\end{abstract}

\section{Introduction}\label{sec:intro}

Conservation laws of dynamical type and topological type
have numerous importance uses in the study of partial differential equations (PDEs).

In local form,
a dynamical conservation law is a continuity equation that holds
for all solutions of a given PDE system on a spatial domain. 
In three spatial dimensions, 
this domain is most often taken to be a spatial volume,
and the continuity equation then states that the time derivative of 
a local density quantity (e.g. mass, energy, charge, momentum, vorticity)
is balanced by the divergence of a local spatial flux vector. 
Domains given by surfaces and curves in three dimensions
can similarly lead to useful continuity equations 
for local flux quantities and local circulation quantities, respectively,
as will be fully developed and explained in the present paper. 

All local conservation laws are an intrinsic coordinate-free aspect of the structure
of a PDE system \cite{Olv-book,BCA-book,Anc-review}. 
They yield potentials and nonlocally-related systems 
\cite{BA-book,BCA-book,CheBlu2010}. 
In the case of volume domains, 
they detect if a PDE system admits an invertible transformation
into a target class of PDE systems
(e.g. nonlinear to linear, or linear variable coefficient to constant coefficient) 
\cite{AncBluWol,Wol}, 
and they typically indicate if a PDE system has integrability structure 
\cite{MikShaSok}. 
They also can be used to construct good discretizations for numerical solution methods
(e.g. conserving energy)
\cite{WanBihNav}. 

In global form,
a dynamical conservation law gives a balance equation in which 
the rate of change of an integral quantity over some given spatial domain
is equal to a net flux measured by an integral quantity over the domain boundary,
holding for all solutions of a given PDE system.
Global conservation laws, especially for volume domains, 
are often the fundamental equations that govern a physical process. 
These conservation laws provide a basic starting point 
in the formulation of mathematical models and equations 
in continuum physical systems, 
such as gas dynamics, fluid mechanics, continuum mechanics, electromagnetism, and magnetohydrodynamics. 
The global form of conservation laws on surface domains and curve domains 
have a similarly important role, but have been less well-studied in general
(apart from some special applications in classical field theory \cite{AndTor1996}). 

All global dynamical conservation laws yield conserved integral quantities 
when suitable boundary conditions are posed for a PDE system. 
These integral quantities provide conserved norms and estimates 
which are central to the analysis of solutions
such as existence and uniqueness, stability, and global behaviour. 
They also allow checking the accuracy of numerical solutions and numerical integration methods. 

Topological conservation laws, in contrast, 
describe an integral quantity that remains conserved for all solutions of a given PDE system
when a spatial domain is deformed in any continuous way that preserves its topology.
These conservation laws typically arise in PDE systems that contain
differential constraints, like spatial divergence or curl equations.
In three spatial dimensions, 
there are two main types of global topological conservation laws, 
given by surface integral quantities and line integral quantities,
which are time-independent counterparts of dynamical conservation laws. 

The primary purpose of this paper is to study 
all of these different types of three-dimensional conservation laws, 
as well as their inter-relationships,
in a unified framework for general PDE systems in $\pdedom$. 
As basic results,
dynamical conservation laws on surface domains and curve domains are formulated 
as continuity equations analogously to dynamical conservation laws on volume domains. 
The formulation is particularly appropriate for dynamical PDE systems 
consisting of evolution equations, with or without spatial constraints, 
in the context of continuum mechanics, 
and it is more general than the standard notion of lower-degree conservation laws \cite{Olv-book,HT-book,BCA-book} 
in which time coordinate is not distinguished from the space coordinates. 
Further basic results are that, for each type of dynamical and topological conservation law in three spatial dimensions, 
an explicit characterization will be provided 
to show when two conservation laws are locally or globally equivalent,
and to show when a conservation law is locally or globally trivial.
In addition,
various relationships between the different types of conservation laws
will be examined.
These results will clarify, in particular, 
the notion of a ``trivial'' conservation law,
especially for the less familiar situation of 
conservation laws on surface domains and curve domains. 
Specifically, it will be shown in what sense a ``trivial'' conservation law is a mathematical identity containing no useful information about the solutions of a given PDE system. 

As interesting new consequences of these results,
it will be shown that under certain conditions
a locally trivial dynamical conservation law formulated in a spatial domain 
can give rise to a globally non-trivial dynamical conservation law on the domain boundary. 
Such boundary conservation laws represent constants of motion
and correspond to continuity equations in which the spatial flux is zero. 
Furthermore, a direct connection will be established between these conservation laws
and differential identities that hold for PDE systems containing both evolution equations and spatial constraint equations. 
This will explain what has been the source of some confusion in the applied mathematics and physics literature
on whether such differential identities are merely ``trivial'' conservation laws.

Throughout, we will formulate conservation laws in the common way
in physics and applied mathematics
by working on the space of solutions of a given PDE system.
We also will state and derive the main results in a concrete form using vector calculus 
that is most useful for physical applications. 

This paper will not deal with the question of how to find 
conservation laws for a given PDE system. 
We remark that a direct constructive method using multipliers 
can be applied to derive local dynamical conservation laws of volume type
for PDE systems
\cite{Olv-book,AncBlu1997,AncBlu2002a,AncBlu2002b,Anc2003,BCA-book,Anc-review}. 
This general multiplier method can be extended for deriving all of the other types of local conservation laws. 
A more abstract approach to the study and computation of conservation laws, 
in the setting of cohomology in the variational bi-complex, 
can be found in Refs.\cite{Vin1984,KraVin}. 

The rest of the present paper is organized as follows.

In Section~\ref{sec:prelims},
we give a brief summary of the PDE systems for 
fluid flow, gas dynamics, electromagnetism, and magnetohydrodynamics,
which will be used to illustrate all of the subsequent main results. 
This summary will establish our notation and show how these physical systems fit into 
a general concrete formulation of PDE systems and conservation laws 
in three spatial dimensions. 
We also discuss some mathematical preliminaries that underlie this framework
and that are essential for doing computations of all types of conservation laws. 
In particular, 
the basic notion of a regular PDE system and its coordinatization via a set of leading derivatives is discussed. 

In Section~\ref{sec:CLs}, 
we discuss the definition, properties, and inter-relationships
for three-dimensional local and global conservation laws of dynamical nature
formulated on volume domains, surface domains, and curve domains. 
The physical meaning of the resulting conserved quantities are explained,
and the conditions under which a dynamical conservation law yields a constant of motion are discussed. 
We similarly discuss time-independent versions of these conservation laws,
which represent three-dimensional topological conserved quantities. 
We also discuss constants of motion arising from dynamical and topological conservation laws. 

In Section~\ref{sec:non-triv}, 
for all of these different types of conservation laws, 
we explain the notions of triviality and equivalence, 
using both local and global formulations. 
We also derive, for all three types of domains, 
the conditions under which a locally trivial dynamical conservation law in a domain 
yields a globally non-trivial dynamical conservation law on the domain boundary.
Additionally, for such boundary conservation laws, 
we give their local formulation and show how they can arise 
from topological conservation laws
as well as from lower-degree dynamical conservation laws with zero flux. 

We will illustrate each different type of conservation law by a physical example
taken from fluid flow, gas dynamics, electromagnetism, and magnetohydrodynamics.

In Section~\ref{sec:examples},
we look systematically at all of these physical examples 
and use them to showcase our main results. 
In particular, for incompressible/irrotational fluid flow, electromagnetism, and magnetohydrodynamics, 
we give physical examples of non-trivial boundary conservation laws that arise from locally trivial dynamical conservation laws in volume domains and surface domains. 
As new applications, two interesting examples will be given for fluid flow with non-vanishing vorticity: 
circulation around closed static curves 
will be shown to be conserved for Beltrami flows,
and net flux of circulatory potential temperature through closed static surfaces 
will be shown to be conserved for flows with diabatic heating in a certain Beltrami state. 
All of the examples also show that how such boundary conservation laws can originate directly from differential identities that hold when a PDE system contains both evolution equations and spatial constraint equations. 

In Section~\ref{sec:potentials},
we explain how non-triviality gets altered when potentials are introduced for a PDE system through a local conservation law. 
We use well-known examples of potentials 
in fluid flow, gas dynamics, electromagnetism, and magnetohydrodynamics 
to illustrate the discussion. 

Finally, we make some concluding remarks in Section~\ref{sec:discuss}.

Some further aspects of our framework and results are given in three appendices.
In Appendix~\ref{sec:interrelations}, 
we state the interrelationships that hold among the different types of conservation laws.
In Appendix~\ref{sec:diffforms},
we transcribe our main new results into the formalism of differential forms
and show how our formulation of dynamical conservation laws on surface domains and curve domains 
is more general than the standard notion of 2-form and 1-form conservation laws.
In Appendix~\ref{sec:jetspace}, 
we give a more rigorous mathematical formulation of conservation laws 
and certain associated technical conditions 
in the setting of jet spaces.

\section{Preliminaries and physical PDE systems}\label{sec:prelims}

We will first summarize in a unified way the PDE systems for 
fluid flow, gas dynamics, electromagnetism, and magnetohydrodynamics
in three spatial dimensions in vector calculus notation. 
(See Refs.\cite{Whi,Jac,HugYou} for more details about the physical derivation of these systems.)
In particular, we distinguish between dynamical equations, constraint equations, and constitutive equations. 
This summary will help to explain the subsequent general formulation of PDE systems
which will we introduce and use throughout our discussion and results on 
conservation laws. 
The general formulation essentially underlies computations of all types of conservation laws. 

We will let $\vec x$ denote the position vector in three dimensions,
and $t$ will denote time. 
The standard divergence, curl, and gradient operators will be denoted 
$\div=\nabla\cdot$, $\curl=\nabla\times$, and $\grad=\nabla$, respectively.

\subsection{Fluid and gas dynamics}\label{sec:fluiddyn}

The flow of fluids in a wide variety of physical situations
is described by the Navier-Stokes equations,
which govern the fluid velocity $\u(t,\vec{x})$, density $\rho(t,\vec{x})$, and pressure $p(t,\vec{x})$. 
In the absence of external forces and for non-adiabatic processes,
the Navier-Stokes equations consist of
a mass continuity equation 
\begin{equation}\label{eq:mass}
\rho_t + \div(\rho\u)=0 
\end{equation}
and a momentum balance equation
\begin{equation}\label{eq:NS:mom}
\rho(\u_t + (\u\cdot\nabla)\u) + \grad p= \mu \,\Delta \u + \nu\grad(\div\u)
\end{equation}
where $\mu,\nu$ are the viscosity coefficients. 
Sometimes the latter equation is written in a shorter form in terms of the variable
$\bar p =p + \nu \div\u$ called the mechanical pressure.

The fluid flow is \emph{inviscid (ideal)} when there is no viscosity, $\mu=\nu=0$. 
Then the momentum balance equation becomes
\begin{equation}\label{eq:Eul:mom}
\rho(\u_t + (\u\cdot\nabla)\u) + \grad p= 0
\end{equation}
which together with the mass continuity equation \eqref{eq:mass} 
constitute the Euler equations for ideal fluids as well as for gas dynamics. 

In addition to the mass and momentum equations, 
in fluid flow an equation of state involving either $\u$, $\rho$, or $p$ 
must be specified to obtain a closed system of equations
and to model particular physical properties of a fluid or gas. 
The most common equations of state are the following:
\begin{align}
& \text{constant density}\quad
\rho=\const;
\label{eq:constdens}
\\
& \text{incompressible flow}\quad
\div\, \u = 0;
\label{eq:incompr}
\\
& \text{barotropic flow}\quad
p=p(\rho) ;
\label{eq:barotropic}
\\
& \text{locally adiabatic flow}\quad
p=p(\rho,S),
\quad 
S_t + \u\cdot\grad S = 0 ;
\label{eq:isentropic}
\end{align}
with $S$ being the local entropy of the fluid. 
For incompressible fluid flow, 
the mass equation \eqref{eq:mass} reduces to a transport equation 
\begin{equation}\label{eq:denstransport}
\rho_t +\u\cdot\grad\rho=0
\end{equation}
for the density. 
Constant-density fluid flow is a special case of incompressible flow,
since the mass equation \eqref{eq:mass} then reduces to the incompressibility equation \eqref{eq:incompr}. 
A flow is \emph{homentropic} when $S$ is constant throughout the fluid. 

In the cases of constant-density fluid flow and incompressible fluid flow, 
the compatibility between the incompressibility equation \eqref{eq:incompr} 
and the momentum equation \eqref{eq:NS:mom} or \eqref{eq:Eul:mom}
yields a Laplace-type equation for the pressure. 
In particular, when the fluid is ideal, the pressure is determined by 
\begin{equation}\label{eq:Eul:press}
\div((1/\rho)\grad p) = -(\nabla\u)\cdot(\nabla\u)^\t 
\end{equation}
(where ``$\t$'' denotes the matrix transpose in Cartesian coordinates). 

The dynamics of gases in the simplest physical situations
is governed by the mass continuity equation \eqref{eq:mass}, 
the inviscid momentum equation \eqref{eq:Eul:mom}, 
and the adiabatic (non-homentropic) process equation \eqref{eq:isentropic}. 
Commonly, the latter equation is inverted to give $S=S(p,\rho)$ 
as a function of pressure and density,
so then the transport equation for $S$ becomes a corresponding transport equation for $p$:
\begin{equation}\label{eq:gas:press}
p_t + \u\cdot\grad p +F(p,\rho)\div\u =0
\end{equation}
with 
\begin{equation}\label{eq:gas:speed}
F(p,\rho) = \frac{\p S/\p (1/\rho)}{\rho\p S/\p p} = \rho c(p,\rho)^2
\end{equation}
which determines the sound speed $c(p,\rho)$ in the gas. 
The sound speed is a constitutive function, which has the role of an equation of state. 

The two most common equations of state are the following:
\begin{align}
& \text{ideal}\quad
c^2 = f(p/(R\rho)), 
\quad
R=\const ;
\label{eq:ideal}
\\
& \text{polytropic (ideal)}\quad
c^2 = \gamma p/\rho, 
\quad
\gamma=\const .
\label{eq:polytrop}
\end{align}
The ideal equation of state \eqref{eq:ideal} 
(in which $f$ is an arbitrary positive function of $p/(R\rho)$)
can be shown to be equivalent to the ideal gas law $p/\rho = R T$
through the thermodynamic relation 
\begin{equation}\label{eq:thermo}
\delta e = T\delta S - p\delta (1/\rho)
\end{equation}
where $T$ is the temperature of the gas. 
In particular, $e=e(T)$ holds when the gas is ideal, 
and $e=c_V T$ with $c_V=\const$ holds in the special case of a polytropic gas. 
This implies $c=c(T)$ for an ideal gas, and $c=\sqrt{\gamma RT}$ when the gas is polytropic.

\subsubsection{Vorticity}\label{sec:vorticity}

The vorticity of a gas/fluid is defined by
\begin{equation}\label{eq:vort}
\vort=\curl\, \u ,
\quad
\div\vort =0 . 
\end{equation}
This allows the momentum equation to be expressed in the form 
\begin{equation}\label{eq:NS:vel:vortform}
\u_t +\vort\times\u = -\tfrac{1}{2}\grad(|\u|^2)  - (1/\rho)\grad \bar p + (\mu/\rho)\Delta\u
\end{equation}
using the identity 
\begin{equation}
(\u\cdot\nabla)\u = \tfrac{1}{2}\grad(|\u|^2) -\u\times\vort .
\end{equation}
The curl of the velocity equation \eqref{eq:NS:vel:vortform}
yields the vorticity transport equation 
\begin{equation}\label{eq:NS:vort}
\vort_t +\curl(\vort\times\u) 
= -\grad(1/\rho)\times(\grad \bar p -\mu \Delta\u) + (\mu/\rho)\Delta \vort . 
\end{equation}
In the case when a fluid is ideal 
and has either constant density \eqref{eq:constdens} 
or barotropic pressure \eqref{eq:barotropic}, 
the velocity and vorticity equations \eqref{eq:NS:vel:vortform} and \eqref{eq:NS:vort}
simplify to the form 
\begin{equation}\label{eq:Eul:vel}
\u_t +\vort\times\u +\grad( \tfrac{1}{2}|\u|^2 + e(\rho)+ p/\rho ) =0
\end{equation}
and 
\begin{equation}\label{eq:Eul:vort}
\vort_t +\curl(\vort\times\u) =0 . 
\end{equation}
Here $e(\rho)$ is the local internal energy density of the fluid/gas,
which is defined through the thermodynamic relation \eqref{eq:thermo}
in the case of adiabatic processes, $\delta S=0$. 
In particular, $e(\rho)=\const$ holds in the constant-density case, 
and $e(\rho)= \int (p(\rho)/\rho^2)d\rho$ holds in the barotropic case. 

A gas/fluid is \emph{irrotational} when there is no vorticity, $\vort=0$. 

The motion of a gas/fluid is a \emph{Beltrami flow} when the vorticity is parallel to the velocity, $\vort\times\u=0$. 

\subsection{Electromagnetism}\label{sec:EM}

The microscopic Maxwell's equations governing 
the electric and magnetic fields $\E(t,\vec{x})$ and $\B(t,\vec{x})$
are given by (in Gaussian units)
\begin{subequations} \label{eq:EM}
\begin{align}
& \E_t-c \curl \B = - 4\pi \J,
\quad
\div \E =4\pi \rho , 
\label{eq:EM:E}
\\
& \B_t+c \curl \E=0,
\quad 
\div \B =0,
\label{eq:EM:B}
\end{align}
\end{subequations}
where $\rho(t,\vec{x})$ is the electric charge density 
and $\J(t,\vec{x})$ is the electric current density. 
Here $c$ denotes the speed of light in vacuum.

In this system \eqref{eq:EM}, 
the charge density $\rho(t,\vec{x})$ is a specified scalar function,
and the current density $\J(t,\vec{x})$ is a specified vector function,
which are related by the continuity equation
\begin{equation}\label{eq:EM:chargecurrent}
\rho_t + \div \,\J =0 . 
\end{equation}
Mathematically, this expresses that the evolution equations for $\E$ and $\B$ 
are each compatible with the divergence equations for $\E$ and $\B$. 

Maxwell's equations \emph{in vacuum} arise when there is 
no charge $\rho=0$ and no current $\J=0$:
\begin{subequations}\label{eq:EM:vac}
\begin{align}
& \E_t-c \curl \B = 0, 
\quad
\div \E = 0 , 
\label{eq:EM:vac:E}
\\
& \B_t+c \curl \E=0,
\quad 
\div \B =0 .
\label{eq:EM:vac:B}
\end{align}
\end{subequations}

\subsection{Magnetohydrodynamics equations}\label{sec:MHD}

The dynamics of plasmas and liquid metals 
in the simplest physical situations that include diffusivity 
are governed by the resistive magnetohydrodynamics (MHD) equations: 
\begin{subequations} \label{eq:MHD}
\begin{align}
& \rho_t + \div(\rho \u) = 0,
\label{eq:MHD:mass}
\\
& \rho (\u_t+ (\vec{u}\cdot \grad)\u) = \J\times \B  -\grad\,p + \mu\, \Delta\u,
\label{eq:MHD:mom}
\\
& \B_t=\curl\,(\u \times \B)+\tfrac{\eta}{\mu_0}\,\Delta \B, 
\label{eq:MHD:B}
\\
& \mu_0\J=\curl \B , 
\label{eq:MHD:J}
\\
& \div \B = 0 
\label{eq:MHD:divB} 
\end{align}
\end{subequations}
for the density $\rho(t,\vec{x})$, velocity $\u(t,\vec{x})$, 
hydrostatic pressure $p(t,\vec{x})$, 
electric current $\J(t,\vec{x})$,
and magnetic field $\B(t,\vec{x})$,
where the constants $\mu_0$, $\mu$, $\eta$ 
are, respectively, 
the magnetic permeability of free space, 
the plasma/liquid-metal viscosity coefficient, 
and the resistivity coefficient. 
The electric field is given by Ohm's law
\begin{equation}\label{eq:MHD:E:Ohm}
\E+\u\times\B=\eta\J. 
\end{equation}
It is useful to note that the magnetic field equation \eqref{eq:MHD:B} 
can be written in the curl form 
\begin{equation}\label{eq:MHD:B:induct}
\B_t=\curl\,(\u \times \B -\tfrac{\eta}{\mu_0}\curl \B) ,
\end{equation}
due to the identity $\curl\,(\curl \B)=\grad(\div\B)-\Delta\B$. 

In addition to these equations, 
an equation of state involving either $\u$, $\rho$, or $p$ 
must be specified to obtain a closed system of equations 
and to model particular physical properties of a plasma/liquid metal. 
The two most common equations of state considered in magnetohydrodynamics 
are given by locally adiabatic flow \eqref{eq:isentropic}
and incompressible flow \eqref{eq:incompr}. 

A plasma/liquid is \emph{inviscid} when there is no fluid viscosity, $\mu=0$,
and is \emph{ideal} when there is no magnetic viscosity, $\eta=0$. 

In incompressible liquid metals, 
the mass equation \eqref{eq:MHD:mass} reduces to a transport equation \eqref{eq:denstransport} for the density. 
Additionally, 
the compatibility between the incompressibility equation \eqref{eq:incompr} 
and the momentum equation \eqref{eq:MHD:mom} 
yields a Laplace-type equation for the pressure. 
In particular, when the liquid is inviscid, the pressure is determined by 
\begin{equation}\label{eq:MHD:press}
\div((1/\rho)\grad p) = -(\nabla\u)\cdot(\nabla\u)^\t + \div((1/\rho)\J\times \B)
\end{equation}
(where ``$\t$'' denotes the matrix transpose in Cartesian coordinates).

\subsection{Preliminaries} 

In three spatial dimensions, 
a general PDE system $\pdesys[\depvars]=0$ can be written as 
\begin{equation}\label{3Dpde}
\pde^a(t, \vec{x}, \depvars, \p\depvars,\ldots,\p^{N_a}\depvars) =0,
\quad
a=1,\ldots,M
\end{equation}
consisting of $M\geq 1$ equations 
with independent variables $t$, $\vec{x}=(x^1, x^2, x^3)$,
and dependent variables $\depvars=\depvars(t,\vec{x}) = (\depvar^1(t,\vec{x}),\ldots,\depvar^m(t,\vec{x}))$, $m \geq 1$,
where $N_a\geq 1$ denotes the differential order of the $a^\th$ equation.
In a given physical system, 
the PDEs \eqref{3Dpde} will comprise all of 
the evolution equations and the spatial constraint equations, if any, 
on the dependent variables,
as well as any compatibility conditions among these equations. 
Any equations of state and any constitutive relations 
will be assumed to have been substituted into the PDEs. 
To illustrate this formalism, 
the dependent variables $\depvars$ and the equations $\pdesys[\depvars]$
for each of the physical systems in Sections~\ref{sec:fluiddyn} to~\ref{sec:MHD}
are shown in table~\ref{table:systems}. 

\begin{table}[htb]
\caption{Examples of physical PDE systems}
\label{table:systems}
\centering
\begin{tabular}{c|c|c|c}
\hline
Physical system 
& Variables $\depvars$
& PDE expressions $\pdesys[\depvars]$
& \tbox{Eqns. of state/}{Constitutive eqns.}
\\
\hline
\hline
gas dynamics
& 
$\rho,\ p,\ \u$
& 
$\begin{aligned}
& \rho_t +\div(\rho\u),\\
& p_t + \u\cdot\grad p +\rho c^2\div\u,\\
& \u_t + (\u\cdot\nabla)\u +(1/\rho)\grad p 
\end{aligned}$
&
$c=c(p,\rho)$
\\
\hline
\tbox{compressible}{fluid flow}
& 
$\rho,\ \u$
& 
$\begin{aligned}
& \rho_t +\div(\rho\u),\\
& \u_t + (\u\cdot\nabla)\u +(1/\rho)\grad p +\mu\Delta\u
\end{aligned}$
&
$p=p(\rho), p(\rho,S)$
\\
\hline
\tbox{incompressible}{ideal fluid flow}
& 
$\rho,\ \u,\ p$
& 
$\begin{aligned}
& \rho_t +\u\cdot\grad \rho,
\quad 
\div\u,\\
& \u_t + (\u\cdot\nabla)\u +(1/\rho)\grad p,\\
& \div((1/\rho)\grad p) +(\grad\u)\cdot(\grad\u)^\t 
\end{aligned}$
&
\\
\hline
electromagnetism
& 
$\E,\ \B$
& 
$\begin{aligned}
& \E_t-c \curl \B +4\pi \J,
\quad
\div \E -4\pi \rho , \\
& \B_t+c \curl \E,
\quad 
\div \B ,
\end{aligned}$
& 
$\begin{aligned}
& \rho(t,\vec{x}), 
\quad
\J(t,\vec{x}),\\
& \rho_t + \div \,\J =0
\end{aligned}$
\\
\hline
MHD
& 
$\rho,\ \u,\ \B$
& 
$\begin{aligned}
& \rho_t + \div(\rho \u), \\
& \rho (\u_t+ (\vec{u}\cdot \grad)\u) - \J\times \B \\
&\quad +\grad\,p -\mu\, \Delta\u,\\
& \B_t -\curl\,(\u \times \B)+\eta\,\Delta \B,
\quad \div \B 
\end{aligned}$
& 
$\begin{aligned}
& p=p(\rho),\\
& \J=\tfrac{1}{\mu_0}\curl \B 
\end{aligned}$
\\
\hline
\tbox{incompressible}{inviscid MHD}
& 
$\rho,\ \u,\ p,\ \B$
&
$\begin{aligned}
& \rho_t + \div(\rho \u), 
\quad
\div\u ,\\
& \rho (\u_t+ (\vec{u}\cdot \grad)\u) - \J\times \B +\grad\,p,\\
& \B_t -\curl\,(\u \times \B)+\eta\,\Delta \B,
\quad \div \B ,\\
& \div((1/\rho)\grad p) +(\grad\u)\cdot(\grad\u)^\t \\
&\quad -\div((1/\rho)\J\times \B)
\end{aligned}$
&
$\J=\tfrac{1}{\mu_0}\curl \B$
\\
\hline
\end{tabular}
\end{table}

The \emph{solution space} of a PDE system \eqref{3Dpde} is the set $\Esp$ 
consisting of all functions $\depvars(t,\vec{x})$ that satisfy each equation in the system. 
Generally, we will consider PDE systems defined on all of $\Rnum^3$. 
Our presentation and results can be easily adjusted to hold on any physically reasonable spatial domain $\Omega\subset\Rnum^3$, 
including any physical boundary conditions posed on solutions $\depvars(t,\vec{x})$. 

Before proceeding to discuss conservation laws, 
we will summarize some notation and preliminaries 
that will be essential for the sequel. 

It is useful to work in the coordinate space 
$(t,\vec{x},\depvars,\p\depvars,\p^2\depvars,\ldots)$, 
called the \emph{jet space},
which is associated with the independent and dependent variables of the PDE system \eqref{3Dpde}.
Here and throughout, 
\[
\p\depvars=(\p_t\depvars,\p_{x^1}\depvars,\p_{x^2}\depvars,\p_{x^3}\depvars)
\]
denotes the set of all first-order partial derivatives of the dependent variables,
and similarly, $\p^k \depvars$, $k\geq 2$, 
denotes the set of all $k$th-order partial derivatives of $\depvars$. 
When $(x^1,x^2,x^3)$ are Cartesian coordinates, 
note 
\[
(\p_{x^1},\p_{x^2},\p_{x^3}) =\nabla =\grad
\] 
is the standard gradient derivative operator,
and
\[
\begin{pmatrix} 
\p_{x^1}^2 & \p_{x^1}\p_{x^2} & \p_{x^1}\p_{x^3} \\
\p_{x^2}\p_{x^1} & \p_{x^2}^2 & \p_{x^2}\p_{x^3} \\
\p_{x^3}\p_{x^1} & \p_{x^3}\p_{x^2} & \p_{x^3}^2 
\end{pmatrix}
= \nabla\otimes\nabla ={\grad}^2 
\] 
is the standard Hessian matrix derivative operator. 
The trace of this matrix operator yields the Laplacian 
\[
\p_{x^1}^2 + \p_{x^2}^2 + \p_{x^3}^2 =\Delta = \div\cdot\grad . 
\]
where $\div=\nabla\cdot$ is the standard divergence operator. 

In jet space, 
smooth functions will be denoted 
\[
f[\depvars]\equiv f(t,\vec{x},\depvars,\p\depvars,\p^2\depvars,\ldots,\p^n\depvars),
\]
where $n\geq 0$ is the maximum order of derivatives of $\depvars$ that appear in the function. 
Such functions are called \emph{differential functions}. 
(Note that a specific differential function need not depend on all partial derivatives of $\depvars$ of order less than or equal to $n$.)
Derivatives of differential functions are defined by total derivatives 
with respect to $t,x^1,x^2,x^3$ acting by the chain rule. 
These derivatives are denoted 
$D=(D_t,\vec{D}_x)$ and $\vec{D}_x=(D_{1},D_{2},D_{3})$. 
Higher total derivatives, $D^k$, $k\geq 2$, are defined in a similar way. 
The variational derivative $\delta/\delta\depvars = (\delta/\delta\depvars^1,\ldots,\delta/\delta\depvars^m)$ (Euler operator) 
with respect to the set of dependent variables $\depvars$ 
is defined in terms of these total derivatives in the standard way. 
Coordinate expressions for the variational derivative, total derivatives, 
as well as for total derivative counterparts of the $\div$, $\curl$, $\grad$ operators, 
will be listed at the end of this subsection. 

In any PDE system \eqref{3Dpde}, 
each PDE itself is given by the vanishing of a differential function;
in particular, $\pde^a[\depvars]=0$ is the $a^\th$ PDE. 
Other important occurrences of differential functions will be the physical densities and fluxes that appear in conservation laws. 

The solution space $\Esp$ of a given PDE system \eqref{3Dpde} 
is represented in jet space by the infinite set of equations 
$\{\pdesys=0,D\pdesys=0,D^2\pdesys=0,\ldots\}$
whenever the PDE system is locally solvable \cite{Olv-book}. 
For finding and verifying conservation laws, 
it is important in practice to have a coordinatization of this set 
in terms of jet-space coordinates. 
We will assume, firstly, that a set of leading derivatives of $\depvars$ 
can be chosen so that each PDE in the system can be put into a solved form
where none of the leading derivatives and none of their differential consequences 
appear on the right hand sides of the solved-form system.
Secondly, we will assume that the differential consequences of each PDE in the system 
can be expressed in an analogous solved form
for the differential consequences of the leading derivatives. 
PDE systems that satisfy these two conditions are known as \emph{regular systems} \cite{Anc-review,AncKar}
and they encompass the PDE systems commonly arising in physical applications,
including those for fluid and gas dynamics, electromagnetism, magnetohydrodynamics 
(cf. Sections~\ref{sec:fluiddyn} to~\ref{sec:MHD}). 
For a regular PDE system, 
the set of equations given by the solved-form PDEs and their solved-form differential consequences 
provide an explicit coordinatization of the solution space 
\[
\Esp=\{\pdesys=0,D\pdesys=0,D^2\pdesys=0,\ldots\}
\subset (t,\vec{x},\depvars,\p\depvars,\p^2\depvars,\ldots)
\]
as a subspace in jet space. 

Given a differential function $f[\depvars]$, 
its evaluation on the solution space $\Esp$ of a PDE system \eqref{3Dpde} 
will be denoted by $f|_\Esp$. 
This notation has the concrete meaning that, in jet space, 
$f|_\Esp$ is given by substituting a solved form of the PDE system (and its differential consequences) into $f$. 

For an example, 
in gas dynamics (cf. section~\ref{sec:fluiddyn}),
the PDE system \eqref{eq:mass}, \eqref{eq:Eul:mom}, \eqref{eq:gas:press}, \eqref{eq:gas:speed} for the dependent variables $(\rho,p,\u)$ 
is an evolution system. 
The simplest choice of a set of leading derivatives consists of the time derivatives 
$\{\p_t\rho,\p_t p,\p_t\u\}$ 
for which this system has the explicit solved form 
\[
\p_t\rho =-\div(\rho\u),
\quad
\p_t p  = -\u\cdot\grad p -\rho\, c(p,\rho)^2\div\u,
\quad
\p_t\u = - (\u\cdot\nabla)\u -(1/\rho)\grad p . 
\]
This set of equations along with their differential consequences with respect to $t,\vec{x}$
comprise the solution space $\Esp$. 
Using this solved form for the PDE system, 
we have that 
\[
\begin{aligned}
& f(t,\vec{x},\rho,p,\u,\p_t\rho,\p_t p,\p_t\u,\grad\rho,\grad p,\grad\u,\ldots)|_\Esp 
\\&\quad
= f(t,\vec{x},\rho,\u,-\div(\rho\u),-(\u\cdot\nabla)\u -(1/\rho)\grad p,-\u\cdot\grad p -\rho\, c(p,\rho)^2\div\u,
\\&\qquad\quad
\grad\rho,\grad p,\grad\u,\ldots)
\end{aligned}
\]
for any differential function $f$. 
There are other possible choices of a set of leading derivatives. 
For instance, in Cartesian coordinates $\vec{x}=(x,y,z)$, 
any one of the sets of first-order spatial coordinate derivatives 
$\{\p_x\rho,\p_x p,\p_x\u\}$, $\{\p_y\rho,\p_y p,\p_y\u\}$, $\{\p_z\rho,\p_z p,\p_z\u\}$ 
can be used. 

As a more involved example, 
in constant-density ideal fluid flow (cf. section~\ref{sec:fluiddyn}),
the pressure and velocity are the dependent variables $(p,\u)$, 
while the density $\rho$ is a constant. 
The PDE system \eqref{eq:Eul:mom}, \eqref{eq:incompr}, \eqref{eq:Eul:press}
governing these variables is not an evolution system, 
and consequently the time derivatives $\{\p_t p,\p_t\u\}$ 
do not provide a set of leading derivatives. 
A leading derivative for the pressure equation \eqref{eq:Eul:press} is given by 
any one of the second-order spatial coordinate derivatives 
$\p_x^2 p,\p_y^2 p,\p_z^2 p$ in Cartesian coordinates $\vec{x}=(x,y,z)$. 
Similarly, 
any one of the first-order spatial coordinate derivatives 
$\p_xu^1,\p_yu^2,\p_zu^3$ provides a leading derivative for the divergence equation \eqref{eq:incompr}. 
For the velocity equation \eqref{eq:Eul:mom}, 
any one of the sets of leading derivatives shown in the gas dynamics example 
can be used, 
other than the leading derivative chosen for the divergence equation. 
For instance, 
$\{\p_x u^1,\p_x^2 p,\p_t\u\}$ provides a set of leading derivatives,
where the PDEs have the explicit solved form 
\[
\begin{aligned}
\p_x u^1 & = -\p_y u^2 -\p_z u^3,
\\
\p_x^2 p & = -\p_y^2 p -\p_z^2 p +2\big( (\p_y u^2)^2 +(\p_z u^z)^2 +(\p_y u^2)(\p_z u^3)
\\&\qquad
+(\p_x u^2)(\p_y u^1)+(\p_y u^3)(\p_z u^y)+(\p_z u^1)(\p_x u^3) \big),
\\
\p_t u^1 & = u^1(\p_y u^2 +\p_z u^3) -u^2\p_y u^1-u^3\p_z u^1 -(1/\rho)\grad p,
\\
\p_t u^2 & = - u^1\p_x u^2 -u^2\p_y u^2 -u^3\p_z u^2 - (1/\rho)\p_y p,
\\
\p_t u^3 & = - u^1\p_x u^3 -u^2\p_y u^3 -u^3\p_z u^3 - (1/\rho)\p_z p.
\end{aligned}
\]
This set of equations along with their differential consequences with respect to $t,x,y,z$
comprise the solution space $\Esp$ of the PDE system.

The notion of evaluating a differential function on the solution space $\Esp$ of a given PDE system 
has an extension to comparing when two differential functions agree on $\Esp$. 
Specifically, 
suppose that two differential functions $f_1[\depvars]$ and $f_2[\depvars]$ 
are not identically equal but satisfy $(f_1[\depvars]-f_2[\depvars])|_\Esp=0$ 
when they are evaluated on $\Esp$. 
Then we say that these functions are \emph{equivalent on solutions} of the PDE system \eqref{3Dpde} and we correspondingly write
\[
f_1[\depvars]|_\Esp = f_2[\depvars]|_\Esp .
\]
Note that we must use the same choice of leading derivatives for evaluating 
both functions. 

An important consideration will be to characterize all differential functions that 
vanish on the solution space $\Esp$ of a given PDE system. 
The following result is known as Hadamard's lemma \cite{Nes-book,Anc-review}.

\begin{lemma}\label{Hadamard}
Let $f[\depvars]$ be a (smooth) differential function satisfying $f[\depvars]|_\Esp=0$
for a PDE system \eqref{3Dpde}. 
Then $f[\depvars]$ is identically equal to a linear combination of the PDEs in the system and their differential consequences,
if the PDE system is regular 
(namely, the system and its differential consequences 
have a solved form with respect to a set of leading derivatives). 
\end{lemma}

More details related to jet spaces, differential functions, the solution space of a PDE system, local solvability, regularity, and Hadamard's lemma, 
are provided in Appendix~\ref{sec:jetspace}. 

To conclude these preliminaries, 
we will now state coordinate expressions for various operators in jet space. 
We begin with total derivatives with respect to $t$ and $x^i$: 
\begin{equation}\label{eq:totD_t}
D_t = \dfrac{\partial}{\partial t} + \depvar^\alpha_{t}\dfrac{\partial}{\partial \depvar^\alpha} + \depvar^\alpha_{tt}\dfrac{\partial}{\partial \depvar^\alpha_{t}} 
+ \depvar^\alpha_{tx^i}\dfrac{\partial}{\partial \depvar^\alpha_{x^i}} +\cdots
\end{equation}
and
\begin{equation}\label{eq:totD_i}
D_{i} = \dfrac{\partial}{\partial x^i} + \depvar^\alpha_{x^i}\dfrac{\partial}{\partial \depvar^\alpha} + \depvar^\alpha_{tx^i}\dfrac{\partial}{\partial \depvar^\alpha_{t}} 
+ \depvar^\alpha_{x^ix^j}\dfrac{\partial}{\partial \depvar^\alpha_{x^j}} +\cdots,
\quad i=1,2,3,
\end{equation}
where the convention of summing over repeated indices is assumed. 
Next we write down the total derivative counterparts of $\div$, $\curl$, $\grad$ operators
when $x^i$, $i=1,2,3$, are Cartesian coordinates. 
The total gradient operator is given by 
\begin{equation}
\Grad f[\depvars]=\vec{D}_x f[\depvars] = (D_{1}f[\depvars],D_{2}f[\depvars],D_{3}f[\depvars])
\end{equation}
acting on scalar differential functions $f[\depvars]$;
the total divergence and curl operators are given by 
\begin{subequations}
\begin{align}
& \Div\vec{f}[\depvars]=\vec{D}_x\cdot\vec{f}[\depvars]
=D_{1}f^1[\depvars] + D_{2}f^2[\depvars]+D_{3}f^3[\depvars],
\\
& \Curl\vec{f}[\depvars]=\vec{D}_x\times\vec{f}[\depvars]
= (D_{2}f^3[\depvars] -D_{3}f^2[\depvars], D_{3}f^1[\depvars] -D_{1}f^3[\depvars], 
D_{1}f^2[\depvars] -D_{2}f^1[\depvars])
\end{align}
\end{subequations}
acting on vector differential functions $\vec{f}[\depvars]=(f^1[\depvars],f^2[\depvars],f^3[\depvars])$. 

All of the vector calculus identities relating ``$\div$'', ``$\curl$'', and ``$\grad$''
hold for their total derivative counterparts:
\begin{equation}\label{divcurlgrad-idents}
\Div\Curl =0,
\qquad
\Curl\Grad = 0 .
\end{equation}
Moreover, the analog of Poincar\'e's lemma in $\Rnum^3$ holds. 

\begin{lemma}\label{divcurlgrad-props}
(i) A differential vector function is a total gradient 
$\vec{f}[\depvars]= \Grad\,F[\depvars]$ 
for some differential function $F[\depvars]$ 
if and only if $\Curl\,\vec{f}[\depvars]= 0$ vanishes identically. 
(ii) A differential vector function is a total curl 
$\vec{f}[\depvars]= \Curl\,\vec{F}[\depvars]$ 
for some differential vector function $\vec{F}[\depvars]$ 
if and only if $\Div\,\vec{f}[\depvars]= 0$ vanishes identically. 
\end{lemma}

Finally, we write down the variational derivative:
\begin{equation}\label{eulerop}
\frac{\delta}{\delta\depvars^\alpha} = 
\frac{\p}{\p\depvars^\alpha} 
- D_t \frac{\p}{\p\depvars^\alpha_t} - D_{i} \frac{\p}{\p\depvars^\alpha_{x^i}} 
+ D_t^2 \frac{\p}{\p\depvars^\alpha_{tt}} +D_tD_{i} \frac{\p}{\p\depvars^\alpha_{tx^i}} 
+D_{i}D_{j} \frac{\p}{\p\depvars^\alpha_{x^ix^j}} 
+ \cdots,
\quad
\alpha=1,\ldots,m. 
\end{equation}
This operator is also known as the Euler operator, which will be denoted $\Eop_{\depvars^\alpha}$. 
It has a purely spatial version given by 
\begin{equation}\label{spatial-eulerop}
\hat \Eop_{\depvars^\alpha} = 
\frac{\p}{\p\depvars^\alpha} 
- D_{i} \frac{\p}{\p\depvars^\alpha_{x^i}} 
+D_{i}D_{j} \frac{\p}{\p\depvars^\alpha_{x^ix^j}} 
+ \cdots,
\quad
\alpha=1,\ldots,m. 
\end{equation}

An important property is that the kernel of the Euler operator consists of 
differential functions that have the form of a total divergence with respect to $t,\vec{x}$. 
Likewise, 
the kernel of the spatial Euler operator consists of 
differential functions given by a total spatial divergence. 

\begin{lemma}\label{Eulerop-props}
(i) $\Eop_{\depvars}(f[\depvars]) =0$ vanishes identically if and only if 
$f[\depvars] = D_t\,\Phi^t[\depvars] + \Div\,\vec{\Phi}[\depvars]$
holds for some scalar differential function $\Phi^t[\depvars]$ 
and vector differential function $\vec{\Phi}[\depvars]=(\Phi^1[\depvars],\Phi^2[\depvars],\Phi^3[\depvars])$. 
(ii) $\hat\Eop_{\depvars}(f[\depvars]) =0$ vanishes identically if and only if 
$f[\depvars] = \Div\,\vec{\Phi}[\depvars]$
holds for some vector differential function $\vec{\Phi}[\depvars]=(\Phi^1[\depvars],\Phi^2[\depvars],\Phi^3[\depvars])$. 
\end{lemma}

\section{Global and local conservation laws}\label{sec:CLs}

We will now discuss the definition and physical meaning of 
three-dimensional local and global conservation laws of dynamical type and topological type 
formulated on volume domains, surface domains, and curve domains. 
For each different type of conservation law, 
we will mention a list of physical examples.

\subsection{Volumetric conservation laws} \label{sec:3D:vol}

A \emph{global volumetric conservation law} of a PDE system \eqref{3Dpde}
in $\pdedom$ is an integral equation of the form
\begin{equation}\label{3Dglobalvolconslaw}
\frac{d}{dt} \int_{\V} \volT[\depvars]\, \dV = -\oint_{\p\V} \volX[\depvars]\cdot\dS
\end{equation}
holding for all solutions $\depvars(t, \vec{x})$ of the system \eqref{3Dpde}. 
Here $\V \subseteq\pdedom$ is a closed volume having a piecewise-smooth boundary surface $\p\V$, 
with $\nor$ being the outward unit normal normal vector 
and $\dS=\nor\,dA$ being the surface element. 
The scalar differential function $\volT[\depvars]$ is called the \emph{conserved density}, 
and the vector differential function $\volX[\depvars]$ is called
the \emph{spatial flux} for the conservation law \eqref{3Dglobalvolconslaw}. 
The density-flux pair
\begin{equation}\label{3Dvolcurrent}
(\volT[\depvars],\volX[\depvars])={\bs\Phi}[\depvars]
\end{equation}
is called a \emph{conserved current}.

For ease of presentation, 
hereafter all volumes $\V\subset\Rnum^3$ 
will be considered to be connected, closed, and have a boundary surface $\p\V$ that is piecewise-smooth. 
Such volumes $\V$ will be called \emph{regular}. 

The physical meaning of a global conservation law \eqref{3Dglobalvolconslaw}
is that the rate of change of the volumetric quantity
\begin{equation}\label{3DvolC}
C[\depvars;\V]= \int_{\V}  \volT[\depvars]\,\dV
\end{equation}
in a regular volume $\V\subseteq\pdedom$ is balanced by the net surface flux
\begin{equation}\label{3Dvolflux}
\flux[\depvars;\p\V] = \oint_{\p\V} \volX[\depvars]\cdot\dS
\end{equation}
escaping through the volume boundary surface $\p\V$,
when $\depvars(t,\vec{x})$ is any solution of the given PDE system. 
In particular,
this balance can be interpreted as an absence of sources or sinks for $\volT[\depvars]$
in the volume $\V$.

A volumetric conservation law \eqref{3Dglobalvolconslaw} that holds 
for all regular volumes $\V$ in $\pdedom$ 
can be formulated as a local continuity equation
\begin{equation}\label{3Dvolconslaw}
D_t \volT[\depvars] + \Div \volX[\depvars] =0
\end{equation}
holding for the space $\Esp$ of all solutions $\depvars(t,\vec{x})$ of the given PDE system. 
This continuity equation \eqref{3Dvolconslaw} is called
a \emph{local volumetric conservation law} of the PDE system \eqref{3Dpde}. 
It is derived from the global conservation law \eqref{3Dglobalvolconslaw} 
by first applying the divergence theorem to the flux integral \eqref{3Dvolflux} to get
$\oint_{\p\V} \volX\cdot\nor \,dA = \int_{\V} \Div\volX\, \dV$,
and then combining this integral and the volumetric integral \eqref{3DvolC} to obtain
$\int_{\V} (D_t \volT +\Div\volX)\, \dV=0$,
which will hold for an arbitrary volume $\V\subseteq\pdedom$ if and only if
$\volT$ and $\volX$ satisfy the local continuity equation \eqref{3Dvolconslaw}. 
Mathematically, this local conservation law \eqref{3Dvolconslaw} is the 
total time-space divergence of the conserved current \eqref{3Dvolcurrent}. 

There are many physical examples of volumetric conservation laws
(cf. Section~\ref{sec:ex:volCLs}):
\begin{itemize}\itemsep=0pt
\item 
mass in fluid flow, gas dynamics, and MHD;
\item 
momentum and angular momentum in fluid flow, gas dynamics, electromagnetism, and MHD;
\item
Galilean momentum in fluid flow, gas dynamics, and MHD;
\item
boost momentum in electromagnetism;
\item
energy in gas dynamics, ideal fluid flow, electromagnetism, and ideal inviscid MHD;
\item
electric charge-current in electromagnetism;
\item
entropy in fluid flow;
\item
helicity in ideal fluid flow;
\item
cross-helicity in ideal inviscid MHD. 
\end{itemize}

\subsection{Surface-flux conservation laws} \label{sec:3D:surf}

Another type of global conservation law
for a PDE system \eqref{3Dpde} in $\pdedom$ 
is a surface-flux equation
\begin{equation}\label{3Dglobalsurfconslaw}
\frac{d}{dt} \int_{\S} \surfT[\depvars]\cdot\dS
= -\oint_{\p\S} \surfX[\depvars]\cdot \ds
\end{equation}
where $\S$ is a connected orientable surface in $\pdedom$
having a piecewise-smooth boundary curve $\p\S$, 
with the line element $\ds = \hat\ell\,ds$ being given in terms of 
the arclength $ds$ and the unit tangent vector $\hat\ell$ along this curve,
and with the surface element $\dS= \nor\,dA$ being given by
the area element $dA$ and the unit normal vector $\nor$ of the surface,
such that $\hat\ell\times\nor$ is pointing outward.
Here $\surfT[\depvars]$ and $\surfX[\depvars]$ are vector differential functions,
which we will respectively refer to as the \emph{conserved flux density}
and the \emph{spatial circulation flux}.
The pair
\begin{equation}\label{3Dsurfcurrent}
(\surfT[\depvars], \surfX[\depvars])={\bs\Phi}[\depvars]
\end{equation}
will be called a \emph{conserved flux current}.

A flux equation \eqref{3Dglobalsurfconslaw} 
holding for all solutions $\depvars(t, \vec{x})$ of the PDE system 
will be called a \emph{global surface-flux conservation law}.
Its physical meaning is that the rate of change of the surface integral quantity
\begin{equation}\label{3DsurfC}
C[\depvars;\S]= \int_{\S} \surfT[\depvars]\cdot\dS
\end{equation}
is balanced by the net flux circulation
\begin{equation}\label{3Dsurfflux}
\flux[\depvars;\p\S] = \oint_{\p\S} \surfX[\depvars]\cdot\ds
\end{equation}
which is trapped around the surface boundary,
for any solution $\depvars(t,\vec{x})$ of the given PDE system. 
When $\S$ is a closed surface, having no boundary $\p\S=\emptyset$,
the quantity \eqref{3DsurfC} defines a time-independent surface flux. 
Conservation in this case can be understood as a consequence of
the flux having nowhere to be trapped.

For ease of presentation, 
hereafter all surfaces $\S\subset\Rnum^3$ 
will be considered to be connected, orientable, and piecewise-smooth.
Such surfaces $\S$ will be called \emph{regular}. 
Consequently, regular surfaces $\S$ that are non-closed 
will have a boundary $\p\S$ that is piecewise smooth. 

A global surface-flux conservation law \eqref{3Dglobalsurfconslaw}
that holds for all regular surfaces $\S\subset \pdedom$
has a formulation as a local vector continuity equation.
First, the surface boundary integral \eqref{3Dsurfflux}
can be converted into a surface integral via Stokes' theorem,
$\oint_{\p\S} \surfX\cdot \ds = \int_{\S} (\Curl\surfX)\cdot\dS$.
Then, this integral can be combined with the surface flux integral \eqref{3DsurfC}
to get
$\int_{\S} (D_t\surfT+ \Curl \surfX)\cdot \dS =0$.
This relation will hold for an arbitrary regular surface $\S\subset\pdedom$ if and only if
$\surfT[\depvars]$ and $\surfX[\depvars]$ satisfy the local continuity equation
\begin{equation}\label{3Dsurfconslaw}
D_t\, \surfT[\depvars] + \Curl\,\surfX[\depvars]  =0,
\end{equation}
holding for all solutions $\depvars(t,\vec{x})$ of the given PDE system.
We will refer to this type of vector continuity equation
as a \emph{local surface-flux conservation law}.

Physical examples of surface-flux conservation laws include 
(cf. Section~\ref{sec:ex:surfCLs}):
\begin{itemize}\itemsep=0pt
\item 
vorticity transport in ideal fluid flow;
\item 
magnetic induction (Faraday's law) in electromagnetism and MHD;
\item
generalized vorticity in electron MHD;
\item
electric displacement current in electromagnetism with a static or vanishing charge distribution.
\end{itemize}
Surface-flux conservation laws also arise in the triviality analysis of
global volumetric conservation laws, as shown in Section~\ref{sec:3D:triv-equiv:volCL}.

\subsection{Circulatory conservation laws} \label{sec:3D:circul}

A third type of global conservation law for a PDE system \eqref{3Dpde} in $\pdedom$ is a line integral equation of the form
\begin{equation}\label{3Dglobalcurvconslaw}
\frac{d}{dt} \int_{\C} \curvT[\depvars]\cdot \ds = -\curvX[\depvars]\Big|_{\p\C}
\end{equation}
in which $\C$ is a piecewise-smooth curve with endpoints $\p\C$.
Here $\curvT[\depvars]$ is a vector differential function,
which we will call the \emph{conserved circulation density},
and $\curvX[\depvars]$ is a scalar differential function, 
which we will call the \emph{spatial endpoint flow}. 
A line integral equation \eqref{3Dglobalcurvconslaw} 
holding for all solutions $\depvars(t, \vec{x})$ of the PDE system 
will be called a \emph{global circulatory conservation law},
and the pair $(\curvT[\depvars],\curvX[\depvars])={\bs\Phi}[\depvars]$
will be called a \emph{conserved circulation current}.

The physical meaning of this type of global conservation law \eqref{3Dglobalcurvconslaw}
is that the rate of change of the line integral quantity
\begin{equation}\label{3DcurvC}
C[\depvars;\C]= \int_{\C} \curvT[\depvars]\cdot\ds
\end{equation}
is balanced by a net flow outward through the two ends of the curve
\begin{equation}\label{3Dcurvflux}
\flux[\depvars;\p\C]= \curvX[\depvars]\Big|_{\p\C} 
\end{equation}
for any solution $\depvars(t,\vec{x})$ of the given PDE system. 
When $\C$ is a closed curve, with no boundary $\p\C=\emptyset$, 
then the line integral quantity \eqref{3DcurvC} 
defines a time-independent total circulation.

For ease of presentation, 
hereafter all curves $\C\subset\Rnum^3$ 
will be considered to be connected and piecewise-smooth.
Such curves $\C$ will be called \emph{regular}. 

A global circulatory conservation law \eqref{3Dglobalcurvconslaw}
that holds for all regular curves $\C\subset \pdedom$ 
can be formulated as a local continuity equation
by applying the fundamental theorem of calculus for line integrals. 
This yields
\begin{equation}\label{3Dcurvconslaw}
D_t\, \curvT[\depvars] + \Grad\,\curvX[\depvars] =0,
\end{equation}
which we will refer to as a \emph{local circulatory conservation law},
holding for all solutions $\depvars(t, \vec{x})$ of the given PDE system.

Some physical examples of circulatory conservation laws are 
(cf. Section~\ref{sec:ex:curvCLs}):
\begin{itemize}\itemsep=0pt
\item 
circulation in ideal fluids with irrotational flow or Beltrami flow;
\item 
density gradient in incompressible fluid flow;
\item
entropy gradient in non-homentropic fluid flow. 
\end{itemize}
Circulatory conservation laws are also shown to arise 
in the triviality analysis of global surface-flux conservation laws in Section \ref{sec:3D:triv-equiv:surfCL}.

\subsection{Topological conservation laws}\label{sec:3D:topCL}

For each of the three types (volumetric, surface-flux, circulatory) of dynamical conservation laws,
there is a corresponding type of time-independent conservation law. 

\subsubsection{Spatial divergence/topological flux conservation laws} 
The first type is given by a local volumetric conservation law \eqref{3Dvolconslaw}
in which the conserved density vanishes, $\volT[\depvars]|_\Esp= 0$,
for all solutions $\depvars(t, \vec{x})$ of a given PDE system \eqref{3Dpde}. 
This yields
\begin{equation}\label{spatial-div-3Dconslaw}
\Div\volX [\depvars]\big|_\Esp= 0,
\end{equation}
which is a purely spatial conservation law
holding on the solution space $\Esp$ of the system. 
We will call equation \eqref{spatial-div-3Dconslaw} 
a \emph{local spatial divergence conservation law}. 
It is sometimes also called a divergence-type conservation law. 
We will refer to the vector differential function $\volX[\depvars]$ 
as the spatial flux vector-density. 

The global form of a  spatial divergence conservation law \eqref{spatial-div-3Dconslaw}
arises by integration of $\Div\volX$ over any regular volume $\V\subseteq\pdedom$. 
After Gauss' theorem is applied to the resulting volume integral,
this yields, for all $t$, a vanishing surface integral 
\begin{equation}\label{spatial-div-3Dglobalconslaw}
\oint_{\S} \volX[\depvars]\cdot\dS\big|_\Esp =0
\end{equation}
on the closed boundary surface(s) $\S=\p\V$, 
where $\dS=\nor\,dA$ is given by the area element $dA$ and the outward unit normal normal vector $\nor$. 
Mathematically, 
this surface integral remains unchanged if $S=\p\V$ is deformed in any continuous way that preserves its topology.  

The physical meaning of a surface integral \eqref{spatial-div-3Dglobalconslaw}
depends on the topological nature of the chosen volume $\V$. 
If the boundary of $\V$ is a single connected surface $\p\V=\S$,
then equation \eqref{spatial-div-3Dglobalconslaw} shows that
the total flux of $\volX[\depvars]$ through $\S$ is zero.
This is due to the absence of sources or sinks, 
which corresponds to the spatial divergence of $\volX[\depvars]$ being zero. 
Hence,
the spatial divergence conservation law \eqref{spatial-div-3Dconslaw}
shows that there is no net flux
\begin{equation}\label{zeroflux:spatial-div-3Dglobalconslaw}
\oint_{\S} \volX[\depvars]\cdot\dS\big|_\Esp =0
\end{equation}
for every closed surface $\S$ that bounds a regular volume in $\pdedom$. 
Alternatively, 
if the boundary $\p\V$ consists of two disjoint surfaces $\S_1$ and $\S_2$, 
then the spatial divergence conservation law \eqref{spatial-div-3Dglobalconslaw}
shows that the total flux of $\volX[\depvars]$ is the same through both surfaces.
Hence,
for any two non-intersecting closed surfaces $\S_1$ and $\S_2$ 
that bound a regular volume $\V\subseteq\pdedom$, 
flux conservation holds:
\begin{equation}\label{topological-3Dglobalconslaw}
\oint_{\S_1} \volX[\depvars]\cdot\dS\big|_\Esp = \oint_{\S_2} \volX[\depvars]\cdot\dS\big|_\Esp
\end{equation}
where the unit normal vectors of $\S_1$ and $\S_2$ are chosen so that 
one is inward-directed and the other is outward-directed with respect to the volume $\V$.
Each of these surface flux integrals will be equal to zero 
due to the spatial divergence conservation law \eqref{spatial-div-3Dconslaw},
unless sources or sinks are present outside of the volume $\V$. 
In particular, if equation \eqref{spatial-div-3Dconslaw} holds only inside $\V$,
then the two surface integrals can be non-zero. 
Both of these surface integrals remain unchanged 
if $\S_1$ and $S_2$ are deformed in any continuous way that preserves their topology.

For the reasons just stated, 
global surface flux integrals of the form \eqref{zeroflux:spatial-div-3Dglobalconslaw} and \eqref{topological-3Dglobalconslaw} 
are usually referred to as \emph{topological flux conservation laws}.
They arise generally in time-independent PDE systems 
containing spatial divergence equations in $\pdedom$. 
They also arise in dynamical PDE systems where
divergence-type constraint equations hold throughout the spatial domain $\pdedom$. 

Physical examples of topological flux conservation laws 
(cf. Section~\ref{sec:ex:divCLs}) are given by:
\begin{itemize}\itemsep=0pt
\item 
vorticity in fluid flow, gas dynamics, and MHD; 
\item 
streamline flux and pressure-gradient flux in incompressible fluid flow and incompressible MHD;
\item
magnetic flux in electromagnetism and MHD;
\item
electric flux in vacuum electromagnetism. 
\end{itemize}

\subsubsection{Spatial curl/topological circulation conservation laws}
A different type of time-independent conservation law 
is given by a local surface-flux conservation law \eqref{3Dsurfconslaw}
in which the flux vector-density vanishes, $\surfT[\depvars]|_\Esp= 0$,
for all solutions $\depvars(t,\vec{x})$ of a given PDE system \eqref{3Dpde} in $\pdedom$. 
This yields
\begin{equation}\label{spatial-curl-3Dconslaw}
\Curl\surfX[\depvars]|_\Esp = 0,
\end{equation}
which we will call a \emph{local spatial curl conservation law}.
We will refer to $\surfX[\depvars]$ as the spatial circulation vector-density. 

The global form of a curl conservation law \eqref{spatial-curl-3Dconslaw}
arises by integration of $\Curl\surfX[\depvars]$ over any non-closed regular surface $\S \subset\pdedom$. 
This yields $\int_{\S} \Curl\surfX[\depvars]\cdot\dS\big|_\Esp=0$
which by Stokes' theorem becomes a vanishing line integral 
around the closed boundary curve(s) $\C=\p\S$, 
\begin{equation}\label{spatial-curl-3Dglobalconslaw}
\int_{\C}  \surfX[\depvars] \cdot \ds\big|_\Esp =0
\end{equation}
holding for all $t$. 
This line integral remains unchanged if $C=\p\S$ is deformed in any continuous way.

The physical meaning of a vanishing line integral \eqref{spatial-curl-3Dglobalconslaw}
depends on the topological nature of the chosen surface $\S$ in $\pdedom$.
If a non-closed regular surface is simply connected, 
then its boundary $\p\S$ consists of a single closed curve $\C$,
in which case equation \eqref{spatial-curl-3Dglobalconslaw} shows that
the net circulation of $\surfX[\depvars]$ around this curve vanishes.
This can be understood as the absence of vorticity of the circulation flux vector $\surfX[\depvars]$. 
In particular, 
the spatial curl conservation law \eqref{spatial-curl-3Dglobalconslaw} 
shows that the circulation flux vector $\surfX[\depvars]$ is irrotational,
whereby there is no net circulation 
\begin{equation}\label{zerocircflux-3Dglobalconslaw}
\oint_{\C}  \surfX[\depvars]\cdot \ds\,\big|_\Esp =0
\end{equation}
around every closed curve $\C$ in $\pdedom$. 

Alternatively, if a non-closed regular surface $\S$ is not simply connected,
then its boundary $\p\S$ consists of two disjoint closed curves.
In this case, 
the spatial curl conservation law \eqref{spatial-curl-3Dglobalconslaw}
shows that the total circulation of $\surfX[\depvars]$ 
around both curves is the same.
Consequently, 
for any two non-intersecting closed curves $\C_1$ and $\C_2$ 
that are the boundary of a non-simply connected surface $\S\subseteq\pdedom$, 
conservation of circulation holds:
\begin{equation}\label{topological-circ-3Dglobalconslaw}
\oint_{\C_1} \surfX[\depvars]\cdot\ds\,\big|_\Esp  
= \oint_{\C_2} \surfX[\depvars]\cdot\ds\,\big|_\Esp
\end{equation}
where the unit tangent vectors of $\C_1$ and $\C_2$ are chosen to have
the same clockwise or counterclockwise orientation with respect to the surface $\S$.
Both circulation integrals remain unchanged if $\C_1$ and $C_2$ are deformed in any continuous way that preserves their topology.
Due to the spatial curl conservation law \eqref{spatial-curl-3Dglobalconslaw}
holding throughout $\pdedom$, 
each of these two circulation integrals will be equal to zero. 
If, however, equation \eqref{spatial-curl-3Dglobalconslaw} holds only 
in some connected volume $\V\subset\pdedom$ that does not contain either curve, 
then the two circulation integrals can be non-zero. 

Accordingly, 
line integral equations of the form \eqref{zerocircflux-3Dglobalconslaw} and \eqref{topological-circ-3Dglobalconslaw}
will be called \emph{global topological circulation conservation laws}.
Local spatial curl conservation laws \eqref{spatial-curl-3Dconslaw} 
arise generally in time-independent PDE systems that contain curl equations,
and in dynamical PDE systems where curl-type constraint equations hold 
throughout the spatial domain $\pdedom$. 

Physical examples of topological circulation conservation laws 
(cf. Section~\ref{sec:ex:curlCLs}) are given by:
\begin{itemize}\itemsep=0pt
\item 
circulation in irrotational fluid flow, irrotational gas dynamics, and irrotational MHD;
\item 
magnetic circulation in magnetostatics;
\item
electric field circulation in electrostatics and equilibrium MHD. 
\end{itemize}

\subsubsection{Spatial gradient conservation laws}

Finally, there is a time-independent version of 
local circulatory conservation laws \eqref{3Dcurvconslaw}
in which the circulation vector-density vanishes, $\curvT[\depvars]|_\Esp= 0$,
for all solutions $\depvars(t,\vec{x})$ of a given PDE system in $\pdedom$.
The resulting spatial conservation law
\begin{equation}\label{spatial-grad-3Dconslaw}
\Grad\curvX[\depvars]|_\Esp = 0
\end{equation}
implies that the flux $\curvX[\depvars]$ has no dependence on $\vec{x}$ 
for all solutions $\depvars(t,\vec{x})$ of the PDE system. 
We will call this equation \eqref{spatial-grad-3Dconslaw} 
a \emph{local spatial gradient conservation law}. 
Its corresponding global form \eqref{3Dglobalcurvconslaw} 
asserts that the difference in the value of flux $\curvX[\depvars]$ 
vanishes at any two points in the spatial domain of the PDE system, 
which has exactly the same content as its local form \eqref{spatial-grad-3Dconslaw}. 

Consequently, 
this type of conservation law \eqref{spatial-grad-3Dconslaw} has no direct topological meaning. 

A physical example of a spatial gradient conservation law is Bernoulli's principle (arising in ideal fluid flow when irrotational, equilibrium flows are considered),
as discussed in Section~\ref{sec:ex:gradCLs}.

\subsection{Constants of motion}\label{sec:3D:com}

For any PDE system \eqref{3Dpde} in $\pdedom$, 
a time-dependent global conservation law 
on a given domain $\Omega\subseteq\pdedom$ 
will yield a constant of motion
\begin{equation}\label{com}
\frac{d}{dt}C[\depvars;\Omega]=0
\end{equation}
if and only if the net flux vanishes, $\flux[\depvars;\p\Omega]=0$,
for all solutions $\depvars(t,\vec{x})$ of the system. 

As already noted, 
any surface-flux conservation law on a closed regular surface $\Omega=\S$, 
and any circulatory conservation law on a closed regular curve $\Omega=\C$, 
automatically yields a constant of motion,
since $\flux[\depvars;\p\Omega]\equiv0$ due to the domain having no boundary, 
$\p\Omega=\emptyset$. 

Some physical examples of non-zero constants of motion are given by:
(cf. Sections~\ref{sec:ex:volCLs} and~\ref{sec:ex:surfCLs}):
\begin{itemize}\itemsep=0pt
\item
mass and entropy in gas dynamics and fluid flow confined to a fixed volume;
\item
energy and helicity in ideal, incompressible or barotropic fluid flow confined to a fixed volume;
\item 
cross-helicity in ideal inviscid MHD confined to a fixed volume;
\item
electric flux on closed surfaces 
in electromagnetism with a static charge distribution. 
\end{itemize}

Constants of motion will also arise from global topological conservation laws
on domains that are simply-connected. 
Specifically, such conservation laws have the form 
\begin{equation}\label{topologicalcom}
C[\depvars;\Omega]=0
\end{equation}
with $\p\Omega=\emptyset$. 
Hence the time derivative of $C[\depvars;\Omega]$ automatically vanishes, 
which thereby yields a constant of motion \eqref{com}.
These constants of motion will be called \emph{topological}. 
In particular, 
any topological flux conservation law \eqref{zeroflux:spatial-div-3Dglobalconslaw}
on a closed connected surface $\S=\Omega$ 
represents a zero-flux constant of motion \eqref{topologicalcom},
and any topological circulation conservation law \eqref{zerocircflux-3Dglobalconslaw}
on a closed connected curve $\C=\Omega$ 
represents a zero-circulation constant of motion \eqref{topologicalcom}. 

In general, topological constants of motion will appear whenever a dynamical PDE system 
contains spatial constraint equations that are compatible with all of the evolution equations in the system. 
The PDEs in such systems satisfy differential identities that correspond to 
locally trivial conservation laws related to topological conservation laws. 
More specifically, 
a spatial constraint equation of divergence type 
represents a spatial divergence conservation law \eqref{spatial-div-3Dconslaw}
whose time derivative yields a local conservation law in which spatial flux is zero. 
Spatial constraint equations of curl or gradient type represent, respectively, 
a spatial curl conservation law \eqref{spatial-curl-3Dconslaw}
or a spatial gradient conservation law \eqref{spatial-grad-3Dconslaw}. 
Their time derivatives yield corresponding local conservation laws 
in which, respectively, the spatial circulation flux and spatial endpoint flow is zero. 

Topological constants of motion also represent one type of non-trivial boundary conservation laws, as discussed in the next section. 
Physical examples of these constants of motion (cf. Section~\ref{sec:ex:temporalCLs}) are given by:
\begin{itemize}\itemsep=0pt
\item 
magnetic flux on closed surfaces 
in electromagnetism and MHD;
\item
electric flux on closed surfaces 
in vacuum electromagnetism; 
\item 
circulation on closed curves 
in irrotational ideal fluid flow. 
\end{itemize}

\section{Non-triviality and equivalence of local and global conservation laws}\label{sec:non-triv}

For any PDE system \eqref{3Dpde} in $\Rnum^3$, 
if a conservation law yields a conserved integral that contains no local information 
about the solutions of the PDE system,
then the conservation law will be called \emph{locally trivial}. 
This occurs when (and only when), for an arbitrary solution of the PDE system, 
the density in the conserved integral on a given spatial domain 
either vanishes or has the form of an exact differential 
given by a divergence, or a curl, or a gradient, 
in the respective cases of a volume domain, or a surface domain, or a curve domain.
If two conservation laws differ by a locally trivial conservation law,
they will be regarded as being \emph{locally equivalent}. 
The widely used definitions of ``trivial conservation laws'' and ``equivalent conservation laws'' 
in the literature \cite{Olv-book,BCA-book} 
coincide with this local notion of triviality. 

Nevertheless, a locally trivial conservation law of a PDE system can sometimes 
contain useful global information about the PDE system and its solutions. 
In general, a local conservation law will called \emph{globally trivial} if 
it yields a conserved integral that contains no information 
(either local or global) about the solutions of the PDE system. 
This will always occur in the case when the conserved integral 
has a vanishing density for all solutions of the system,
since then the conserved integral clearly contains no information at all 
about the PDE system and its solutions. 
Thus, this type of locally trivial conservation law is globally trivial. 

In the case when the density in the conserved integral of a locally trivial conservation law 
instead has the form of an exact differential, 
the conserved integral reduces to a conserved boundary integral 
whose triviality is determined by the topological nature of the boundary domain, 
and by whether the given conservation law is dynamical or topological. 

If a given spatial domain is closed, 
then its boundary is empty and hence any boundary integral is identically zero.
In this situation, 
any conserved boundary integral arising from a locally trivial conservation law 
will be globally trivial, since its vanishing is entirely due to the topology of the domain. 
Consequently, 
as all topological conservation laws are formulated on closed spatial domains, 
any topological conservation law that is locally trivial is thereby globally trivial. 
Likewise, any dynamical conservation law that is locally trivial will be globally trivial
on a closed spatial domain. 

In contrast, 
if a given spatial domain is non-closed, 
then any boundary integral arising from a dynamical conservation law that is locally trivial
will not vanish identically. 
In this case, 
if the conserved boundary integral is time-dependent for an arbitrary solution of the PDE system, 
then the dynamical conservation law essentially becomes a mathematical identity 
which has no useful information about the solutions of the PDE system. 
This type of locally trivial dynamical conservation law is therefore globally trivial. 
If instead the conserved boundary integral is time-independent and non-vanishing
for an arbitrary solution of the PDE system,
then this integral is a non-trivial constant of motion 
which corresponds to a globally non-trivial dynamical conservation law 
on a lower-dimensional (boundary) domain. 
In particular, 
the local form of such boundary conservation laws consists of a purely temporal conservation law in which the spatial flux is zero. 

Note that the vanishing of a conserved integral does not itself necessarily imply 
that a conservation law is globally trivial. 
In particular, 
such a conserved integral can be measuring a physical net flux or circulation that 
vanishes due to an absence of sources and sinks of flux or circulation.

In general, 
the conservation laws of primary interest for a given PDE system
are the globally non-trivial ones. 
Note that a non-trivial dynamical conservation law on a given spatial domain 
can be changed by the addition of a topological conservation law on 
the domain boundary,
since this affects only the spatial flux and not the conserved integral. 
Apart from this possibility, 
two non-trivial global conservation laws, whether dynamical or topological, 
will have the same physical content for a PDE system if and only if 
they yield, up to a constant multiple, the same conserved integral 
for all solutions of the system.

We will now give a complete discussion of non-triviality 
for all of the different types of conservation laws in three dimensions. 
In particular, we will formulate necessary and sufficient local conditions
under which a conservation law is locally non-trivial, 
and under which two conservation laws are locally equivalent. 
We will also formulate necessary and sufficient conditions
under which a conservation law is globally non-trivial. 
These formulations provide a substantial improvement of the widely used notion of ``trivial conservation laws'' 
\cite{Olv-book,BCA-book} in the literature. 
We will call a conservation law \emph{non-trivial} 
if and only if it is both locally and globally non-trivial. 
Likewise, we will call a constant of motion \emph{non-trivial} 
if and only if it arises from a non-trivial conservation law. 

We will show in detail how, under certain conditions,  
a locally trivial conservation law on a spatial domain can yield 
a constant of motion given by a non-trivial global conservation law 
on the boundary of the domain. 
In particular, 
non-trivial surface-flux constants of motion arise from 
locally trivial volumetric conservation laws, 
while non-trivial circulatory constants of motion arise from 
locally trivial surface-flux conservation laws.
Moreover, 
we will also show how these kinds of constants of motion 
appear when a dynamical PDE system contains compatible spatial constraint equations. 
Specifically, 
the compatibility between the constraint equations and the evolution equations 
in such a PDE system takes the form of differential identities which correspond to 
locally trivial conservation laws. 

These main results clarify some confusing statements in the literature, 
especially for PDE systems that possess differential identities. 
We will mention physical examples to illustrate each result.

\subsection{Non-triviality of topological conservation laws}\label{sec:3D:triv-equiv:topCL}

The notions of global and local triviality and equivalence 
for the two main kinds of topological conservation laws 
presented in Section~\ref{sec:3D:topCL}
will be formulated first. 
These topological conservation laws will turn out to enter into the subsequent 
formulation of necessary and sufficient local conditions 
for dynamical conservation laws to be locally trivial.

\subsubsection{Spatial divergence/topological flux conservation laws}  
To begin, 
we consider global topological flux conservation laws \eqref{spatial-div-3Dglobalconslaw}
on a surface $\S=\p\V$ that bounds a regular volume $\V$ 
within the spatial domain of a given PDE system \eqref{3Dpde} in $\Rnum^3$.

\begin{definition}\label{defn:3Dspatial-div-conslaw-loctriv}
A topological flux conservation law \eqref{spatial-div-3Dglobalconslaw} of a PDE system \eqref{3Dpde} is called \emph{locally trivial} 
if the flux vector-density $\volX[\depvars]$ 
has the form of a curl
\begin{equation}\label{spatial-div-triv}
\volX[\depvars]|_\Esp = \Curl\vec{\Theta}[\depvars]|_\Esp 
\end{equation}
in terms of a vector differential function $\vec{\Theta}[\depvars]$, 
holding on the solution space $\Esp$ of the given system. 
\end{definition}

By Lemma~\ref{Hadamard}, 
this curl condition \eqref{spatial-div-triv} is equivalent to requiring that $\surfX$ is identically given by
\begin{equation}\label{spatial-div-triv-offsolns}
\volX[\depvars] = \Curl\vec{\Theta}[\depvars] + \vec{\Gamma}_\triv[\depvars]
\end{equation}
for arbitrary (sufficiently smooth) functions $\depvars(t, \vec{x})$,
where $\vec{\Gamma}_\triv$ is any vector differential function 
vanishing on the solution space $\Esp$ of the system:
\begin{equation}\label{triv-vector}
\vec{\Gamma}_\triv[\depvars]|_\Esp =0.
\end{equation}
Note that the flux vector-density \eqref{spatial-div-triv-offsolns}
identically satisfies 
\begin{equation}\label{spatial-div-trivcond}
\Div\volX= \Div \vec{\Gamma}_\triv . 
\end{equation}
Thus, the resulting local conservation law 
$(\Div \volX[\depvars])|_\Esp = (\Div\vec{\Gamma}_\triv[\depvars])|_\Esp =0$
contains neither local nor global information about the PDE system and its solutions. 
In particular, a local conservation law of this form is merely an identity,
and the corresponding conserved flux integral is identically zero by Stokes' theorem, 
\begin{equation}\label{spatial-div-triv-global}
\oint_{\S} \Curl\vec{\Theta}[\depvars]\cdot\dS\, \big|_\Esp \equiv 0 
\end{equation} 
due to the surface $\S=\p\V$ being closed, 
$\p\S=\p^2\V=\emptyset$. 
Moreover, this is the only way that a topological flux conservation law 
can hold as an identity. 

Therefore, we have the following result. 

\begin{proposition}\label{prop:spatial-div-conslaw-globaltriv}
A topological flux conservation law \eqref{spatial-div-3Dglobalconslaw} is 
\emph{globally trivial} for an arbitrary regular surface $\S\subset \pdedom$ 
if and only if it is locally trivial \eqref{spatial-div-triv}. 
\end{proposition}
As a consequence, 
a necessary and sufficient local condition 
for two topological flux conservation laws \eqref{spatial-div-3Dconslaw}
to yield the same global surface-flux integral \eqref{spatial-div-3Dglobalconslaw} 
for any given regular $\V\subset\pdedom$
is that they can differ only by a locally trivial topological flux conservation law \eqref{spatial-div-triv-offsolns}. 
This provides a precise notion of local and global equivalence for topological flux conservation laws. 

\begin{proposition}
Two topological flux conservation laws \eqref{spatial-div-3Dglobalconslaw} 
are \emph{equivalent} if and only if they differ by 
a locally trivial topological flux conservation law \eqref{spatial-div-triv}. 
\end{proposition}

A sufficient condition for triviality can be formulated 
by using the properties of the total divergence operator 
(cf. Lemma~\ref{divcurlgrad-props}). 

\begin{proposition}\label{prop:3Ddivconslaw-triv}
A topological flux conservation law \eqref{spatial-div-3Dglobalconslaw} 
of a PDE system \eqref{3Dpde} in $\Rnum^3$ is trivial 
if its flux density is identically divergence-free 
$\Div\,\volX[\depvars]=0$
off of the solution space of the PDE system. 
\end{proposition}

As will be shown in Section~\ref{sec:ex:divCLs}, 
physical examples of non-trivial topological flux conservation laws
are given by 
streamline flux and pressure-gradient flux in incompressible fluid flow and incompressible MHD;
magnetic flux in electromagnetism and MHD;
electric flux in vacuum electromagnetism. 
One example of a trivial topological flux conservation law 
is connected with vorticity in fluid flow, gas dynamics, and MHD.

\subsubsection{Spatial curl/topological circulation conservation laws} 
We next consider topological circulation conservation laws \eqref{spatial-curl-3Dglobalconslaw} 
on a closed curve $\C=\p\S$ that bounds a non-closed regular surface $\S$ 
within the spatial domain of a given PDE system \eqref{3Dpde} in $\Rnum^3$. 
The discussion of triviality is very similar to that for topological flux conservation laws. 

\begin{definition}\label{defn:3Dspatial-curl-conslaw-loctriv}
A topological circulation conservation law \eqref{spatial-curl-3Dglobalconslaw} of a PDE system \eqref{3Dpde} is called \emph{locally trivial} 
if the circulation vector-density $\surfX[\depvars]$ 
has the form of a gradient 
\begin{equation}\label{spatial-curl-triv}
\surfX[\depvars]|_\Esp = \Grad\,\Theta[\depvars]|_\Esp 
\end{equation}
in terms of a scalar differential function $\Theta[\depvars]$, 
holding on the solution space $\Esp$ of the given system. 
\end{definition}

Lemma~\ref{Hadamard} shows that this gradient condition \eqref{spatial-curl-triv} holds
if and only if $\surfX$ is identically given by
\begin{equation}\label{spatial-curl-triv-offsolns}
\surfX[\depvars] = \Grad\,\Theta[\depvars] + \vec{\Gamma}_\triv[\depvars]
\end{equation}
for arbitrary (sufficiently smooth) functions $\depvars(t,\vec{x})$, 
where $\vec{\Gamma}_\triv$ is any vector differential function
having the property \eqref{triv-vector}. 
A circulation vector-density of this form \eqref{spatial-curl-triv-offsolns}
identically satisfies 
\begin{equation}\label{spatial-curl-trivcond}
\Curl\surfX \equiv \Curl \vec{\Gamma}_\triv, 
\end{equation}
whereby the resulting local conservation law 
$(\Curl \surfX[\depvars])|_\Esp = (\Curl\vec{\Gamma}_\triv[\depvars])|_\Esp =0$
contains neither local nor global information about the PDE system and its solutions. 
Any such local conservation law holds as an identity,
and the corresponding conserved circulation integral is identically zero 
due to the fundamental theorem of calculus for line integrals,
\begin{equation}\label{spatial-curl-triv-global}
\oint_{\C} \Grad\,\Theta[\depvars]|_\Esp\cdot\ds \equiv 0 . 
\end{equation} 
where the curve $\C=\p\S$ is closed, 
$\p\C=\p^2\S=\emptyset$. 
Moreover, this is the only way that a topological circulation conservation law 
can hold as an identity. 

Therefore, we have the following result. 

\begin{proposition}\label{prop:spatial-curl-conslaw-globaltriv}
A topological circulation conservation law \eqref{spatial-curl-3Dglobalconslaw} is 
\emph{globally trivial} for an arbitrary regular curve $\C\subset \pdedom$ 
if and only if it is locally trivial \eqref{spatial-curl-triv}. 
\end{proposition}

Consequently, 
a necessary and sufficient local condition 
for two topological circulation conservation laws \eqref{spatial-curl-3Dconslaw}
to yield the same global circulation-flux integral \eqref{spatial-curl-3Dglobalconslaw} 
for any given non-closed regular surface $\S$ 
is that they can differ only by a locally trivial topological circulation conservation law \eqref{spatial-curl-triv}. 
This provides a precise notion of local and global equivalence for topological circulation conservation laws. 

\begin{proposition}
Two topological circulation conservation laws \eqref{spatial-curl-3Dglobalconslaw} 
are \emph{equivalent} if and only if they differ by 
a locally trivial topological circulation conservation law \eqref{spatial-curl-triv}. 
\end{proposition}

A sufficient condition for triviality can be formulated 
by using the properties of the total curl operator 
(cf Lemma~\ref{divcurlgrad-props}). 

\begin{proposition}\label{prop:3Dglobalcircconslaw-triv}
A topological circulation conservation law \eqref{spatial-curl-3Dglobalconslaw} 
of a PDE system \eqref{3Dpde} in $\Rnum^3$ is trivial 
if its circulation density is identically curl-free 
$\Curl\,\surfX[\depvars] =0$
off of the solution space of the PDE system. 
\end{proposition}

As will be discussed in Section~\ref{sec:ex:curlCLs}, 
physical examples of non-trivial topological circulation conservation laws 
arise from 
circulation in irrotational fluid flow, irrotational gas dynamics, and irrotational MHD;
magnetic circulation in magnetostatics;
and electric field circulation in electrostatics and equilibrium MHD.

\subsection{Non-triviality and equivalence of time-dependent circulatory conservation laws}\label{sec:3D:triv-equiv:curvCL}

We will now formulate the notions of local and global triviality and equivalence
for dynamical conservation laws, 
beginning with time-dependent circulatory conservation laws. 

\begin{definition}\label{defn:3Dcurvconslaw-loctriv}
A time-dependent circulatory conservation law \eqref{3Dcurvconslaw} of a PDE system \eqref{3Dpde} in $\Rnum^3$ is called \emph{locally trivial} 
if, on the solution space $\Esp$ of the system, 
the conserved circulation density $\curvT$ and the spatial endpoint flow $\curvX$
have the respective forms 
\begin{subequations}\label{3Dcurvconslaw-triv}
\begin{align}
& \curvT[\depvars]|_\Esp = \Grad\,\Theta[\depvars]|_\Esp,
\label{3Dcurvconslaw-trivT}\\
& \curvX[\depvars]|_\Esp  = -D_t\,\Theta[\depvars]|_\Esp, 
\label{3Dcurvconslaw-trivX}
\end{align}
\end{subequations}
in terms of a scalar differential function $\Theta[\depvars]$.
Any two time-dependent circulatory conservation laws \eqref{3Dcurvconslaw} that differ 
only by a locally trivial circulatory conservation law are \emph{locally equivalent}.
\end{definition}

By applying Lemma~\ref{Hadamard}, 
we see that any conserved circulation current 
$(\curvT,\curvX)$ of the form \eqref{3Dcurvconslaw-triv} 
can be expressed off of the solution space $\Esp$ of the PDE system 
by the equivalent formulation 
\begin{equation}\label{3Dcurvconslaw-triv-offsolns}
\curvT[\depvars] = \Grad\,\Theta[\depvars] + \vec{\Gamma}_\triv[\depvars], 
\quad
\curvX[\depvars] = -D_t\,\Theta[\depvars]  + \Xi_\triv[\depvars]
\end{equation}
holding identically for all (sufficiently smooth) functions $\depvars(t,\vec{x})$, 
where $\Xi_\triv$  and $\vec{\Gamma}_\triv$ are (scalar and vector) differential functions 
vanishing on the solution space $\Esp$ of the system:
$\Xi_\triv[\depvars]|_\Esp =0$
and
$\vec{\Gamma}_\triv[\depvars]|_\Esp =0$.

The key aspect of local triviality is that 
the circulation current \eqref{3Dcurvconslaw-triv-offsolns} identically satisfies 
$D_t\, \curvT[\depvars] + \Grad\,\curvX[\depvars] \equiv 
D_t\,\vec{\Gamma}_\triv[\depvars] + \Grad\,\Xi_\triv[\depvars]$
which automatically vanishes when $\depvars(t,\vec{x})$ 
is any solution of the given PDE system \eqref{3Dpde}. 
Correspondingly, 
for any regular curve $\C\subset \pdedom$ 
within the spatial domain of a given PDE system \eqref{3Dpde},
the resulting time-dependent global circulation conservation law \eqref{3Dglobalcurvconslaw} 
becomes the line integral identity
\begin{equation}\label{3Dglobalcurvconslaw-triv}
\frac{d}{dt} \int_\C \Grad\,\Theta[\depvars]\cdot \ds 
= \int_\C \Grad\, (D_t\,\Theta[\depvars])\cdot \ds 
=  (D_t\,\Theta[\depvars]) \Big|_{\p\C}
= \frac{d}{dt}\Big( \Theta[\depvars] \Big|_{\p\C} \Big)
\end{equation}
apart from trivial terms that vanish on the solution space $\Esp$ of the PDE system.
This identity contains no local information 
about the given PDE system or its solutions. 

For discussing global triviality, 
we will call a circulatory conserved current \eqref{3Dcurvconslaw-triv-offsolns}
type I trivial if $\Theta[\depvars]|_\Esp$ is constant, 
and otherwise type II trivial if $\Theta[\depvars]|_\Esp$ is non-constant. 
In particular:
\begin{align}
\text{type I trivial}\quad 
&\curvT[\depvars]|_\Esp = 0, 
\quad
\curvX[\depvars]|_\Esp  = 0 ;
\label{3Dcurvconslaw-triv-typeI}
\\
\text{type II trivial}\quad 
&\curvT[\depvars]|_\Esp = \Grad\,\Theta[\depvars]|_\Esp \neq 0,
\quad
\curvX[\depvars]|_\Esp  = -D_t\,\Theta[\depvars]|_\Esp \neq 0 . 
\label{3Dcurvconslaw-triv-typeII}
\end{align}
Clearly, a type I trivial circulatory conservation law 
contains no global information about solutions of the given PDE system,
and hence this type of locally trivially conservation law is globally trivial. 
A type II trivial circulatory conservation law 
will likewise contain no global information about the given PDE system 
when the curve $\C$ is closed,
since then the circulation integral 
$\int_\C \Grad\,\Theta[\depvars]\cdot \ds$
vanishes (by the fundamental theorem of line integrals) 
due to the curve having no boundary, $\p\C=\emptyset$. 
In contrast, when the curve $\C$ is non-closed, 
a type II trivial circulatory conservation law 
can contain some useful global information 
if the net flux term $\flux[\depvars;\p\C]=(D_t\,\Theta[\depvars])\big|_{\p\C}$ 
in the integral identity \eqref{3Dglobalcurvconslaw-triv} 
vanishes when $\depvars(t,\vec{x})$ is an arbitrary solution of the PDE system,
since this yields 
\begin{equation}\label{3Dglobalendpointconslaw}
\frac{d}{dt} \Big( \Theta[\depvars] \Big|_{\p\C} \Big) =0
\end{equation}
with $\Theta[\depvars]|_\Esp$ being non-constant. 
To see the content of this equation, 
we observe that it holds for an arbitrary non-closed curve $\C$ 
if and only if 
$D_t\Grad\,\Theta[\depvars]|_\Esp=0$, 
which is a purely temporal conservation law with 
$\Grad\Theta[\depvars]|_\Esp = \curvT[\depvars]|_\Esp \neq 0$
being the conserved density. 

\begin{proposition}\label{prop:3Dcurvconslaw-nontrivflow}
(i) A locally trivial time-dependent circulatory conservation law \eqref{3Dcurvconslaw-triv} of 
any PDE system \eqref{3Dpde} in $\Rnum^3$ is globally trivial 
for a regular curve $\C\subset\pdedom$ if and only if
either the curve surface is closed, 
or the curve is non-closed and the net circulation potential $\Theta[\depvars]|_{\p\C}$ 
is either time dependent or zero, 
for an arbitrary solution $\depvars(t,\vec{x})$ of the system.
(ii) When a locally trivial surface-flux conservation law \eqref{3Dsurfconslaw-triv} 
is globally non-trivial for a non-closed regular surface $\S\subset\pdedom$, 
it yields a global pointwise conservation law of the form \eqref{3Dglobalendpointconslaw}
on the endpoints of the curve, 
with $\Theta[\depvars]|_\Esp$ being spatially non-constant. 
\end{proposition}

We will now establish a converse for the first part of Proposition~\ref{prop:3Dcurvconslaw-nontrivflow}. 
Suppose a time-dependent global circulatory conservation law \eqref{3Dglobalcurvconslaw} 
holding for a regular curve $\C\subset \pdedom$ 
in the spatial domain of a given PDE system \eqref{3Dpde} 
contains no global information about the solutions of the PDE system. 
Firstly, 
the circulation integral $C[\depvars;\C]=\int_\C \curvT[\depvars]\cdot \ds$ 
must reduce to endpoint terms by the fundamental theorem of line integrals,
which requires that the circulation density is a gradient, 
$\curvT[\depvars]|_\Esp = \Grad\,\Theta[\depvars]|_\Esp$. 
This condition will be sufficient when the curve $\C$ is closed, 
since the net flux $\flux[\depvars;\p\C]$ will then vanish identically, 
without any condition being necessary 
on the spatial endpoint flow $\curvX[\depvars]|_\Esp$. 
When the curve is non-closed, 
we must additionally have 
$\int_C D_t\Grad\,\Theta[\depvars]\cdot\ds|_\Esp = -\int_C\Grad\,\curvX[\depvars]\cdot\ds|_\Esp$,
as this is necessary for the global conservation law \eqref{3Dglobalcurvconslaw} to hold. 
The equality of these two line integrals for an arbitrary curve $\C$
requires $\Grad(D_t\Theta[\depvars] +\curvX[\depvars])|_\Esp=0$,
which implies $(D_t\Theta[\depvars] +\curvX[\depvars])|_\Esp$ is the density 
for a local spatial gradient conservation law. 
For this conservation law to contain no global information, 
it must be trivial,
whereby we must have $\curvX[\depvars]|_\Esp = -D_t\Theta[\depvars]|_\Esp$.
Hence the conserved circulatory current is locally trivial \eqref{3Dcurvconslaw-triv-offsolns},
and consequently the global conservation law \eqref{3Dglobalcurvconslaw} 
reduces to the form \eqref{3Dglobalcurvconslaw-triv}. 
Finally, for the net flux $\flux[\depvars;\p\C]=(D_t\,\Theta[\depvars])\big|_{\p\C}$ 
to contain no global information about solutions $\depvars(t,\vec{x})$, 
the net circulation potential $\Theta[\depvars]|_{\p\C}$ 
must either be time dependent whereby the global conservation law holds as an identity, 
or be spatially constant whereby both $C[\depvars;\C]=0$ and $\flux[\depvars;\p\C]=0$
are trivial. 

Thus, we have the following result. 

\begin{theorem}\label{thm:3Dcurvconslaw-globaltriv}
A time-dependent global circulatory conservation law \eqref{3Dglobalcurvconslaw} of a PDE system \eqref{3Dpde} 
is globally trivial for an arbitrary regular curve $\C\subset \pdedom$
if and only if the conserved circulation density $\curvT[\depvars]$ 
is locally trivial \eqref{3Dcurvconslaw-trivT}
and, when the curve is non-closed, 
the spatial endpoint flow $\curvX[\depvars]$ is locally trivial \eqref{3Dcurvconslaw-trivX}
such that the net circulation potential $\Theta[\depvars]|_{\p\C}$ 
is either time dependent or zero, 
for an arbitrary solution $\depvars(t,\vec{x})$ of the system.
Consequently, 
in a globally non-trivial circulatory conservation law \eqref{3Dglobalcurvconslaw} 
holding on a regular curve $\C\subset\Rnum^3$,
the total circulation $C[\depvars;\C]$ 
either is given by a line integral that essentially depends on $\depvars(t,\vec{x})$ at all points along $\C$, 
or reduces to a pointwise constant of motion \eqref{3Dglobalendpointconslaw} that essentially depends on $\depvars(t,\vec{x})$ only at the end points $\p\C$ of $\C$. 
\end{theorem}

This leads to a corresponding notion of global equivalence. 

\begin{corollary}
Two time-dependent global circulatory conservation laws of a PDE system \eqref{3Dpde}
are \emph{equivalent} 
in the sense of containing the same information about the solutions of the given system 
if and only if they differ by a globally trivial circulatory conservation law. 
\end{corollary}

The same notions of equivalence and (non-) triviality extend to circulatory constants of motion. 
Specifically, 
a circulatory constant of motion 
\begin{equation}\label{3Dcurvcom}
\frac{d}{dt} \int_{\C} \curvT[\depvars]\cdot \ds = 0
\end{equation}
is trivial if and only if 
the conserved density $\curvT[\depvars]$ is locally trivial \eqref{3Dcurvconslaw-trivT}
and, when the curve $\C$ is non-closed, 
the spatial flow vanishes, $\curvX[\depvars]|_\Esp=0$. 

Finally, 
using the differential identities \eqref{divcurlgrad-idents}, 
we can formulation a curl condition for a conserved circulation current 
$(\curvT[\depvars],\curvX[\depvars])$
to be trivial. 

If $(\curvT[\depvars],\curvX[\depvars])$ has the locally trivial form \eqref{3Dcurvconslaw-triv-offsolns},
then $\Curl\,\curvT[\depvars]= \Curl\,\vec{\Gamma}_\triv[\depvars]$
satisfies 
\begin{equation}\label{3Dcurvdensity-curlfree}
\Curl\,\curvT[\depvars]|_\Esp =0 .
\end{equation}
This represents a necessary condition. 
Now consider the converse. 
If $\Curl\,\curvT[\depvars]$ vanishes on $\Esp$, 
then $\curvT[\depvars]$ represents a conserved density for a local spatial curl conservation law \eqref{spatial-curl-3Dconslaw}. 
Supposing that the given PDE system admits no non-trivial conservation laws of that type,
then Proposition~\ref{prop:spatial-curl-conslaw-globaltriv}
shows that $\curvT[\depvars]$ will be locally trivial,
and hence $\curvT[\depvars]|_\Esp = \Grad\,\Theta[\depvars]|_\Esp$
holds from Definition~\ref{defn:3Dcurvconslaw-loctriv}. 
This implies 
$D_t \curvT[\depvars]|_\Esp = \Grad\,D_t\Theta[\depvars]|_\Esp = -\Grad\,\Psi[\depvars]|_\Esp$
whereby $\Grad(\Psi[\depvars] + D_t\Theta[\depvars])|_\Esp =0$ 
is a local spatial gradient conservation law \eqref{spatial-grad-3Dconslaw}. 
If the only conservation laws of this type admitted by the given PDE system are trivial, 
then we have $(\Psi[\depvars] + D_t\Theta[\depvars])|_\Esp =0$,
and thus $\Psi[\depvars]|_\Esp$ has the locally trivial form \eqref{3Dcurvconslaw-triv}. 

This formulation of local triviality establishes the following result. 

\begin{proposition}\label{prop:3Dcurvconslaw-localtriv}
A necessary condition for 
a time-dependent local circulatory conservation law \eqref{3Dcurvconslaw} 
of a PDE system \eqref{3Dpde} in $\Rnum^3$ 
to be locally trivial is that 
its conserved density $\curvT[\depvars]$ is curl-free \eqref{3Dcurvdensity-curlfree}
for an arbitrary solution $\depvars(t,\vec{x})$ of the PDE system. 
This curl-free condition is sufficient 
if the PDE system has no non-trivial topological circulation conservation laws \eqref{spatial-curl-3Dglobalconslaw} 
and no non-trivial spatial gradient conservation laws \eqref{spatial-grad-3Dconslaw}. 
\end{proposition}

The physical meaning of the curl-free condition \eqref{3Dcurvdensity-curlfree}
in the spatial domain of the given PDE system 
is simply that the conserved circulation density $\curvT[\depvars]$ is irrotational, 
or equivalently that it has vanishing vorticity. 

Combining Proposition~\ref{prop:3Dcurvconslaw-localtriv} and Theorem~\ref{thm:3Dcurvconslaw-globaltriv}, 
we obtain a simple sufficient condition for non-triviality of circulatory conservation laws. 

\begin{corollary}\label{cor:3Dglobalcurvconslaw-suff-nontriv}
If the conserved density $\curvT[\depvars]$ of 
a time-dependent (local or global) circulatory conservation law 
satisfies $\Curl\,\curvT[\depvars]|_\Esp \neq 0$,
then the conservation law is locally and globally non-trivial. 
\end{corollary}

We will see in the next subsection 
how non-trivial time-dependent global circulatory conservation laws
can arise from locally trivial surface-flux conservation laws. 

As will be shown in Section~\ref{sec:ex:curvCLs}, 
a physical example of a non-trivial circulatory conservation law is 
circulation in irrotational ideal fluid flow; 
two physical examples of trivial circulatory conservation laws
arise from the density gradient in incompressible fluid flow,
and the entropy gradient in non-homentropic fluid flow.

\subsection{Non-triviality and equivalence of time-dependent surface-flux conservation laws}\label{sec:3D:triv-equiv:surfCL}

The notions of triviality and equivalence of time-dependent surface-flux conservation laws
are similar to those for time-dependent circulation conservation laws. 
One main difference, however, is that 
any topological circulation conservation law \eqref{spatial-curl-3Dglobalconslaw}
can be added to a time-dependent surface-flux conservation law, 
without affecting the surface-flux conserved density. 
Another difference is that some gauge freedom arises in the form of a trivial surface-flux conserved current. 

\begin{definition}\label{defn:3Dsurfconslaw-loctriv}
A time-dependent surface-flux conservation law \eqref{3Dsurfconslaw} of a PDE system \eqref{3Dpde} in $\Rnum^3$ is called \emph{locally trivial} 
if, on the solution space $\Esp$ of the system, 
the conserved surface-flux density $\surfT$ and the spatial circulation flux $\surfX$
have the respective forms 
\begin{subequations}\label{3Dsurfconslaw-triv}
\begin{align}
& \surfT[\depvars]|_\Esp = \Curl\,\vec{\Theta}[\depvars]|_\Esp,
\label{3Dsurfconslaw-trivT}
\\
& 
\surfX[\depvars]|_\Esp  = -D_t\,\vec{\Theta}[\depvars]|_\Esp + \Grad\,\Lambda[\depvars]|_\Esp,
\label{3Dsurfconslaw-trivX}
\end{align}
\end{subequations}
in terms of a scalar differential function $\Lambda[\depvars]$ 
and a vector differential function $\vec{\Theta}[\depvars]$. 
Any two time-dependent surface-flux conservation laws \eqref{3Dsurfconslaw} that differ
only by a locally trivial surface-flux conservation law are \emph{locally equivalent}.
\end{definition}

Note that the form \eqref{3Dsurfconslaw-triv} 
of a locally trivial conserved surface-flux current $(\surfT,\surfX)$ is not unique, 
since it is preserved by 
\begin{equation}\label{3Dsurfconslaw-triv-gauge}
\vec{\Theta}[\depvars]|_\Esp \rightarrow (\vec{\Theta}[\depvars]+\Grad\,\Xi[\depvars])|_\Esp,
\quad
\Lambda[\depvars]|_\Esp \rightarrow (\Lambda[\depvars] + D_t\Xi[\depvars])|_\Esp
\end{equation}
where $\Xi[\depvars]$ is an arbitrary scalar differential function. 

Moving off of the solution space $\Esp$ of the given PDE system, 
we can apply Lemma~\ref{Hadamard} to see that 
any locally trivial conserved surface-flux current \eqref{3Dsurfconslaw-triv} 
has the equivalent formulation 
\begin{equation}\label{3Dsurfconslaw-triv-offsolns}
\surfT[\depvars] = \Curl\,\vec{\Theta}[\depvars] + \vec{\Gamma}_\triv[\depvars], 
\quad
\surfX[\depvars] = -D_t\,\vec{\Theta}[\depvars]  +\Grad\,\Lambda[\depvars] + \vec{\Phi}_\triv[\depvars]
\end{equation}
holding identically for all (sufficiently smooth) functions $\depvars(t,\vec{x})$, 
where $\vec{\Gamma}_\triv$ and $\vec{\Phi}_\triv$
are any (vector) differential functions 
vanishing on the solution space $\Esp$ of the system. 

Local triviality expresses the property that the surface-flux current \eqref{3Dsurfconslaw-triv-offsolns} 
identically satisfies 
$D_t\, \surfT[\depvars] + \Curl\,\surfX[\depvars] \equiv
D_t\,\vec{\Gamma}_\triv[\depvars] + \Curl\,\vec{\Phi}_\triv[\depvars]$
which automatically vanishes when $\depvars(t,\vec{x})$ 
is any solution of the given PDE system \eqref{3Dpde}. 
Correspondingly, 
for any regular surface $\S\subset \pdedom$ 
within the spatial domain of a given PDE system \eqref{3Dpde},
the resulting time-dependent global surface-flux conservation law \eqref{3Dglobalsurfconslaw} 
reduces by Stokes' theorem to an integral identity
\begin{equation}\label{3Dglobalsurfconslaw-triv}
\begin{aligned}
& \frac{d}{dt} \int_\S \Curl\,\vec{\Theta}[\depvars]\cdot \dS
= \frac{d}{dt} \oint_{\p\S} \vec{\Theta}[\depvars]\cdot \ds
\\
& = \int_\S \Curl\, (D_t\,\vec{\Theta}[\depvars])\cdot \dS
=  \oint_{\p\S}D_t\,\vec{\Theta}[\depvars]\cdot\ds
\end{aligned}
\end{equation}
apart from trivial terms that vanish on the solution space $\Esp$ of the PDE system.
This is an identity which contains no local information 
about the given PDE system or its solutions. 

We will refer to a locally trivial time-dependent surface-flux conserved current \eqref{3Dsurfconslaw-triv-offsolns}
as being 
either type I trivial if both $\vec{\Theta}[\depvars]|_\Esp$ and $\Lambda[\depvars]|_\Esp$ 
are zero, 
or type IIa trivial if $\Lambda[\depvars]|_\Esp$ is non-zero 
while $\vec{\Theta}[\depvars]|_\Esp$ is zero, 
or type IIb trivial if $\vec{\Theta}[\depvars]|_\Esp$ is non-zero,
modulo the gauge freedom \eqref{3Dsurfconslaw-triv-gauge} in each case. 
In particular:
\begin{align}
\text{type I trivial}\quad
& \surfT[\depvars]|_\Esp = 0, 
\quad
\surfX[\depvars]|_\Esp  = 0 ;
\label{3Dsurfconslaw-triv-typeI}
\\
\text{type IIa trivial}\quad
& \surfT[\depvars]|_\Esp = 0, 
\quad
\surfX[\depvars]|_\Esp  = \Grad\,\Lambda[\depvars]|_\Esp \neq 0;
\label{3Dsurfconslaw-triv-typeIIa}
\\
\text{type IIb trivial}\quad
&\begin{aligned}
&\surfT[\depvars]|_\Esp = \Curl\,\vec{\Theta}[\depvars]|_\Esp \neq 0,
\\
&\surfX[\depvars]|_\Esp  = -D_t\,\vec{\Theta}[\depvars]|_\Esp + \Grad\,\Lambda[\depvars]|_\Esp \neq 0. 
\end{aligned}
\label{3Dsurfconslaw-triv-typeIIb}
\end{align}
This distinction is useful for discussing global triviality. 

Clearly, a type I trivial surface-flux conservation law contains neither global nor local information about solutions of the given PDE system. 
Hence, this type of locally trivially conservation law is globally trivial. 

A type IIa trivial surface-flux conservation law also contains no information about the given PDE system,
since the integral identity \eqref{3Dglobalsurfconslaw-triv} 
only involves  $\vec{\Theta}[\depvars]$ 
modulo an arbitrary gradient $\Grad\,\Xi[\depvars]$, 
whereby each integral in the identity vanishes identically. 
Likewise a type IIb trivial surface-flux conservation law also contains 
no information about the given PDE system 
when the surface $\S$ is closed,
since then the surface integral $\int_\S \Curl\,\vec{\Theta}[\depvars]\cdot \dS$
vanishes (by Stokes' theorem) due to $\p\S$ being empty. 

When the surface $\S$ is non-closed, 
a type IIb trivial surface-flux conservation law 
can contain some useful global information 
if the net flux integral 
$\flux[\depvars;\p\S]=\oint_{\p\S}D_t\,\vec{\Theta}[\depvars]\cdot\ds$
in the identity \eqref{3Dglobalsurfconslaw-triv} vanishes 
when $\depvars(t,\vec{x})$ is an arbitrary solution of the PDE system. 
In this situation, 
the identity reduces to a global circulatory conservation law \eqref{3Dglobalcurvconslaw}
\begin{equation}\label{3Dglobalboundarycurvconslaw}
\frac{d}{dt} \oint_{\p\S} \vec{\Theta}[\depvars]\cdot \ds\,\big|_\Esp =0 
\end{equation}
on the boundary curve(s) $\C=\p\S$ of the non-closed surface $\S$, 
with no spatial (endpoint) flow because $\p\C=\p^2\S=\emptyset$ is empty. 
Since here we have $\vec{\Theta}[\depvars]|_\Esp \neq \Grad\,\Xi[\depvars]|_\Esp$
by type IIb triviality, 
we conclude from Theorem~\ref{thm:3Dcurvconslaw-globaltriv} 
that the circulatory conservation law \eqref{3Dglobalboundarycurvconslaw}
is globally non-trivial. 

This discussion establishes the following interesting result showing how 
a locally trivial surface-flux conservation law \eqref{3Dsurfconslaw-triv-typeIIb}
can yield a globally non-trivial circulation conservation law \eqref{3Dglobalboundarycurvconslaw}
on the boundary of a non-closed surface. 

\begin{proposition}\label{prop:3Dsurfconslaw-nontrivcirc}
(i) A locally trivial time-dependent surface-flux conservation law \eqref{3Dsurfconslaw-triv} of a PDE system \eqref{3Dpde} in $\Rnum^3$ 
is globally trivial for a regular surface $\S\subset\Rnum^3$ if and only if 
either the surface is closed,
or the surface is non-closed 
and the circulation integral $\oint_{\p\S}\vec{\Theta}[\depvars]\cdot\ds$
is either time dependent or identically zero,
for an arbitrary solution $\depvars(t,\vec{x})$ of the given PDE system. 
(ii) When a locally trivial surface-flux conservation law \eqref{3Dsurfconslaw-triv} 
is globally non-trivial for an arbitrary non-closed regular surface $\S\subset\Rnum^3$, 
it corresponds to a purely temporal curl-type conservation law 
\begin{equation}\label{3Dtemporal-curlconslaw}
D_t\Curl\,\vec{\Theta}[\depvars]|_\Esp=0
\end{equation}
that yields a non-trivial circulatory constant of motion \eqref{3Dcurvcom}
on the boundary curve(s) $\C=\p\S$. 
\end{proposition}

The second part of this result has a converse. 
Every curl-type temporal conservation law \eqref{3Dtemporal-curlconslaw}
yields a circulatory constant of motion  
on any curve $\C=\p\S$ given by the boundary of a non-closed regular surface $\S$ within the spatial domain of the PDE system. 
Through Stokes' theorem,
the circulatory constant of motion can be expressed as 
a surface-flux conservation law
\begin{equation}
0= \frac{d}{dt} \oint_{\p\S} \vec{\Theta}[\depvars] \cdot\ds\,\big|_\Esp
= \int_{\p\S} D_t \vec{\Theta}[\depvars] \cdot\ds\,\big|_\Esp
= \frac{d}{dt} \int_{\S} \Curl\,\vec{\Theta}[\depvars] \cdot\dS\,\big|_\Esp
\end{equation}
whose conserved current $\surfT[\depvars]|_\Esp = \Curl\,\vec{\Theta}[\depvars]|_\Esp \neq 0$ 
is locally trivial of type IIb \eqref{3Dsurfconslaw-triv-typeIIb}. 

Curl-type temporal conservation laws \eqref{3Dtemporal-curlconslaw}
can be obtained from local circulatory conservation laws 
and local spatial curl conservation laws. 
Specifically, 
the curl of any locally non-trivial circulatory conservation law \eqref{3Dcurvconslaw}, 
written as $(D_t\,\vec{\Theta}[\depvars]+ \Grad\,\Lambda[\depvars])|_\Esp=0$,
yields a temporal conservation law \eqref{3Dtemporal-curlconslaw},
and the time derivative of any locally non-trivial spatial curl conservation law \eqref{spatial-curl-3Dconslaw}, 
written as $(\Curl\,\vec{\Theta}[\depvars])|_\Esp=0$,
also yields a temporal conservation law \eqref{3Dtemporal-curlconslaw}.

We will now establish a converse for the first part of Proposition~\ref{prop:3Dsurfconslaw-nontrivcirc},
in a similar way to the proof of Theorem~\ref{thm:3Dcurvconslaw-globaltriv}. 

Suppose a time-dependent global surface-flux conservation law \eqref{3Dglobalsurfconslaw} 
holding for a regular surface $\S\subset \pdedom$ 
in the spatial domain of a given PDE system \eqref{3Dpde} 
contains no global information about the solutions of the PDE system. 
Firstly, 
the surface-flux integral $C[\depvars;\S]=\int_\S \surfT[\depvars]\cdot \dS$ 
must reduce to a boundary line integral by Stokes' theorem, 
which requires the surface-flux density to be a curl, 
$\surfT[\depvars]|_\Esp = \Curl\,\vec{\Theta}[\depvars]|_\Esp$. 
This condition will be sufficient when the surface $\S$ is closed, 
since the net flux $\flux[\depvars;\p\S]$ will then vanish identically, 
without any condition being necessary 
on the spatial circulation flux $\surfX[\depvars]|_\Esp$. 
When the surface has a boundary, 
we must additionally have 
$\oint_{\p\S} D_t \vec{\Theta}[\depvars]\cdot \ds\,|_\Esp = -\oint_{\p\S} \surfX[\depvars]\cdot \ds\,|_\Esp$
so that the global conservation law \eqref{3Dglobalsurfconslaw} holds. 
The equality of these two line integrals for an arbitrary boundary $\p\S$
requires $\Curl(D_t\vec{\Theta}[\depvars] +\surfX[\depvars])|_\Esp=0$,
which implies $(D_t\vec{\Theta}[\depvars] +\surfX[\depvars])|_\Esp$ is the density 
for a local spatial curl conservation law \eqref{spatial-curl-3Dconslaw}. 
This conservation law must be locally trivial,
as otherwise it will contain global information. 
Therefore, we must have 
$(\surfX[\depvars] + D_t\vec{\Theta}[\depvars])|_\Esp = \Grad\,\Lambda[\depvars]|_\Esp$
for some differential scalar function $\Lambda[\depvars]$. 
Hence the conserved surface-flux current is locally trivial \eqref{3Dsurfconslaw-triv-offsolns},
and consequently the global conservation law \eqref{3Dglobalsurfconslaw} 
reduces to the form \eqref{3Dglobalsurfconslaw-triv}. 
Finally, for the net flux 
$\flux[\depvars;\p\S]=\oint_{\p\S}D_t\,\vec{\Theta}[\depvars]\cdot\ds$
to contain no global information about solutions $\depvars(t,\vec{x})$, 
the circulation integral $\oint_{\p\S}\vec{\Theta}[\depvars]\cdot\ds$
must either be time dependent so that the global conservation law holds as an identity, 
or be identically zero so that both $C[\depvars;\S]=0$ and $\flux[\depvars;\p\S]=0$
are trivial. 

Hence, we obtain the following main result, 
extending the first part of Proposition~\ref{prop:3Dsurfconslaw-nontrivcirc}. 

\begin{theorem}\label{thm:3Dsurfconslaw-globaltriv}
A time-dependent global surface-flux conservation law \eqref{3Dglobalsurfconslaw} of a PDE system \eqref{3Dpde} 
is globally trivial for an arbitrary regular surface $S\subset\pdedom$ 
if and only if its conserved density $\surfT[\depvars]$ is locally trivial \eqref{3Dsurfconslaw-trivT}
and, when the surface is non-closed, 
the spatial circulation flux $\surfX[\depvars]$ is locally trivial \eqref{3Dsurfconslaw-trivX}
such that the associated circulation integral $\oint_{\p\S}\vec{\Theta}[\depvars]\cdot\ds$ 
is either time dependent or identically zero, 
for an arbitrary solution $\depvars(t,\vec{x})$ of the system.
Consequently, 
in a globally non-trivial surface-flux conservation law \eqref{3Dglobalsurfconslaw} 
holding on a regular surface $\S\subset\pdedom$, 
the total surface-flux $C[\depvars;\S]$ 
either is given by a surface integral that essentially depends on $\depvars(t,\vec{x})$ at all points on $\S$, 
or reduces to a circulatory constant of motion that essentially depends on $\depvars(t,\vec{x})$ only at the boundary $\p\S$ of $\S$. 
\end{theorem}

This result leads to a notion of global equivalence for surface-flux conservation laws. 

\begin{corollary}\label{cor:3Dsurfconslaw-globequiv}
Two global time-dependent surface-flux conservation laws of a PDE system \eqref{3Dpde}
are \emph{equivalent} 
in the sense of containing the same information about the solutions of the given system 
if and only if they differ by a globally trivial surface-flux conservation law. 
\end{corollary}

The same notions of equivalence and (non-) triviality extend to surface-flux constants of motion. 
Specifically, 
a surface-flux constant of motion 
\begin{equation}\label{3Dsurfcom}
\frac{d}{dt} \int_{\S} \surfT[\depvars]\cdot \dS = 0
\end{equation}
is said to be trivial if 
the conserved density $\surfT[\depvars]$ is locally trivial \eqref{3Dsurfconslaw-trivT}
and, when the surface $\S$ is non-closed, 
the spatial circulation flux vanishes, $\surfX[\depvars]|_\Esp=0$. 

The conditions in Theorem~\ref{thm:3Dsurfconslaw-globaltriv}
and Proposition~\ref{prop:3Dsurfconslaw-nontrivcirc} 
for triviality can be formulated in terms of a divergence condition, 
similarly to the curl condition for triviality of conserved circulation currents. 
If a conserved surface-flux current $(\surfT[\depvars],\surfX[\depvars])$ 
is locally trivial \eqref{3Dsurfconslaw-triv-offsolns},
then $\Div\,\surfT[\depvars]= \Div\,\vec{\Gamma}_\triv[\depvars]$
holds by the differential identities \eqref{divcurlgrad-idents}. 
Hence $\surfT[\depvars]$ satisfies 
\begin{equation}\label{3Dsurfdensity-divfree}
\Div\,\surfT[\depvars]|_\Esp =0 .
\end{equation}
This represents a necessary condition. 
Conversely, if $\Div\,\surfT[\depvars]$ vanishes on the solution space $\Esp$, 
then $\surfT[\depvars]$ represents a conserved density 
for a local spatial divergence conservation law \eqref{spatial-div-3Dconslaw}. 
Supposing that the given PDE system admits no non-trivial conservation laws of that type,
then Proposition~\ref{prop:spatial-div-conslaw-globaltriv}
shows that $\surfT[\depvars]$ will be locally trivial,
$\surfT[\depvars]|_\Esp = \Curl\,\vec{\Theta}[\depvars]|_\Esp$. 
This implies 
$D_t \surfT[\depvars]|_\Esp = \Curl\,D_t\vec{\Theta}[\depvars]|_\Esp = -\Curl\,\vec{\Psi}[\depvars]|_\Esp$
whereby $\Curl(\vec{\Psi}[\depvars] + D_t\vec{\Theta}[\depvars])|_\Esp =0$ 
is a local spatial curl conservation law \eqref{spatial-curl-3Dconslaw}. 
If the given PDE system admits only trivial conservation laws of that type, 
then from Definition~\ref{defn:3Dspatial-curl-conslaw-loctriv} 
we have $(\surfX[\depvars] + D_t\Theta[\depvars])|_\Esp$ is a gradient, 
which implies $\surfX[\depvars]|_\Esp$ has the locally trivial form \eqref{3Dsurfconslaw-triv-offsolns}. 

This argument establishes the following result. 

\begin{proposition}\label{prop:3Dglobalsurfconslaw-triv}
A necessary condition for 
a time-dependent surface-flux conservation law \eqref{3Dsurfconslaw} 
of a PDE system \eqref{3Dpde} in $\Rnum^3$ 
to be locally trivial \eqref{3Dsurfconslaw-triv}
is that its conserved density $\surfT[\depvars]$ is divergence-free \eqref{3Dsurfdensity-divfree}
for an arbitrary solution $\depvars(t,\vec{x})$ of the PDE system. 
This divergence-free condition is sufficient if 
the PDE system has no non-trivial topological flux conservation laws \eqref{spatial-div-3Dglobalconslaw} 
and no non-trivial topological circulation conservation laws \eqref{spatial-curl-3Dglobalconslaw}. 
\end{proposition}

The physical meaning of the divergence-free condition 
$\Div\,\surfT[\depvars]|_\Esp =0$ 
is that there are no sources and sinks for $\surfT[\depvars]$
in the spatial domain of the given PDE system. 

We will now derive a similar formulation of the conditions in Theorem~\ref{thm:3Dsurfconslaw-globaltriv} 
for global triviality. 
The formulation differs depending on whether surfaces with or without boundaries 
are considered. 

In the case of an arbitrary regular surface $\S$ without a boundary, 
global triviality is equivalent to local triviality. 
In contrast, the case of an arbitrary regular surface $\S$ with a boundary 
requires a further argument as follows. 

Suppose the circulation integral $\oint_{\p\S}\vec{\Theta}[\depvars]\cdot\ds\,\big|_\Esp$ 
vanishes identically for an arbitrary non-closed regular surface $\S$. 
Then, from the fundamental theorem for line integrals, 
$\vec{\Theta}[\depvars]|_\Esp = \Grad\,\Xi[\depvars]|_\Esp$ 
holds for a scalar differential function $\Xi[\depvars]$,
and thus we have $\Curl\,\vec{\Theta}[\depvars]|_\Esp = 0$ 
by the second of the differential identities \eqref{divcurlgrad-idents}. 
Conversely, $\Curl\,\vec{\Theta}[\depvars]|_\Esp = 0$ 
implies that $\vec{\Theta}[\depvars]|_\Esp = \Grad\,\Xi[\depvars]|_\Esp$ holds, 
if the given PDE system has no non-trivial local spatial curl conservation laws. 
Hence, under this condition, 
the vanishing of $\oint_{\p\S}\vec{\Theta}[\depvars]\cdot\ds|_\Esp$ 
due to the fundamental theorem for line integrals 
will hold when and only when 
$0=\Curl\,\vec{\Theta}[\depvars]|_\Esp = \surfT[\depvars]|_\Esp$. 

Suppose instead the circulation integral $\oint_{\p\S}\vec{\Theta}[\depvars]\cdot\ds\,\big|_\Esp$ 
is time dependent for an arbitrary non-closed regular surface $\S$. 
This is equivalent to the surface integral 
$\int_\S \Curl\,\vec{\Theta}[\depvars]\cdot\dS|_\Esp$ 
being time dependent. 
Since $\S$ is arbitrary,
an equivalent condition is that $\Curl\,\vec{\Theta}[\depvars]|_\Esp$ itself must be time dependent. 
Consequently, time dependence of $\oint_{\p\S}\vec{\Theta}[\depvars]\cdot\ds\,\big|_\Esp$
will hold when and only when 
$\Curl\,\vec{\Theta}[\depvars]|_\Esp = \surfT[\depvars]|_\Esp$ is time dependent. 

The preceding argument establishes the following result. 

\begin{theorem}\label{thm:3Dglobalsurfconslaw-nontriv}
A necessary condition for a time-dependent local surface-flux conservation law \eqref{3Dsurfconslaw} of a PDE system \eqref{3Dpde} 
to yield a globally trivial surface-flux conservation law \eqref{3Dglobalsurfconslaw} 
for an arbitrary regular (closed or non-closed) surface $\S\subset\pdedom$
is that the conserved surface-flux density $\surfT[\depvars]|_\Esp$ 
is divergence free \eqref{3Dsurfdensity-divfree}. 
This divergence-free condition is sufficient 
in the case of closed surfaces $\S$
if the PDE system has no non-trivial topological flux conservation laws \eqref{spatial-div-3Dglobalconslaw}. 
In the case of non-closed surfaces $\S$, 
a sufficient condition is that $\surfT[\depvars]|_\Esp$ 
either is both divergence-free and time dependent, or is zero, 
and also that the PDE system has no non-trivial topological flux conservation laws \eqref{spatial-div-3Dglobalconslaw}. 
\end{theorem}

This result gives a simple sufficient condition for global non-triviality. 

\begin{corollary}\label{cor:3Dglobalsurfconslaw-suff-nontriv}
If the conserved density $\surfT[\depvars]$ of 
a time-dependent (local or global) surface-flux conservation law 
satisfies $\Div\,\surfT[\depvars]|_\Esp \neq 0$,
then the conservation law is locally and globally non-trivial. 
\end{corollary}

In the next subsection, 
we will see how non-trivial time-dependent global surface-flux conservation laws
can arise from locally trivial volumetric conservation laws. 

As shown in Section~\ref{sec:ex:surfCLs}, 
physical examples of non-trivial surface-flux conservation laws are given by 
magnetic induction (Faraday's law) in electromagnetism and ideal MHD,
electric field flux in vacuum electromagnetism, 
and generalized vorticity in electron MHD. 
A physical example of a trivial surface-flux conservation law 
arises from the vorticity transport equation in ideal fluid flow. 

Physical examples of locally trivially surface-flux conservation laws 
connected with curl-type temporal conservation laws that yield 
globally non-trivial circulatory constants of motion 
arise in fluid dynamics and gas dynamics for irrotational flows and Beltrami flows, 
which are discussed in Section~\ref{sec:ex:temporalCLs}.

\subsection{Non-triviality and equivalence of volumetric conservation laws}\label{sec:3D:triv-equiv:volCL}

Last, we discuss triviality and equivalence of time-dependent volumetric conservation laws. 
These notions will be very similar to those for time-dependent surface-flux conservation laws. 
Note that any topological flux conservation law \eqref{spatial-div-3Dglobalconslaw}
can be added to a time-dependent volumetric conservation law,
without affecting the volumetric conserved density. 

\begin{definition}\label{defn:3Dvolconslaw-loctriv}
A time-dependent volumetric conservation law \eqref{3Dvolconslaw} of a PDE system \eqref{3Dpde} in $\Rnum^3$ is called \emph{locally trivial} 
if, on the solution space $\Esp$ of the system, 
the conserved density $\volT$ and the spatial flux $\volX$
have the respective forms 
\begin{subequations}\label{3Dvolconslaw-triv}
\begin{align}
& \volT[\depvars]|_\Esp = \Div\,\vec{\Theta}[\depvars]|_\Esp,
\label{3Dvolconslaw-trivT}
\\
&
\volX[\depvars]|_\Esp  = -D_t\,\vec{\Theta}[\depvars]|_\Esp +\Curl\,\vec{\Lambda}[\depvars]|_\Esp,
\label{3Dvolconslaw-trivX}
\end{align}
\end{subequations}
in terms of a pair of vector differential functions $\vec{\Lambda}[\depvars]$ and $\vec{\Theta}[\depvars]$. 
Any two time-dependent volumetric conservation laws \eqref{3Dvolconslaw} that differ
only by a locally trivial volumetric conservation law are \emph{locally equivalent}.
\end{definition}

Note that the form \eqref{3Dvolconslaw-triv} 
of a locally trivial conserved volumetric current $(\volT,\volX)$ is not unique, 
since it is preserved by 
\begin{equation}\label{3Dvolconslaw-triv-gauge}
\vec{\Theta}[\depvars]|_\Esp \rightarrow (\vec{\Theta}[\depvars]+\Curl\,\vec{\Xi}[\depvars])|_\Esp,
\quad
\vec{\Lambda}[\depvars]|_\Esp \rightarrow (\vec{\Lambda}[\depvars] + D_t\vec{\Xi}[\depvars])|_\Esp
\end{equation}
where $\vec{\Xi}[\depvars]$ is an arbitrary vector differential function. 

When we move off of the solution space $\Esp$ of the given PDE system, 
Lemma~\ref{Hadamard} shows that 
any locally trivial conserved volumetric current \eqref{3Dvolconslaw-triv} 
has the equivalent formulation 
\begin{equation}\label{3Dvolconslaw-triv-offsolns}
\volT[\depvars] = \Div\,\vec{\Theta}[\depvars] + \Gamma_\triv[\depvars], 
\quad
\volX[\depvars] = -D_t\,\vec{\Theta}[\depvars]  -\Curl\,\vec{\Lambda}[\depvars] + \vec{\Phi}_\triv[\depvars],
\end{equation}
holding identically for all (sufficiently smooth) functions $\depvars(t,\vec{x})$, 
where $\Gamma_\triv$ and $\vec{\Phi}_\triv$ are (scalar and vector) differential functions 
vanishing on the solution space $\Esp$ of the system. 

Local triviality expresses the property that the volumetric current \eqref{3Dvolconslaw-triv-offsolns} 
identically satisfies 
$D_t\, \volT[\depvars] + \Div\,\volX[\depvars] \equiv
D_t\,\Gamma_\triv[\depvars] + \Div\,\vec{\Phi}_\triv[\depvars]$
which automatically vanishes when $\depvars(t,\vec{x})$ 
is any solution of the given PDE system \eqref{3Dpde}. 
Correspondingly, 
for any regular volume $\V\subset \pdedom$ 
within the spatial domain of a given PDE system \eqref{3Dpde},
the resulting time-dependent global volumetric conservation law \eqref{3Dglobalvolconslaw} 
reduces by Gauss' theorem to an integral identity
\begin{equation}\label{3Dglobalvolconslaw-triv}
\begin{aligned}
& \frac{d}{dt} \int_\V \Div\,\vec{\Theta}[\depvars]\dV
= \frac{d}{dt} \oint_{\p\V} \vec{\Theta}[\depvars]\cdot \dS
\\
& = \int_\V \Div\, (D_t\,\vec{\Theta}[\depvars])\dV
=  \oint_{\p\V}D_t\,\vec{\Theta}[\depvars]\cdot\dS
\end{aligned}
\end{equation}
apart from trivial terms that vanish on the solution space $\Esp$ of the PDE system.
This identity contains no local information 
about the given PDE system or its solutions. 

To discuss global triviality, 
we will refer to a locally trivial time-dependent volumetric conservation law \eqref{3Dvolconslaw-triv-offsolns}
as being 
either type I trivial if both $\vec{\Theta}[\depvars]|_\Esp$ and $\vec{\Lambda}[\depvars]|_\Esp$ 
are zero, 
or type IIa trivial if $\vec{\Lambda}[\depvars]|_\Esp$ is non-zero 
while $\vec{\Theta}[\depvars]|_\Esp$ is zero, 
or type IIb trivial if $\vec{\Theta}[\depvars]|_\Esp$ is non-zero,
modulo the gauge freedom \eqref{3Dvolconslaw-triv-gauge} in each case. 
In particular:
\begin{align}
\text{type I trivial}\quad
& \volT[\depvars]|_\Esp = 0,
\quad
\volX[\depvars]|_\Esp  = 0;
\label{3Dvolconslaw-triv-typeI}
\\
\text{type IIa trivial}\quad
& \volT[\depvars]|_\Esp = 0,
\quad
\volX[\depvars]|_\Esp  = \Curl\,\vec{\Lambda}[\depvars]|_\Esp ;
\label{3Dvolconslaw-triv-typeIIa}
\\
\text{type IIb trivial}\quad
&\begin{aligned}
&\volT[\depvars]|_\Esp = \Div\,\vec{\Theta}[\depvars]|_\Esp \neq 0,
\\
&\volX[\depvars]|_\Esp  = -D_t\,\vec{\Theta}[\depvars]|_\Esp +\Curl\,\vec{\Lambda}[\depvars]|_\Esp \neq 0 .
\end{aligned}
\label{3Dvolconslaw-triv-typeIIb}
\end{align}
This is a direct analog of the three types of local triviality for time-dependent surface-flux conservation laws. 

Similarly to the situation for time-dependent surface-flux conservation laws,
volumetric conservation laws that are type I trivial or type IIa trivial 
contain neither global nor local information about the given PDE system. 
Type IIb trivial volumetric conservation laws, in contrast, 
can contain some useful global information 
whenever the net flux integral 
$\flux[\depvars;\p\V]=\oint_{\p\V}D_t\,\vec{\Theta}[\depvars]\cdot\dS$
in the identity \eqref{3Dglobalvolconslaw-triv} vanishes
for an arbitrary solution $\depvars(t,\vec{x})$ of the PDE system. 
The identity thereby becomes a global surface-flux conservation law \eqref{3Dglobalsurfconslaw} 
holding on the boundary surface $\S=\p\V$ of the volume $\V$: 
\begin{equation}\label{3Dglobalboundarysurfconslaw}
\frac{d}{dt} \oint_{\p\V} \vec{\Theta}[\depvars]\cdot \dS =0 
\end{equation}
with no net spatial flux because the surface $\p\V$ is closed, $\p^2\V=\emptyset$.
Since type IIb triviality implies $\vec{\Theta}[\depvars]\neq \Curl\,\vec{\Xi}[\depvars]$,
we conclude from Theorem~\ref{thm:3Dsurfconslaw-globaltriv} 
that the surface-flux conservation law \eqref{3Dglobalboundarysurfconslaw} 
is globally non-trivial. 

We have now established the following interesting result which shows 
how a locally trivial volumetric conservation law \eqref{3Dvolconslaw-triv-typeIIb}
can yield a globally non-trivial surface-flux conservation law \eqref{3Dglobalboundarysurfconslaw}
on the boundary of a volume. 

\begin{proposition}\label{prop:3Dvolconslaw-nontrivflux}
(i) A locally trivial time-dependent volumetric conservation law \eqref{3Dvolconslaw-triv} 
of a PDE system \eqref{3Dpde} in $\Rnum^3$ 
is globally trivial for a regular volume $\V\subset\Rnum^3$ if and only if 
the flux integral $\oint_{\p\V}\vec{\Theta}[\depvars]\cdot\dS$ 
is either time dependent or identically zero, 
for an arbitrary solution $\depvars(t,\vec{x})$ of the system.
(ii) When a locally trivial volumetric conservation law \eqref{3Dvolconslaw-triv} 
is globally non-trivial for a regular volume $\V\subset\Rnum^3$, 
it corresponds to a divergence-type temporal conservation law 
\begin{equation}\label{3Dtemporal-divconslaw}
D_t\Div\,\vec{\Theta}[\depvars]|_\Esp=0
\end{equation}
that yields a non-trivial surface-flux constant of motion \eqref{3Dsurfcom}
on the boundary surface(s) $\S=\p\V$. 
\end{proposition}

There is a converse for the second part of this result. 
Every divergence-type temporal conservation law \eqref{3Dtemporal-divconslaw}
yields a surface-flux constant of motion  
on any surface $\S=\p\V$ given by the boundary of a regular volume $\V$ within the spatial domain of the PDE system. 
The surface-flux constant of motion can be expressed, by Gauss' theorem,
as a volumetric conservation law
\begin{equation}
0= \frac{d}{dt} \oint_{\p\V} \vec{\Theta}[\depvars] \cdot\dS\,\big|_\Esp
= \frac{d}{dt} \int_{\V} \Div\,\vec{\Theta}[\depvars]\, \dV\,\big|_\Esp
\end{equation}
whose conserved current $\volT[\depvars]|_\Esp = \Div\,\vec{\Theta}[\depvars]|_\Esp \neq 0$ 
is locally trivial of type IIb \eqref{3Dvolconslaw-triv-typeIIb}. 

Similarly to the situation for curl-type temporal conservation laws, 
we can obtain divergence-type temporal conservation laws \eqref{3Dtemporal-divconslaw}
from local surface-flux conservation laws 
and local spatial divergence conservation laws. 
In particular, 
the divergence of any locally non-trivial surface-flux conservation law \eqref{3Dsurfconslaw}, 
written as $(D_t\,\vec{\Theta}[\depvars]+ \Curl\,\vec{\Lambda}[\depvars])|_\Esp=0$,
yields a temporal conservation law \eqref{3Dtemporal-divconslaw},
and the time derivative of any locally non-trivial spatial divergence conservation law \eqref{spatial-div-3Dconslaw}, 
written as $(\Div\,\vec{\Theta}[\depvars])|_\Esp=0$,
also yields a temporal conservation law \eqref{3Dtemporal-divconslaw}.

A converse for the first part of Proposition~\ref{prop:3Dvolconslaw-nontrivflux}
will now be established, 
similarly to Theorem~\ref{thm:3Dsurfconslaw-globaltriv} for surface-flux conservation laws. 

Suppose a time-dependent global volumetric conservation law \eqref{3Dglobalvolconslaw} 
holding for a regular volume $\V\subset \pdedom$ 
in the spatial domain of a given PDE system \eqref{3Dpde} 
contains no global information about the solutions of the PDE system. 
Firstly, 
we must have that $\volT[\depvars]|_\Esp = \Div\,\vec{\Theta}[\depvars]|_\Esp$ is a divergence, 
so the volumetric integral $C[\depvars;\V]=\int_\V \volT[\depvars]\, \dV$ 
reduces to a boundary surface integral by Gauss's theorem. 
Secondly, 
for the global conservation law \eqref{3Dglobalvolconslaw} to hold, 
we must additionally have 
$\oint_{\p\V} D_t \vec{\Theta}[\depvars]\cdot \dS\,|_\Esp = -\oint_{\p\V} \volX[\depvars]\cdot \ds\,|_\Esp$. 
The equality of these two surface integrals for an arbitrary boundary $\p\V$
requires 
$\Div(D_t \vec{\Theta}[\depvars] + \surfX[\depvars])|_\Esp=0$,
which implies $(D_t \vec{\Theta}[\depvars] + \surfX[\depvars])|_\Esp$
is the density for a local spatial divergence conservation law \eqref{spatial-div-3Dconslaw}. 
Thirdly, for this conservation law to be locally trivial, 
we must have 
$(D_t \vec{\Theta}[\depvars] + \surfX[\depvars])|_\Esp = (\Curl\,\vec{\Lambda}[\depvars])|_\Esp$
for some vector differential function $\vec{\Lambda}[\depvars]$. 
This shows that the conserved volumetric current is locally trivial \eqref{3Dvolconslaw-triv},
whereby the global conservation law \eqref{3Dglobalvolconslaw} 
reduces to the form \eqref{3Dglobalvolconslaw-triv}. 
Finally, for the net flux 
$\flux[\depvars;\p\V]=\oint_{\p\V}D_t\,\vec{\Theta}[\depvars]\cdot\dS$
to contain no global information about solutions $\depvars(t,\vec{x})$, 
the flux integral $\oint_{\p\V}\vec{\Theta}[\depvars]\cdot\dS$
must either be time dependent so that the global conservation law holds as an identity, 
or be identically zero so that both $C[\depvars;\V]=0$ and $\flux[\depvars;\p\V]=0$
are trivial. 

Hence, we obtain the following main result, 
which extends the first part of Proposition~\ref{prop:3Dvolconslaw-nontrivflux}. 

\begin{theorem}\label{thm:3Dvolconslaw-globaltriv}
A time-dependent global volumetric conservation law \eqref{3Dglobalvolconslaw} of a PDE system \eqref{3Dpde} 
is globally trivial for an arbitrary regular volume $\V\subset\Rnum^3$ 
if and only if its conserved density $\volT[\depvars]$ is locally trivial \eqref{3Dvolconslaw-trivT}
and the associated flux integral $\oint_{\p\V}\vec{\Theta}[\depvars]\cdot\dS|_\Esp$ 
is either time dependent or identically zero, 
for an arbitrary solution $\depvars(t,\vec{x})$ of the system.
Consequently, 
in a globally non-trivial volumetric conservation law \eqref{3Dglobalvolconslaw} 
holding for a regular volume $\V\subset\pdedom$, 
the volumetric quantity $C[\depvars;\V]$ 
either is given by a volume integral that essentially depends on $\depvars(t,\vec{x})$ at all points in $\V$, 
or reduces to a surface-flux constant of motion that essentially depends on $\depvars(t,\vec{x})$ only at the boundary $\p\V$ of $\V$. 
\end{theorem}

This result leads to a notion of global equivalence for volumetric conservation laws. 

\begin{corollary}\label{cor:3Dvolconslaw-globequiv}
Two global time-dependent volumetric conservation laws of a PDE system \eqref{3Dpde}
are \emph{equivalent} 
in the sense of containing the same information about the solutions of the given system 
if and only if they differ by a globally trivial volumetric conservation law. 
\end{corollary}

The preceding notions of equivalence and (non-) triviality extend to volumetric constants of motion. 
Specifically, 
a volumetric constant of motion 
\begin{equation}\label{3Dvolcom}
\frac{d}{dt} \int_{\V} \volT[\depvars]\,\dV = 0
\end{equation}
is said to be trivial if its conserved density $\volT[\depvars]$ is locally trivial \eqref{3Dvolconslaw-trivT}
(with the spatial flux vanishing, $\volX[\depvars]|_\Esp=0$). 

Unlike the situation for surface-flux and circulatory conservation laws, 
the triviality conditions 
in Theorem~\ref{thm:3Dvolconslaw-globaltriv} and Proposition~\ref{prop:3Dvolconslaw-nontrivflux}
have no general formulation involving a local condition on the conserved density $\volT[\depvars]$, 
because total divergences are not vector differential functions to which the differential identities \eqref{divcurlgrad-idents} can be applied. 
Nevertheless, when the form of $\volT[\depvars]$ off of the solution space 
has a lower differential order than that of the given PDE system, 
a local condition for triviality can be formulated using 
the spatial Euler operator (cf. Lemma~\ref{Eulerop-props}). 

\begin{proposition}\label{prop:3Dglobalvolconslaw-triv}
Suppose the conserved density $\volT[\depvars]$ 
in a time-dependent local volumetric conservation law \eqref{3Dvolconslaw} of a PDE system \eqref{3Dpde} in $\Rnum^3$ 
has a lower differential order off of the solution space 
than the differential order of all PDEs in the given system. 
Then a necessary condition for the conservation law to be locally trivial \eqref{3Dvolconslaw-triv}
is that $\volT[\depvars]$ identically satisfies 
\begin{equation}\label{3Dvoldensity-eulercond}
\hat \Eop_{\depvars}(\volT[\depvars]) \equiv 0 
\end{equation}
where $\hat \Eop_{\depvars}$ is the spatial Euler operator \eqref{spatial-eulerop}. 
This condition \eqref{3Dvoldensity-eulercond} is sufficient 
if the PDE system has no non-trivial topological flux conservation laws \eqref{spatial-div-3Dglobalconslaw}. 
\end{proposition}

The proof of this result goes as follows. 
When the differential function $\volT[\depvars]$ has a lower differential order 
than the differential order of every PDE in the given system, 
the definition of local triviality \eqref{3Dvolconslaw-triv-offsolns} of $\volT[\depvars]$ 
off of the solution space takes the form 
$\volT[\depvars] = \Div\,\vec{\Theta}[\depvars]$,
with $\Gamma_\triv[\depvars] \equiv 0$. 
Then, because total spatial divergences comprise the kernel of the spatial Euler operator
(cf Lemma~\ref{Eulerop-props}),
we conclude that the condition \eqref{3Dvoldensity-eulercond} is necessary and sufficient 
for $\volT[\depvars]$ to be locally trivial. 
We thereby have 
$D_t \volT[\depvars]|_\Esp = \Div\,D_t\vec{\Theta}[\depvars]|_\Esp = -\Div\,\vec{\Psi}[\depvars]|_\Esp$,
which implies $\Div(\vec{\Psi}[\depvars] + D_t\vec{\Theta}[\depvars])|_\Esp =0$ 
is a local spatial divergence conservation law \eqref{spatial-div-3Dconslaw}. 
If the given PDE system admits only trivial conservation laws of that type, 
then from Definition~\ref{defn:3Dspatial-div-conslaw-loctriv} 
we have $(\volX[\depvars] + D_t\vec{\Theta}[\depvars])|_\Esp$ is a curl, 
and consequently $\volX[\depvars]$ is locally trivial \eqref{3Dvolconslaw-triv-offsolns}. 
This completes the proof. 

We can derive a similar formulation of the conditions in Theorem~\ref{thm:3Dvolconslaw-globaltriv} 
for global triviality
by the following argument. 

Suppose the flux integral $\oint_{\p\V}\vec{\Theta}[\depvars]\cdot\dS\,\big|_\Esp$ 
vanishes identically for an arbitrary regular volume $\V$. 
Then, Stokes' theorem shows that 
$\vec{\Theta}[\depvars]|_\Esp = \Curl\,\vec{\Xi}[\depvars]|_\Esp$ 
holds for a vector differential function $\vec{\Xi}[\depvars]$,
whereby we have $\Div\,\vec{\Theta}[\depvars]|_\Esp = 0$ 
by the first of the differential identities \eqref{divcurlgrad-idents}. 
Conversely, $\Div\,\vec{\Theta}[\depvars]|_\Esp = 0$ 
implies that $\vec{\Theta}[\depvars]|_\Esp = \Curl\,\vec{\Xi}[\depvars]|_\Esp$ holds, 
if the given PDE system has no non-trivial local spatial divergence conservation laws. 
Consequently, under this condition, 
the vanishing of $\oint_{\p\V}\vec{\Theta}[\depvars]\cdot\dS|_\Esp$ due to Stokes' theorem
will hold when and only when 
$0=\Div\,\vec{\Theta}[\depvars]|_\Esp = \volT[\depvars]|_\Esp$. 

Suppose instead the flux integral $\oint_{\p\V}\vec{\Theta}[\depvars]\cdot\dS\,\big|_\Esp$ 
is time dependent for an arbitrary regular volume $\V$. 
This is equivalent to the volume integral 
$\int_\V \Div\,\vec{\Theta}[\depvars]\dV|_\Esp$ 
being time dependent. 
Since $\V$ is arbitrary,
an equivalent condition is that $\Div\,\vec{\Theta}[\depvars]|_\Esp$ itself must be time dependent. 
Consequently, time dependence of $\oint_{\p\V}\vec{\Theta}[\depvars]\cdot\dS\,\big|_\Esp$
will hold when and only when 
$\Div\,\vec{\Theta}[\depvars]|_\Esp = \volT[\depvars]|_\Esp$ is time dependent. 

Thus we have established the following result. 

\begin{theorem}\label{thm:3Dglobalvolconslaw-nontriv}
Suppose the conserved density $\volT[\depvars]$ 
in a time-dependent local volumetric conservation law \eqref{3Dvolconslaw} of a PDE system \eqref{3Dpde} in $\Rnum^3$ 
has a lower differential order off of the solution space 
than the differential order of all PDEs in the given system. 
Then a necessary condition for a time-dependent local volumetric conservation law \eqref{3Dvolconslaw} of a PDE system \eqref{3Dpde} 
to yield a globally trivial volumetric conservation law \eqref{3Dglobalvolconslaw} 
for an arbitrary regular volume $\V\subset\pdedom$
is that the conserved volumetric density $\volT[\depvars]|_\Esp$ 
identically vanishes under the spatial Euler operator \eqref{3Dvoldensity-eulercond}. 
A sufficient condition is that the conserved density $\volT[\depvars]$ 
satisfies the necessary condition \eqref{3Dvoldensity-eulercond} and is time dependent, 
or vanishes identically, 
and also that the PDE system has no non-trivial topological flux conservation laws \eqref{spatial-div-3Dglobalconslaw}. 
\end{theorem}

This provides a simple sufficient condition for global non-triviality. 

\begin{corollary}\label{cor:3Dglobalvolconslaw-suff-nontriv}
If the conserved density $\volT[\depvars]$ of 
a time-dependent (local or global) volumetric conservation law 
has a lower differential order off of the solution space 
than the differential order of all PDEs in the given 
and satisfies the variational condition $\hat \Eop_{\depvars}(\volT[\depvars]) \neq 0$, 
then the conservation law is locally and globally non-trivial. 
\end{corollary}

Numerous physical examples of volumetric conservation laws are locally and globally non-trivial: 
\begin{itemize}\itemsep=0pt
\item 
mass in fluid flow, gas dynamics, and MHD;
\item 
momentum and angular momentum in fluid flow, gas dynamics, electromagnetism, and MHD;
\item
Galilean momentum in fluid flow, gas dynamics, and MHD;
\item
boost momentum in electromagnetism;
\item
energy in gas dynamics, ideal fluid flow, electromagnetism, and ideal inviscid MHD;
\item
entropy in fluid flow;
\item
helicity in ideal fluid flow;
\item
cross-helicity in ideal inviscid MHD. 
\end{itemize}

Two examples of a locally and globally trivial conservation law 
are electric charge-current conservation in vacuum electromagnetism
and Ertel's theorem in ideal fluid flow. 
shown in Section~\ref{sec:ex:volCLs}. 

Physical examples of locally trivially volumetric conservation laws 
connected with divergence-type temporal conservation laws that yield 
globally non-trivial surface-flux constants of motion 
arise in electromagnetism, MHD, incompressible fluid flow, 
and compressible fluid flow with diabatic heating, 
which are discussed in Section~\ref{sec:ex:temporalCLs}.

\section{Physical examples of topological and dynamical conservation laws}\label{sec:examples}

We will now present examples of conservation laws of physical significance,
and their interrelationships, that arise in the physically important PDE systems
for fluid flow, gas dynamics, electromagnetism, and magnetohydrodynamics.
For a survey of all known conservation laws for these systems, 
the reader is referred to Refs.
\cite{CRC-book,Bat-book,MajBer-book,MarPul-book,CavMor1987,CavMor1989,GusYum,AncDar2009,AncDar2010,Anc2013,KelCheObe,CheObe2014,Che2014,AncDar2015}
for fluid flow and gas dynamics; 
Refs.
\cite{AncPoh2001,AncPoh2004,AncBlu,AncThe}
for electromagnetism;
and Refs.
\cite{BerFie,FinAnt,SagTurYan,TurYan,Kats2003,Kats2004,CheAnc2008,WebDasMcKHuZan,Yah2017,WebAnc2017,AncPshWol}
for magnetohydrodynamics. 

We emphasize that the conservation law examples here will not be exhaustive; 
they have been chosen to illustrate all of the different types of local and global conservation laws that arise in three dimensions. 
In particular, all of the physical examples of these conservation laws 
mentioned in Sections~\ref{sec:CLs} and~\ref{sec:non-triv} will be discussed.

\subsection{Topological conservation laws}\label{sec:ex:topolCLs}

First, examples of topological flux conservation laws and topological circulation conservation laws will be presented in both their local and global forms. 

\subsubsection{Spatial divergence/topological flux conservation laws}\label{sec:ex:divCLs}

There are several physical examples of topological flux conservation laws \eqref{spatial-div-3Dglobalconslaw}
that are (locally and globally) non-trivial. 

One main example is magnetic flux in electromagnetism and MHD. 
In local form, the conserved flux density is $\volX=\B$ 
which physically describes the absence of magnetic charges, $\div\B =0$. 
Because the differential order of $\volX = \B$ is zero, 
it clearly cannot be expressed as 
a curl of a differential function in terms of the dynamical variables 
$(\E,\B)$ in electromagnetism, or $(\rho,\u,\B)$ in MHD. 
Hence, $\div\B =0$ is a locally non-trivial spatial divergence conservation law,
by Definition~\ref{defn:3Dspatial-div-conslaw-loctriv}. 

The global form of this local conservation law 
consists of a topological flux conservation law 
holding on any connected volume 
within the physical domain of the electromagnetic or MHD system. 
If a volume has a single boundary surface $\S=\p\V$, 
then the topological flux is given by a vanishing magnetic flux integral 
\begin{equation}\label{ex:EM:magneticcharge:zero}
\oint_{\S} \B\cdot\dS =0 . 
\end{equation}
Alternatively, if the boundary of the volume is given by two disjoint surfaces $\S_1$ and $\S_2$, 
then the topological flux has the form 
\begin{equation}\label{ex:EM:magneticflux}
\oint_{\S_1} \B\cdot\dS = \oint_{\S_2} \B\cdot\dS 
\end{equation}
which may be non-zero. 
The physical meaning of these topological flux integrals is
that the total flux of magnetic field lines measured through a closed surface is invariant
under continuous deformations of the surface, 
and that this flux vanishes for surfaces that bound connected domains, 
even when the magnetic field is time-dependent. 
These global conservation laws \eqref{ex:EM:magneticcharge:zero} and \eqref{ex:EM:magneticflux}
are non-trivial from Proposition~\ref{prop:spatial-div-conslaw-globaltriv}. 

A similar physical example is the electric flux in a connected volume $\V$ 
within the physical domain of an electromagnetic system
that contains no electric charges: $\div\E =0$ in $\V$. 
This is a locally non-trivial spatial divergence conservation law
which yields a topological flux conservation law 
\begin{equation}\label{ex:EM:electriccharge:zero}
\oint_{\S} \E\cdot\dS =0 
\end{equation}
whose conserved flux density is $\volX=\E$. 

Another physical example of a local spatial divergence conservation law \eqref{spatial-div-3Dconslaw} 
is given by the incompressibility equation \eqref{eq:incompr} 
in fluid flow and in MHD. 
The conserved density $\volX=\u$ is the velocity,
which clearly cannot be expressed as 
a curl of a differential function in terms of the dynamical variables 
$(\u,\rho,p)$ in fluid flow, or $(\rho,\u,\B)$ in MHD. 
Hence, this conservation law is locally non-trivial, 
by Definition~\ref{defn:3Dspatial-div-conslaw-loctriv}. 

The global form of the incompressibility conservation law 
consists of a topological flux conservation law 
holding on any static connected volume 
within the physical domain of the fluid or MHD system. 
For a volume $\V$ with a single boundary surface $\S=\p\V$, 
the topological flux is given by
\begin{equation}\label{ex:incompress:streamlineflux:zero}
\oint_{\p\V} \u \cdot\dS =0 . 
\end{equation}
Physically, this integral expresses that there are no sources or sinks of streamline flux. 
For a volume whose boundary of consists of two disjoint closed surfaces $\S_1$ and $\S_2$, 
the topological flux expresses 
\begin{equation}\label{ex:incompress:streamlineflux}
\oint_{\S_1} \u\cdot\dS = \oint_{\S_2} \u\cdot\dS
\end{equation}
showing that the total streamline flux is invariant
under continuous deformations of a static closed surface
within the fluid or MHD system. 
This has the physical meaning that the fluid/plasma volume is preserved. 
Both of these global conservation laws \eqref{ex:incompress:streamlineflux:zero} and \eqref{ex:incompress:streamlineflux}
are non-trivial from Proposition~\ref{prop:spatial-div-conslaw-globaltriv}. 

Another example of a local spatial divergence conservation law 
arises in incompressible fluid flow and incompressible MHD,
where the pressure satisfies the respective Laplace-type equations 
\eqref{eq:Eul:press} and \eqref{eq:MHD:press}. 
These equations arise from the compatibility between the velocity equation and the incompressibility equation,
and consequently they have a spatial divergence form \eqref{spatial-div-3Dconslaw} 
in which 
$\volX=(1/\rho)\grad p + (\u\cdot\nabla)\u$ for fluid flow,
and $\volX=(1/\rho)\grad p + (\u\cdot\nabla)\u  - (1/\rho)\B\times(\tfrac{1}{\mu_0}\curl\B)$
for MHD. 
In both cases, 
$\volX = -\u_t$ and $\Div\,\volX = -\div\u_t= -D_t\div\u=0$ 
holds for solutions of the fluid system and the MHD system. 
Consequently, these spatial divergence conservation laws are locally non-trivial. 

In global form, the resulting topological flux conservation laws 
for any static closed surface $\S$ are given by 
\begin{equation}\label{ex:incompressFD:streamlineflux}
\int_S \big( (1/\rho)\grad p + (\u\cdot\nabla)\u \big)\cdot\dS 
= - \dfrac{d}{dt} \int_S \u\cdot\dS 
=0
\end{equation}
in incompressible fluid flow,
and 
\begin{equation}\label{ex:incompressMHD:streamlineflux}
\int_S \big( (1/\rho)\grad p + (\u\cdot\nabla)\u - (1/\rho)\B\times(\tfrac{1}{\mu_0}\curl\B) \big)\cdot\dS 
= - \dfrac{d}{dt} \int_S \u\cdot\dS 
=0
\end{equation}
in incompressible MHD. 
Thus, these global conservation laws are consequences of 
the streamline flux conservation law \eqref{ex:incompress:streamlineflux:zero}. 
A similar result holds for disjoint static closed surfaces $\S_1$ and $\S_2$
that bound any static connected volume within the physical domain of the fluid or MHD system. 

An example of a (locally and globally) trivial spatial divergence conservation law 
is the vorticity relation $\div\vort =0$ in fluid flow and gas dynamics,
where $\vort=\curl\u$ is a curl. 
The global form of this conservation law for any static closed surface $\S=\p\V$ 
within the fluid or gas 
is an identity
\begin{equation}
\oint_{\S} (\nabla\times\u)\cdot\dS \equiv 0
\end{equation}
by Stokes' theorem, since $\p\S=\emptyset$.

\subsubsection{Spatial curl/topological circulation conservation laws}\label{sec:ex:curlCLs}

A main physical example of a (locally and globally) non-trivial 
topological circulation conservation law \eqref{spatial-curl-3Dglobalconslaw}
arises in irrotational gas dynamics and fluid flow. 
These physical systems have vanishing vorticity $\vort= \curl\u=0$ 
everywhere in the gas or fluid,
and thus $\surfX = \u$ is a conserved circulation density
for a local spatial curl conservation law  \eqref{spatial-curl-3Dconslaw}. 
Because the differential order of $\u$ is zero, 
it clearly cannot be expressed as 
a gradient a differential function in terms of the dynamical variables 
$(\u,\rho,p)$ in gas dynamics, or $(\u,\rho)$ in fluid flow. 
Consequently, the spatial curl conservation law $\curl\u =0$ is locally non-trivial
by Definition~\ref{defn:3Dspatial-curl-conslaw-loctriv}. 

The global form of this conservation law is a topological circulation integral \eqref{spatial-curl-3Dglobalconslaw}
holding on any static connected non-closed surface $\S$ 
within the physical domain of the gas or fluid system. 
If the boundary of the surface consists of a single closed curve $\C=\p\S$,
then the circulation integral is given by
\begin{equation}\label{ex:FD:circ:closed}
\oint_{\C} \u\cdot \ds =0 .
\end{equation}
Likewise, if instead the boundary $\p\S$ of the surface is given by 
two disjoint closed curves $\C_1$ and $\C_2$,
then the two corresponding circulation integrals are equal, 
\begin{equation}\label{ex:FD:circ}
\oint_{\C_1} \u\cdot\ds = \oint_{\C_2} \u\cdot\ds . 
\end{equation}
The physical meaning of these topological conservation laws is that 
the net circulation of streamlines measured around a static closed curve 
is invariant under continuous deformations of the curve, 
and that this circulation vanishes for all static closed curves 
within the spatial domain of the physical system. 
Both conservation laws \eqref{ex:FD:circ:closed} and \eqref{ex:FD:circ}
are non-trivial from Proposition~\ref{prop:spatial-curl-conslaw-globaltriv}. 

Another physical example is magnetic circulation in a magnetostatics system:
$\curl\B=0$ and $\div\B=0$. 
Here $\surfX=\B$ is the conserved circulation density 
for the local spatial curl conservation law $\curl\B=0$. 
Clearly, since $\B$ is the dynamic variable, 
it cannot be expressed as a gradient a differential function in terms of itself. 
Hence this conservation law is locally non-trivial
by Definition~\ref{defn:3Dspatial-curl-conslaw-loctriv}. 

For any closed connected curve $\C$ 
within the physical domain the magnetostatics system, 
the resulting global conservation law is given by 
\begin{equation}\label{ex:magnetostatics:magneticcirc:closed}
\oint_{\C} \B\cdot \ds =0 .
\end{equation}
This is a topological circulation integral, 
expressing that the net magnetic circulation vanishes. 
Similarly, for any two disjoint closed curves $\C_1$ and $\C_2$, 
the two corresponding circulation integrals are equal, 
\begin{equation}\label{ex:magnetostatics:magneticcirc}
\oint_{\C_1} \B\cdot\ds = \oint_{\C_2} \B\cdot\ds . 
\end{equation}
These global conservation laws \eqref{ex:magnetostatics:magneticcirc:closed} and \eqref{ex:magnetostatics:magneticcirc}
are non-trivial from Proposition~\ref{prop:spatial-curl-conslaw-globaltriv}. 

A counterpart of the previous example is 
the electric field circulation in an electrostatics system:
$\curl\E=0$ and $\div\E=4\pi\rho$,
where $\surfX=\E$ is the conserved density 
for the local spatial curl conservation law $\curl\E=0$.
This conservation law is locally non-trivial 
and yields topological circulation conservation laws analogous to 
the magnetic circulation integrals  \eqref{ex:magnetostatics:magneticcirc:closed} and \eqref{ex:magnetostatics:magneticcirc}. 

There is a similar example arising in equilibrium ideal MHD, 
where the velocity and magnetic field satisfy the curl equation $\curl(\u\times\B)=0$
due to $\u_t=0$ and $\B_t=0$. 
This curl equation is a spatial curl conservation law that is locally non-trivial 
because $\surfX=\u\times\B$ has differential order zero. 
The global form of this conservation law is given by 
the non-trivial topological circulation integral 
\begin{equation}\label{ex:MHD:ideal:equil:circ}
\oint_{\C}  (\u \times \B) \cdot \ds =0
\end{equation}
for any static closed connected curve $\C$ within the physical domain of the MHD system.
Physically, this global conservation law expresses that no net circulation is produced
by the electric field $\E= -\u\times\B$ around static closed curves.  

In the important special case \cite{Bog}
when the MHD equilibrium is field-aligned, $\u\times\B=0$,  
the previous conservation law becomes trivial.

\subsubsection{Spatial gradient conservation laws}\label{sec:ex:gradCLs}

Irrotational equilibria of ideal fluids provide the main physical example of 
a local spatial gradient conservation law \eqref{spatial-curl-3Dglobalconslaw},
where $\u_t=0$, $\rho_t=0$, and $\vort=0$. 
The fluid velocity equation \eqref{eq:Eul:vel} for this time-independent physical system 
reduces to $\grad( \tfrac{1}{2}|\u|^2 + e(\rho)+ p/\rho )=0$
when the fluid is either constant-density \eqref{eq:constdens} or barotropic \eqref{eq:barotropic}. 
Hence $\curvX = \tfrac{1}{2}|\u|^2 + e(\rho)+ p/\rho$ is the conserved density. 
The resulting local conservation law is known as Bernoulli's principle 
\cite{Mil-Tho}.

\subsection{Dynamical conservation laws}\label{sec:ex:dynamCLs}

Examples of dynamical conservation laws of volumetric, surface-flux, circulatory type
in both their local and global forms 
will be presented next. 
Since the results in Appendix~\ref{sec:interrelations}
show that circulatory conservation laws give rise to 
surface-flux and volumetric conservation laws, 
and also that surface-flux conservation laws give rise to volumetric conservation laws,
we arrange the examples in this order: circulatory; surface-flux; volumetric. 

\subsubsection{Circulatory conservation laws}\label{sec:ex:curvCLs}

The main physical example of a circulatory conservation law that is locally and globally non-trivial 
is circulation in irrotational ideal fluid flow. 

In local form the fluid circulation conservation law is given by 
the velocity equation \eqref{eq:Eul:vel}
when the fluid flow has no vorticity, $\vort=0$,
and is either constant-density \eqref{eq:constdens} or barotropic \eqref{eq:barotropic}. 
This yields a local circulatory conservation law \eqref{3Dcurvconslaw}
where the conserved density is $\curvT = \u$,
and the spatial flow is $\curvX = \tfrac{1}{2}|\u|^2 + e(\rho)+ p/\rho$,
with $e(\rho)=\const$ in the constant-density case, 
and $e(\rho)= \int (p(\rho)/\rho^2)d\rho$ in the barotropic case. 
Because the differential order of $\curvT = \u$ is zero, 
it clearly cannot be expressed as 
a gradient of a differential function in terms of the dynamical variables 
$(\u,p)$ in the constant-density case, or $(\u,\rho)$ in the barotropic case. 
Hence this conservation law is locally non-trivial from Definition~\ref{defn:3Dcurvconslaw-loctriv}. 

For an arbitrary fixed (static) curve $\C$ within the fluid, 
the net circulation $\int_{\C} \u\cdot\ds$ 
satisfies the global circulatory conservation law \eqref{3Dglobalcurvconslaw}:
\begin{equation}\label{ex:irrFD:circulation}
\frac{d}{dt} \int_{\C} \u\cdot\ds = -\big( \tfrac{1}{2}|\u|^2 + e(\rho)+ p/\rho \big)\Big|_{\p\C}.
\end{equation}
Here the net spatial flow physically measures the amount of circulation 
escaping through the endpoints $\p\C$ of the curve. 
For closed curves, $\p\C=\emptyset$,
this global conservation law becomes 
\begin{equation}\label{ex:irrFD:circulation:closedcurv}
\frac{d}{dt} \oint_{\C} \u\cdot \ds =0 ,
\end{equation}
showing that the total circulation for any fixed closed curve within the fluid 
is a constant of motion. 
This is a static counterpart of Kelvin's circulation theorem 
for moving closed curves transported in a fluid
and it holds as a consequence of the topological circulation conservation law \eqref{ex:FD:circ:closed}. 
(We will discuss this further in Section~\ref{sec:ex:temporalCLs}.) 
The global conservation laws \eqref{ex:irrFD:circulation} and \eqref{ex:irrFD:circulation:closedcurv}
are globally non-trivial from Theorem~\ref{thm:3Dcurvconslaw-globaltriv}. 

Two examples of locally trivial circulatory conservation laws arise from 
the density gradient and entropy gradient in fluid flow. 

First, for incompressible fluids, 
the gradient of the density transport equation \eqref{eq:denstransport} 
yields $(\grad\rho)_t + \grad( \u\cdot\grad \rho) = 0$. 
This has the form of a local circulatory conservation law \eqref{3Dcurvconslaw},
where the conserved density is a gradient $\curvT=\grad\rho$, 
and the spatial flow is $\curvX=\u\cdot\grad \rho = - \rho_t$. 
Hence the conserved current $(\curvT,\curvX)=(\grad\rho,\u\cdot\grad \rho) =(\grad\rho, - \rho_t)$
is locally trivial \eqref{3Dcurvconslaw-triv}. 
The global form of this circulatory conservation law is given by 
\begin{equation}
\frac{d}{dt} \int_{\C} \grad\rho\cdot\ds 
= \int_{\C} \grad\rho_t\cdot\ds 
= \rho_t \Big|_{\p\C} 
= -\big( \u\cdot\grad \rho )\Big|_{\p\C} 
\end{equation}
which holds as an integral identity. 
Its global triviality is a consequence of Proposition~\ref{prop:3Dcurvconslaw-nontrivflow},
since $\rho$ is time dependent for an arbitrary solution of the irrotational fluid system. 

Second, for locally adiabatic fluids, 
the gradient of the entropy transport equation \eqref{eq:isentropic} 
yields $(\grad S)_t + \grad( \u\cdot\grad S) = 0$,
which has the form of a local circulatory conservation law \eqref{3Dcurvconslaw}, 
with $\curvT=\grad S$ and $\curvX=\u\cdot\grad S = - S_t$. 
This conservation law has the same properties as the locally and globally trivial one 
just discussed for the density gradient. 

\subsubsection{Surface-flux conservation laws}\label{sec:ex:surfCLs}

One primary physical example of a surface-flux conservation law that is locally and globally non-trivial 
is magnetic induction (Faraday's law) in electromagnetism and ideal MHD. 

The magnetic induction equation \eqref{eq:MHD:B:induct} in MHD 
yields a local surface-flux conservation law \eqref{3Dsurfconslaw}
where the conserved density is $\surfT = \B$,
and the spatial circulation flux is $\surfX = \u \times \B -\tfrac{\eta}{\mu_0}\curl \B$. 
Because the differential order of $\surfT = \B$ is zero, 
it cannot be expressed as 
a curl of a differential function in terms of the dynamical variables $(\rho,\u,\B)$,
and thus this conservation law is locally non-trivial from Definition~\ref{defn:3Dsurfconslaw-loctriv}. 

For an arbitrary fixed non-closed surface $\S$, 
the net magnetic flux $\int_{\S} \B\cdot\dS$ 
satisfies the global surface-flux conservation law \eqref{3Dglobalsurfconslaw}
which is given by 
\begin{equation}\label{ex:MHD:Faraday}
\frac{d}{dt} \int_{\S} \B\cdot\dS = - \int_{\p\S} (\B\times\u+\tfrac{\eta}{\mu_0}\curl \B)\cdot\ds . 
\end{equation}
Its physical meaning in the case $\eta=0$ of ideal MHD is that 
the rate of change of magnetic flux enclosed by a fixed surface in a plasma/liquid metal 
is balanced exactly by the transport of the magnetic field through the surface by the flow. 
In general, for closed surfaces, $\p\S=\emptyset$,
this global conservation law \eqref{ex:MHD:Faraday} becomes 
\begin{equation}\label{ex:MHD:Faraday:closedsurf}
\frac{d}{dt} \oint_{\S} \B\cdot \dS =0 ,
\end{equation}
which holds as a consequence of the topological magnetic flux conservation law \eqref{ex:EM:magneticcharge:zero}.
(We will discuss this further in Section~\ref{sec:ex:temporalCLs}.) 
These global conservation laws \eqref{ex:MHD:Faraday:closedsurf} and \eqref{ex:MHD:Faraday}
are globally non-trivial from Proposition~\ref{prop:3Dsurfconslaw-nontrivcirc}. 

The same local and global conservation laws for magnetic flux arise in electromagnetism, 
where the magnetic induction equation has the form \eqref{eq:EM:B}. 
In global form, this is a statement of Faraday's law
\begin{equation}\label{ex:EM:Faraday}
\frac{d}{dt} \int_{\S} \B\cdot\dS = -c\int_{\p\S} \E\cdot\ds ,
\end{equation}
which gives the amount of electromotive force circulating 
around the boundary of a fixed surface when the net magnetic flux is varying in time. 

An analogous surface-flux conservation law is given by the electric field equation \eqref{eq:EM:vac:E}
in vacuum electromagnetism. 
The conserved density is $\surfT = \E$,
and the spatial circulation flux is $\surfX = -c\B$. 
For an arbitrary fixed non-closed surface $\S$, 
the global form of this conversation law states that 
a time-varying electric flux through the surface 
generates a magnetic circulation along the boundary of the surface, 
\begin{equation}\label{ex:EM:magneticcirc}
\frac{d}{dt} \int_{\S} \E\cdot\dS = c\int_{\p\S} \B\cdot\ds 
\end{equation}
(where $\E_t$ is the electric displacement current).
When closed surfaces, $\p\S=\emptyset$, are considered, 
the resulting global conservation law 
\begin{equation}\label{ex:EM:magneticcirc:closedsurf}
\frac{d}{dt} \oint_{\S} \E\cdot \dS =0 
\end{equation}
holds as a consequence of the topological electric flux conservation law \eqref{ex:EM:electriccharge:zero}. 
These conservation laws are locally and globally non-trivial
by the same argument explained for the magnetic induction conservation laws. 

A more general situation where a non-trivial surface-flux conservation law occurs 
is for the microscopic electric field given by Maxwell's equation \eqref{eq:EM:E}
when the electric charge distribution in a volume $\V$ is static but non-zero, 
namely $\rho_t=0$, $\grad\rho\not\equiv 0$. 
In this situation, the electric current will be source-free, $\div\J=0$,
which implies that it can be expressed in a curl form $\J=\curl\vec{K}$
if the volume $\V$ is topologically trivial. 
As a consequence, Maxwell's equation \eqref{eq:EM:E} takes the form of 
a local surface-flux conservation law $\E_t = \curl\,(c \B - 4\pi \vec{K})$,
where the conserved density is $\surfT = \E$ 
and the spatial circulation flux is $\surfX = -c\B + 4\pi \vec{K}$. 
The global form of this conversation law 
for the closed boundary surface $\S=\p\V$ shows that the net electric flux 
is a constant of motion \eqref{ex:EM:magneticcirc:closedsurf}. 
Unlike the vacuum case (when $\rho\equiv 0$), 
here this constant of motion can be non-zero 
as it measures the total charge contained in $\V$:
\[
\int_{\S} \E\cdot\dS = 4\pi \int_{\V} \rho\,\dV . 
\]

Another example of a non-trivial surface-flux conservation law occurs in incompressible electron MHD \cite{GorKinRud}
when the generalized vorticity of the flow of electrons is considered \cite{Lyu}.
In this model, the electron velocity $\u$ is generated by the magnetic field $\B$ via 
\[
\u= -\tfrac{c}{n\,e}\curl\B
\]
where $n$ is the constant electron density. 
The generalized vorticity which is given by 
\begin{equation}\label{ex:EMHD:vorteqn}
\vort_e = \curl\u - \tfrac{e}{c m_e}\B,
\quad
\div\vort_e=0
\end{equation}
satisfies the dynamical equation 
\begin{equation}\label{ex:EMHD:vort:eqn}
\vort_e{}^{\mathstrut}_{t} = \curl(\u\times\vort_e).
\end{equation}
This equation constitutes a surface-flux conversation law. 
Its global form states that the vorticity flux 
through an arbitrary fixed non-closed surface $\S$ 
generates a vorticity circulation along the boundary of the surface, 
\begin{equation}\label{ex:EMHD:vortcirc}
\frac{d}{dt} \int_{\S} \vort_e \cdot\dS = \int_{\p\S} (\u\times\vort_e)\cdot\ds . 
\end{equation}
For closed surfaces, $\p\S=\emptyset$, 
the net vorticity circulation is a constant of motion,
\[
\frac{d}{dt} \oint_{\S} \vort_e\cdot \dS =0 . 
\]

An example of a trivial surface-flux conservation law is given by 
the vorticity transport equation \eqref{eq:Eul:vort}
in an ideal fluid that has either constant density or a barotropic pressure. 
Here the conserved density is a curl, $\surfT=\vort=\curl\u$, 
while the spatial circulation flux is given by 
$\surfX=\vort \times \u = -u_t -\grad( \tfrac{1}{2}|\u|^2 + e(\rho)+ p/\rho )$
due to the velocity equation \eqref{eq:Eul:vel}. 
The conserved current $(\surfT,\surfX)$ is locally trivial \eqref{3Dsurfconslaw-triv}, 
from Definition~\ref{defn:3Dsurfconslaw-loctriv}. 
In global form, for an arbitrary fixed (static) surface $\S$ within the fluid, 
this surface-flux conservation law is given by 
\begin{equation}
\frac{d}{dt} \int_{\S} \vort\cdot\dS
= \frac{d}{dt} \int_{\p\S} \u\cdot\ds 
= \int_{\p\S} \u_t\cdot\ds 
= -\oint_{\p\S} (\vort \times \u) \cdot \ds 
\end{equation}
which holds as an integral identity. 
Its global triviality is a consequence of Proposition~\ref{prop:3Dsurfconslaw-nontrivcirc},
since $\u$ is time dependent for an arbitrary solution of the fluid system. 

A related physical example of a trivial surface-flux conservation law 
arises in irrotational constant-density ideal fluid flow with a non-constant (locally adiabatic) entropy density,
where the fluid velocity and the entropy density obey the equations
\[
\u_t + \grad(\tfrac{1}{2}|\u|^2 + p/\rho) =0,
\quad
S_t + \u\cdot\grad\, S = 0 .
\]
By taking the cross-product of the velocity equation with $\grad\,S$,
and adding the cross-product of the velocity with the gradient of the entropy transport equation, 
we obtain a local surface-flux conservation law \eqref{3Dsurfconslaw}
having the conserved density 
\[
\surfT =\u\times\grad\,S = -\curl(S\u)
\]
and the spatial circulation flux 
\[
\surfX = (\tfrac{1}{2}|\u|^2 + p/\rho)\grad\, S - (\u\cdot\grad\, S)\u
= (S\u)_t -\grad( (\tfrac{1}{2}|\u|^2 + p/\rho)S ). 
\]
This yields a conserved current $(\surfT,\surfX)$ that is locally trivial \eqref{3Dsurfconslaw-triv}, 
from Definition~\ref{defn:3Dsurfconslaw-loctriv}. 
The global form of this surface-flux conservation law 
on an arbitrary fixed (static) surface $\S$ within the fluid
is given by the identity 
\[
\begin{aligned}
& \frac{d}{dt} \int_{\S} (\u\times\grad\, S)\cdot\dS
= -\frac{d}{dt} \int_{\S} \curl(S\u)\cdot\dS
\\
& 
= -\oint_{\p\S} (S\u)_t\cdot \ds 
= -\oint_{\p\S} \big( (\tfrac{1}{2}|\u|^2 + p/\rho)\grad\, S - (\u\cdot\grad\, S)\u \big)\cdot \ds . 
\end{aligned}
\]
This conservation law is globally trivial,
since when $\u\times\grad\,S =\surfT \neq 0$,
we have that $\u$ and $\grad\,S$ are not collinear,
whereby $(\tfrac{1}{2}|\u|^2 + p/\rho)\grad\, S - (\u\cdot\grad\, S)\u = \surfX\neq 0$. 

\subsubsection{Volumetric conservation laws}\label{sec:ex:volCLs}

We begin by presenting the main physical examples of non-trivial volumetric conservation laws:
mass, entropy, momentum, energy, helicity and cross-helicity. 
In table~\ref{table:nontriv-volCLs},
these conservation laws are written in local form \eqref{3Dvolconslaw}
for the volumetric density $\volT$ and spatial flux $\volX$,
which allows the similarities among the various conservation laws to be seen. 
(The abbreviations FD, GD, EM, MHD to refer to the respective physical systems 
for fluid flow, gas dynamics, electromagnetism, magnetohydrodynamics.)

Each of these conservation laws is both locally and globally non-trivial. 
The non-triviality of mass, entropy, momentum, energy, and cross-helicity
is a consequence of Corollary~\ref{cor:3Dglobalvolconslaw-suff-nontriv}. 
For helicity, its non-triviality can be shown by a generalization of the proof of this Corollary
adapted to the specific form of the conserved density. 

The physical meaning of the global form of the momentum and angular momentum conservation laws,
as well as the conservation laws for mass, entropy, and energy, 
is very well-known. 
Helicity has the physical meaning that the volumetric quantity $\int_\V \u\cdot\vort\,\dV$
measures the linkage of vortex lines in any fixed (static) volume $\V$ within the fluid. 
Similarly for cross-helicity, 
the volumetric quantity $\int_\V \u\cdot\B\,\dV$ in a fixed volume $\V$ 
physically measures  the collinearity between the velocity and the magnetic field within the plasma/liquid. 

For a gas or fluid confined to a fixed volume $\V$, 
the flow velocity vector at the boundary will be tangential to boundary surface $\p\V$,
$\u|_{\p\V}\cdot\nor=0$. 
As a consequence, the flux of mass will vanish at the boundary. 
Likewise, the flux of entropy will vanish when the flow is locally adiabatic,
and the flux of energy will vanish when the flow has no viscosity. 
In these situations, the total mass, entropy, and energy are constants of motion. 
Similarly, for an ideal, inviscid plasma or liquid metal confined to a fixed volume $\V$, 
if the magnetic field is also confined to $\V$ then 
both the energy flux and cross-helicity flux will vanish at the boundary $\p\V$
(as seen from table~\ref{table:nontriv-volCLs}), 
due to $\u|_{\p\V}\cdot\nor=0$ and $\B|_{\p\V}\cdot\nor=0$. 
Consequently, the total energy and the total cross-helicity will be constants of motion. 

In contrast, 
the helicity for an ideal (incompressible or barotropic) fluid confined to a fixed volume $\V$ 
is not in general a constant of motion. 
The boundary condition for the vorticity vector $\vort$ in the flow is that 
$\vort|_{\p\V}\, \parallel\, \nor$, 
and thus the total helicity satisfies
$\frac{d}{dt}\int_\V \u\cdot\vort\,\dV = -\oint_{\p\V} (p/\rho + e -|\u|^2)\vort\cdot\nor\,dA$ 
which is generally non-vanishing. 
In particular, the rate of change in total helicity in the volume $\V$ is balanced by vorticity within the boundary surface $\p\V$, 
$\vort|_{\p\V}\cdot\nor\neq 0$. 

Other well-known examples of non-trivial volumetric conservation laws are
angular momentum, Galilean momentum, and boost momentum. 

\begin{table}[htb]
\caption{Non-trivial volumetric conservation laws}
\label{table:nontriv-volCLs}
\centering
\begin{tabular}{c|c|c|c}
\hline
Name & Density $\volT$ & Flux $\volX$ & Physical system 
\\
\hline
\hline
mass
& 
$\rho$
&
$\rho\u$
& 
FD, GD, MHD
\\
\hline
entropy
&
$S$
&
$S\u$
&
locally adiabatic FD
\\
\hline
\tbox{momentum}{($\hat{\xi}=\mathbf{i,j,k}$)}
& 
$\rho\hat{\xi}\cdot\u$
&
$\rho(\hat{\xi}\cdot\u)\u + \bar p\hat{\xi}-\mu(\hat{\xi}\cdot\nabla)\u$
&
FD, GD
\\
&
$\rho\hat{\xi}\cdot\u$
&
$\begin{aligned}
&\rho(\hat{\xi}\cdot\u)\u -\tfrac{1}{\mu_0}(\hat{\xi}\cdot\B)\B\\
&\quad  + \big(\bar p+\tfrac{1}{2\mu_0}|\B|^2\big)\hat{\xi} -\mu(\hat{\xi}\cdot\nabla)\u
\end{aligned}$
&
MHD
\\
&
$\tfrac{1}{c}\hat{\xi}\cdot(\E\times\B)$ 
&
$\begin{aligned}
&(\hat{\xi}\cdot\E)\E + (\hat{\xi}\cdot\B)\B\\
&\quad -\tfrac{1}{2}(|\E|^2 +|\B|^2)\hat{\xi}
\end{aligned}$
&
vacuum EM
\\
\hline
energy 
& 
$\rho(\tfrac{1}{2}|\u|^2 +e)$
&
$(\tfrac{1}{2}\rho|\u|^2 +\rho e +p)\u$
&
ideal FD, ideal GD
\\
&
$\begin{aligned}
& \tfrac{1}{2}\rho|\u|^2 + \tfrac{1}{\gamma}p\\
&\quad + \tfrac{1}{2\mu_0}|\B|^2
\end{aligned}$
&
$\begin{aligned}
&\big(\tfrac{1}{2}\rho|\u|^2 + (1+\tfrac{1}{\gamma})p\big)\u\\
&\quad -\tfrac{1}{\mu_0}(u\times\B)\times\B
\end{aligned}$
&
ideal inviscid MHD
\\
&
$\tfrac{1}{2}(|\E|^2 +|\B|^2)$
&
$c\E\times\B$ 
&
vacuum EM
\\
\hline
helicity
&
$\u\cdot\vort$
&
$(p/\rho + e)\vort + (\vort\times\u)\times\u$ 
&
\tbox{ideal barotropic FD}{ideal incompressible FD}
\\
\hline
cross-helicity
&
$\u\cdot\B$
&
$\begin{aligned}
&\big(\tfrac{1}{2}|\u|^2 + (1+\tfrac{1}{\gamma})p/\rho\big)\B\\
&\quad  -(\B\times\u)\times\u
\end{aligned}$
&
ideal inviscid MHD
\\
\hline
\end{tabular}
\end{table}

One physical example of a locally trivial volumetric conservation law is 
the charge-current continuity equation \eqref{eq:EM:chargecurrent}
in the non-vacuum Maxwell's equations. 
The conserved density is given by $\volT=\rho = \tfrac{1}{4\pi}\div\E$ 
which is a spatial divergence,
while the spatial flux is given by $\volX = 4\pi \J = -\tfrac{1}{4\pi}\E_t +\tfrac{c}{4\pi}\curl\B$. 
Hence, the total charge in a fixed volume $\V$ can expressed as a surface-flux integral 
$\int_\V \rho\dV=\tfrac{1}{4\pi}\int_{\p\V} \E\cdot\dS$
whose time derivative is given by 
\[
\dfrac{d}{dt}\int_{\V} \rho\dV
= \dfrac{d}{dt}\Big(\frac{1}{4\pi}\int_{\p\V} \E\cdot\dS\Big)
= \frac{1}{4\pi}\int_{\p\V} \E_t\cdot\dS
= -\int_{\p\V} \J\cdot\dS
\]
through Gauss' theorem. 
This represents a global conservation law. 
It holds as an identity for the surface-flux integral $\tfrac{1}{4\pi}\int_{\p\V} \E\cdot\dS$,
and hence is globally trivial in this form, 
which can also be deduced from Proposition~\ref{prop:3Dvolconslaw-nontrivflux}.
Nevertheless, the integral equalities 
$\frac{1}{4\pi}\int_{\p\V} \E\cdot\dS = \int_{\V} \rho\,\dV$
and $\frac{1}{4\pi}\int_{\p\V} \E_t\cdot\dS = -\int_{\p\V} \J\cdot\dS = -\int_{\V} \div\J\,\dV$
themselves are non-trivial in a both a mathematical and physical sense. 
In the vacuum case, where $\rho=0$ and $\J=0$, 
the global conservation law is globally non-trivial 
but it holds as a consequence of the non-trivial topological electric flux conservation law \eqref{ex:EM:electriccharge:zero}. 

Another example of a locally trivial volumetric conservation law 
arises in ideal fluid flow that has either constant density or a barotropic pressure, 
with the entropy density being non-constant (locally adiabatic). 
By combining the entropy transport equation \eqref{eq:isentropic}
and the vorticity transport equation \eqref{eq:Eul:vort}, 
we obtain a local volumetric conservation law
\begin{equation}\label{ex:fluidflow:Ertel}
\p_t(\vort\cdot\grad\,S) + \div \left( (\vort\cdot\grad\,S)\u\right)=0 .
\end{equation}
This result is known as \emph{Ertel's theorem} \cite{Ped,Sal1988,Che2014}. 
In this conservation law, 
the conserved density is equal to a spatial divergence
\[
\volT = \vort\cdot\grad\,S = \div (\u\times\grad\,S) , 
\]
while the spatial flux can be expressed as 
\[
\volX = (\vort\cdot\grad\,S)\u
= -(\u\times\grad\, S)_t -\Curl\left( (S\u)_t + S\vort\times\u \right) . 
\]
Thus the conserved current $(\volT,\volX)$ is locally trivial, 
by Definition~\ref{defn:3Dvolconslaw-loctriv}. 
The global form of the conservation law \eqref{ex:fluidflow:Ertel}
in any fixed (static) volume $\V$ within the fluid 
is given by
\begin{equation}\label{ex:fluidflow:Ertel:global}
\begin{aligned}
& \frac{d}{dt} \int_{\V} \vort\cdot\grad\,S\,\dV 
= \frac{d}{dt} \int_{\V} \div (\u\times\grad\,S)\dV
\\
& = \oint_{\p\V} (\u\times\grad S)_t\cdot\dS
= -\oint_{\p\V} (\vort\cdot\grad\,S)\u\cdot\dS 
\end{aligned}
\end{equation}
which holds as an integral identity. 
This conservation law is globally trivial,
since when $\vort\cdot\grad\,S = \volT \neq 0$,
we have $(\vort\cdot\grad\,S)\u=\volX \neq 0$ provided $\u\neq0$. 

An analogous trivial volumetric conservation law will hold 
for any scalar function $f[\rho,S,p,\u]$ that satisfies the transport equation
$D_t f + \u\cdot\Grad\, f=0$.

A related trivial conservation law is the generalization of Ertel's theorem 
for potential vorticity \cite{HayMcI1987,HayMcI1990}
in diabatic heated ideal fluids and gases that are either barotropic or incompressible. 
The potential vorticity is given by $\vort\cdot\grad\,\theta$ 
in terms of the potential temperature $\theta = T (p_0/p)^\kappa$, 
where $T$ is the local temperature, 
$p_0$ is any chosen constant reference pressure,
and $\kappa$ is the Poisson constant. 
Potential vorticity satisfies the conservation law
\begin{equation}\label{ex:fluidflow:potentialvort}
\p_t(\vort\cdot\grad\,\theta) + \div \left( (\vort\cdot\grad\,\theta)\u -(Q/\rho)\vort\right)=0 
\end{equation}
where $Q$ is the diabatic heating rate given by 
\[
\theta_t + \u\cdot\grad\, \theta = Q
\]
The conserved density and the spatial flux can be expressed in the form of 
a locally trivial conserved current similarly to the expressions in Ertel's theorem 
with $S$ replaced by $\theta$. 
The global form of the conservation law \eqref{ex:fluidflow:potentialvort}
in any fixed (static) volume within the fluid/gas  
is a mathematical identity. 

\subsubsection{Globally non-trivial boundary conservation laws}\label{sec:ex:temporalCLs}

Non-trivial boundary conservation laws in global form 
are given by surface-flux constants of motion \eqref{3Dsurfcom} 
on a closed boundary surface $\S=\p\V$ of a volume, 
and circulatory constants of motion \eqref{3Dcurvcom} 
on a closed boundary curve $\C=\p\S$ of a surface. 
The local form of these conservation laws consists of 
divergence-type temporal conservation laws \eqref{3Dtemporal-divconslaw}
and curl-type temporal conservation laws \eqref{3Dtemporal-curlconslaw}, 
respectively. 
Physical examples arise in PDE systems that possess evolution equations and spatial constraint equations:
electromagnetism, MHD, 
incompressible fluid flow, 
irrotational fluid flow,
and compressible fluid flow in a Beltrami state with non-vanishing vorticity. 
In particular, 
the latter example is new 
and will provide two interesting applications of the results in Section~\ref{sec:non-triv}.

In electromagnetism and MHD, 
the magnetic field is divergence free, $\div\B=0$. 
Since this equation is compatible with the evolution equations in these PDE systems,
we have $D_t(\div\B)=0$,
which is a divergence-type temporal conservation law for the conserved density $\volT=\div\B$. 
This local conservation law is trivial because $\div\B=0$ 
holds for all solutions of both systems. 
More specifically, it has the form of a locally trivial volumetric conservation law 
with vanishing spatial flux, $\volX=0$. 
Its global form for any volume $\V$ within the physical domain of each system 
is given by the magnetic flux conservation law \eqref{ex:MHD:Faraday:closedsurf}
for the closed boundary surface $\S=\p\V$. 
In particular, this is a non-trivial global surface-flux conservation law, 
which arises from a locally trivial volumetric conservation law. 

Similarly, in incompressible fluid flow and incompressible MHD, 
since the velocity is divergence free, $\div\u=0$, 
we obtain the divergence-type temporal conservation law $D_t(\div\u)=0$
which has the form of a locally trivial volumetric conservation law $\volT=\div\u$
with vanishing spatial flux, $\volX=0$. 
It has the global form \eqref{ex:incompressFD:streamlineflux} and \eqref{ex:incompressMHD:streamlineflux}
which states the vanishing of the net streamline flux through the closed boundary surface $\S=\p\V$ of any static volume $\V$ 
within the physical domain of these respective systems 
is conserved. 

Another similar example is irrotational fluid flow or gas dynamics,
where $\curl\u=0$. 
The time derivative of this equation gives $D_t(\curl\u)=0$,
which is compatible with the evolution equation for $\u$ in these systems. 
This equation is a curl-type temporal conservation law. 
It has the form of a locally trivial surface-flux conservation law,
in which $\surfT=\curl\u$ is the conserved density 
and $\surfX=0$ is the spatial circulation flux. 
The global form of this conservation law
states that the vanishing of the net circulation \eqref{ex:irrFD:circulation:closedcurv} 
around any static closed curve 
within the physical domain of the irrotational systems 
is conserved. 

A more interesting example of a curl-type temporal conservation law 
arises in ideal barotropic fluid flow with non-vanishing vorticity 
when Beltrami flows are considered.
Since a Beltrami flow is characterized by the relation $\vort\times\u=0$, 
the vorticity transport equation \eqref{eq:Eul:vort} then becomes
$\vort_t=0$. 
This has the form of a locally trivial surface-flux conservation law,
and the resulting global conservation law states that the net circulation \eqref{ex:irrFD:circulation:closedcurv} 
around any static closed curve within the physical domain of the Beltrami flow 
is conserved. 
Importantly, here the fluid can be irrotational, 
in which case the conserved circulation can be non-zero. 
In particular, 
this physical situation provides an example of 
a non-trivial global circulatory conservation law 
arising from a locally trivial surface-flux conservation law. 

A different interesting example from fluid flow with non-vanishing vorticity 
comes from the conservation law for potential vorticity \eqref{ex:fluidflow:potentialvort}
in diabatic heated fluids. 
If the flow is in a Beltrami state such that the relation 
$(Q/\rho)\vort = (\vort\cdot\grad\,\theta)\u$ holds,
then the potential vorticity satisfies $\p_t(\vort\cdot\grad\,\theta)=0$.
This constitutes a locally trivial volumetric conservation law
because the conserved density 
$T=\vort\cdot\grad\,\theta = \div (\theta\vort)$ is a divergence. 
The resulting global conservation law in any fixed (static) volume $\V$ within the fluid 
is given by
\begin{equation}\label{ex:fluidflow:potvort:Beltrami:global}
\frac{d}{dt} \int_{\V} \vort\cdot\grad\,\theta\,\dV 
= \frac{d}{dt} \oint_{\p\V} \theta\,\vort\cdot\dS 
= \frac{d}{dt} \oint_{\p\V} (\u\times\grad\,\theta)\cdot\dS 
=0
\end{equation}
which states the net flux of potential vorticity through the boundary surface $\p\V$ 
is a constant of motion. 
It can also be interpreted as conservation of circulatory potential temperature $\u\times\grad\,\theta$. 
This boundary conservation law is globally non-trivial whenever the fluid is irrotational.

\section{Potentials and non-triviality}\label{sec:potentials}

All non-trivial conservation laws admitted by a given physical PDE system $\pdesys[\depvars]=0$
contain local and global information about the solutions $\depvars(t,\vec{x})$ of the system. 
It is often useful to convert a non-trivial conservation law into an identity 
by the introduction of a set of potentials. 
The information contained in the conservation law 
then becomes transferred to the set of equations 
that relate the set of potentials to the dynamical variables in the conservation law. 
Part of the information may also reside in the topological properties of 
the spatial domain in which the potential is defined. 

From a mathematical viewpoint, 
the introduction of potentials involves moving 
from the original jet space $(t,\vec{x},\depvars,\p\depvars,\p^2\depvars,\ldots)$ 
to a new jet space in which some of the dynamical variables (and their derivatives) 
get replaced by the set of potentials (and their derivatives). 
In terms of the variables in the new jet space, 
the conservation law in local form holds as an identity which no longer contains 
any local information about solutions of the PDE system. 
Thus, the conservation law becomes locally trivial. 

To illustrate this relationship between 
potentials and triviality of conservation laws, 
we will give a few important physical examples from 
fluid flow, gas dynamics, electromagnetism, and MHD.

\subsection{Potentials in irrotational fluid flow and gas dynamics}

One important physical example of a potential 
arises in irrotational fluid flow and gas dynamics,
where the vorticity $\vort= \curl\u=0$ vanishes everywhere in the spatial domain 
$\Omega\subseteq\pdedom$ of the fluid or gas. 
This equation represents a locally non-trivial spatial curl conservation law \eqref{spatial-curl-3Dconslaw},
as discussed in Section~\ref{sec:ex:curlCLs}. 
A corresponding potential is given by 
\begin{equation}\label{vel-potential}
\u = \grad\varphi
\end{equation}
where $\varphi$ is called the velocity potential. 
When the spatial domain $\Omega$ has trivial topology 
(such that all closed loops in $\Omega$ are contractible), 
the velocity potential will be a smooth function in $\Omega$. 
Then the vorticity equation holds as an identity
\begin{equation}\label{vel-identity}
\curl\u= \curl\grad\varphi\equiv 0 
\end{equation}
everywhere in $\Omega$. 

While this conservation law $\curl\u=0$ is non-trivial 
with respect to the jet space of dynamical variables $(\u,\rho,p)$
in the fluid or gas system, 
its formulation as an identity \eqref{vel-identity} 
means that it is locally trivial with respect to the new jet space of variables $(\varphi,\rho,p)$. 

For any static closed curve $\C=\p\S$ bounding a connected surface $\S\in\Omega$, 
the global form of the conservation law $\curl\u=0$ is given by 
the vanishing circulation integral \eqref{ex:FD:circ:closed},
which has the physical meaning that the fluid or gas is irrotational. 
Once the velocity potential is introduced, 
the resulting global conservation law becomes a line integral identity 
\begin{equation}
\oint_{\C} \u\cdot \ds = \oint_{\C} \grad\varphi\cdot \ds \equiv 0
\end{equation}
(holding due to the fundamental theorem of calculus for line integrals) 
since the curve has no boundary, $\p\C=\emptyset$. 
This form of the conservation law contains no information about the physical system,
while the physical information about the fluid or gas being irrotational resides instead 
in the equation \eqref{vel-potential} relating the velocity potential $\varphi$ to $\u$.

\subsection{Potentials in incompressible fluid flow}

Another important physical example of a potential 
arises in incompressible fluid flow as well as in incompressible MHD, 
where the velocity is divergence free, $\div\u=0$, 
everywhere in the spatial domain 
$\Omega\subseteq\pdedom$ of the fluid or MHD system. 
This incompressibility equation represents a locally non-trivial spatial divergence conservation law \eqref{spatial-div-3Dconslaw},
as discussed in Section~\ref{sec:ex:divCLs}. 
Its has the physical content that the fluid volume or liquid metal volume 
in the flow is not expanding or contracting. 
Mathematically, it states that there are no sources or sinks of velocity streamlines. 

In two spatial dimensions, 
the velocity $\u$ can be expressed in terms of a single scalar potential, 
known as the stream function, 
whose gradient is orthogonal to $\u$. 
A generalization to three spatial dimensions 
involves introducing a pair of scalar potentials $(\alpha,\beta)$ given by 
\begin{equation}\label{incompress-potential}
\u = \grad\alpha\times\grad\beta . 
\end{equation}
Since $\u$ is orthogonal to the gradient of each potential, 
these potentials $(\alpha,\beta)$ are analogous to stream functions
and are known as Clebsch variables \cite{Lam,Mil-Tho}. 
In particular, $\u$ lies in the intersection of the corresponding potential surfaces 
$\alpha=\const$ and $\beta=\const$. 

When the spatial domain $\Omega$ has trivial topology 
(such that all closed surfaces in $\Omega$ are contractible), 
both potentials will be smooth functions in $\Omega$. 
Then the incompressibility equation holds as an identity
\begin{equation}\label{incompress-identity}
\div\u= \div(\grad\alpha\times\grad\beta) 
= (\curl\grad\alpha)\times\grad\beta - \grad\alpha\times(\curl\grad\beta) 
\equiv 0 
\end{equation}
everywhere in $\Omega$. 
In this situation, the potentials $(\alpha,\beta)$ exist for any divergence-free $\u$. 
Specifically, by Poincar\'e's lemma, 
$\u = \curl \vec{\psi}$ holds for some smooth vector field $\vec{\psi}$. 
The Clebsch representation theorem \cite{Lam} states that 
$\vec{\psi}= \grad h + f\grad g$ 
holds for some smooth functions $f,g,h$,
whence $\curl \vec{\psi} = \curl( f\grad g ) = \grad f\times\grad g = \u$. 

With respect to the jet space of dynamical variables $(\u,\rho,p)$ in fluid flow, 
or $(\u,\rho,\B)$ in MHD, 
the conservation law $\div\u=0$ is non-trivial,
but it becomes locally trivial with respect to the new jet space of variables $(\alpha,\beta,\rho,p)$. 
The physical and mathematical content of this conservation law 
then resides in the equation \eqref{incompress-potential} 
relating the Clebsch potentials $(\alpha,\beta)$ to $\u$. 

For any static closed surface $\S=\p\V$ bounding a connected volume $\V\in\Omega$, 
the global form of the conservation law $\div\u=0$ is given by 
the vanishing streamline-flux integral \eqref{ex:incompress:streamlineflux:zero}. 
Once the velocity potential is introduced, 
the resulting global conservation law becomes a surface integral identity 
\begin{equation}
\oint_{\S} \u\cdot \dS = \oint_{\S} (\grad\alpha\times\grad\beta)\cdot \dS 
= \oint_{\S} \curl(\alpha\grad\beta)\cdot \dS 
\equiv 0
\end{equation}
(holding due to Stokes' theorem)
since the surface has no boundary, $\p\S=\emptyset$. 
Hence this form of the conservation law is globally trivial.

\subsection{Potentials in electromagnetism}

One more important physical example comes from Maxwell's equations \eqref{eq:EM} 
for electromagnetism,
where the magnetic field $\B$ is divergence free 
and its time derivative is given by the curl of the electric field. 
The divergence equation $\div\B=0$ represents 
a locally non-trivial spatial divergence conservation law,
while the evolution equation $\B_t= -c\curl\E$ represents 
a locally non-trivial surface-flux conservation law,
which are discussed in Sections~\ref{sec:ex:divCLs} and~\ref{sec:ex:surfCLs} respectively. 
Note that these two equations for $\B$ are compatible: 
the time derivative of $\div\B$ is equal to the divergence of $\B_t+c\curl\E$
(namely, these equations are related by a differential identity). 
In global form, 
the resulting conservation laws \eqref{ex:EM:magneticcharge:zero} and \eqref{ex:EM:Faraday}
for any closed surface $\S=\p\V$ bounding a volume $\V$ 
within the physical domain $\Omega\subseteq\pdedom$ of the electromagnetic system
state that the net magnetic flux through $\S$ is zero
and that this flux is time-independent,
due to the absence of magnetic charges. 
Both of these conservation laws are globally non-trivial
with respect to the jet space of dynamical variables $(\E,\B)$. 

First, a potential for the divergence equation $\div\B=0$ is given by 
\begin{equation}\label{ex:EM:B-potential}
\B = \curl\A
\end{equation}
where $\A(t,\vec{x})$ is called the (magnetic) vector potential. 
When the spatial domain $\Omega$ has trivial topology 
(such that all closed surfaces in $\Omega$ are contractible), 
the vector potential will be a smooth vector function in $\Omega$. 
The divergence equation then holds as an identity
\begin{equation}\label{ex:EM:divB-identity}
\div\B= \div\curl\A \equiv 0 
\end{equation}
everywhere in $\Omega$. 
Consequently, this conservation law becomes locally trivial with respect to the jet space of the variables $(\E,\A)$. 
The corresponding global conservation law \eqref{ex:EM:magneticcharge:zero} 
then holds as surface integral identity 
\begin{equation}
\oint_\S \B\cdot\dS = \oint_\S (\curl\A)\cdot\dS \equiv 0 
\end{equation}
by Stokes' theorem since the surface is closed, $\p\S=\emptyset$. 
The physical information that there are no magnetic charges 
resides instead in the equation \eqref{ex:EM:B-potential} 
relating the magnetic field $\B$ to the vector potential $\A$. 

When the evolution equation $\B_t= -c\curl\E$ is expressed in terms of the vector potential, 
it becomes a curl equation 
\begin{equation}\label{ex:EM:eq:AE}
\curl(\A_t +c\E)=0
\end{equation}
which represents a locally non-trivial spatial curl conservation law. 
The global form of this conservation law relates the net electric circulation (electromotive force) 
around any closed curve $\C$  
to the rate of change of circulation of the vector potential:
\begin{equation}\label{ex:EM:EA-circ}
\oint_\C \E\cdot\ds = -\frac{1}{c}\frac{d}{dt}\oint_\C \A\cdot\ds . 
\end{equation}

Next, a potential for the curl equation \eqref{ex:EM:eq:AE} is given by 
\begin{equation}\label{E-potential}
\E = \grad\phi - \tfrac{1}{c}\A_t
\end{equation}
where $\phi(t,\vec{x})$ is called the (electric) scalar potential. 
When the spatial domain $\Omega$ has trivial topology 
(such that all closed loops in $\Omega$ are contractible), 
the scalar potential will be a smooth function in $\Omega$. 
The curl equation then holds as an identity
\begin{equation}\label{ex:ME:curlE-identity}
\curl(\A_t+c\E) = \curl\grad\phi \equiv 0 
\end{equation}
everywhere in $\Omega$. 
As a consequence, this conservation law becomes locally trivial with respect to the jet space of the potentials $(\phi,\A)$. 
Likewise, the corresponding global conservation law \eqref{ex:EM:EA-circ}
becomes a line integral identity 
\begin{equation}
\oint_\C \E\cdot\ds + \frac{1}{c}\frac{d}{dt}\oint_\C \A\cdot\ds 
= \oint_\C \grad\phi\cdot\ds \equiv 0 
\end{equation}
since $\p\C=\emptyset$. 

One difference between this electromagnetic example and the previous examples in fluid flow
is that the pair of electromagnetic potentials $(\phi,\A)$ 
have gauge freedom given by 
\begin{equation}
\phi \rightarrow \phi + \tfrac{1}{c}\chi\strut_t,
\quad
\A \rightarrow \A + \grad\chi
\end{equation}
in terms of an arbitrary scalar function $\chi(t,\vec{x})$. 
The physical variables $(\E,\B)$, along with all of the preceding conservation laws, 
are invariant under this gauge freedom.

\section{Conclusion}\label{sec:discuss}

We have explored the properties and relationships of 
the different types of dynamical and topological conservation laws 
for PDE systems in three spatial dimensions. 
These types are distinguished by the dimensionality of the domain 
on which the global form of the conservation law is formulated:
volumes, surfaces, and curves, in the case of dynamical conservation laws;
surfaces and curves, in the case of topological conservation laws. 

We have introduced both global and local formulations of all of these different conservation laws within a unified framework,
and we have also explained the conditions under which these conservation laws
yield constants of motion. 

Our main results consist of providing an explicit and systematic characterization
for when two conservation laws are locally or globally equivalent,
and for when a conservation law is locally or globally trivial,
as well as deriving relationships among the different types of conservation laws.
These results significantly clarify and improve the notion of a ``trivial'' conservation law. 

We have used these results to show how
if a trivial local conservation law on a domain has zero flux
then under certain conditions
it can yield a non-trivial global conservation law on the domain boundary. 
These boundary conservation laws are found to be related to constants of motion 
that arise from differential identities holding in a PDE system 
when it contains both evolution equations and spatial constraint equations. 
This demonstrates that such differential identities are not merely ``trivial'' conservation laws, 
which has been the source of some confusion in the applied mathematics and physics literature.

Additionally, 
we have explained how non-triviality of a conservation law gets altered 
when potentials are introduced. 

Throughout, physical examples from fluid flow, gas dynamics, electromagnetism, and magnetohydrodynamics have been used to illustrate the results. 
Because the examples are formulated within a unified framework, 
they shed light on the similarities and connections among various conservation laws 
in all of these physical systems.

In subsequent papers, 
we plan first to extend all of these results to conservation laws that are formulated on 
moving spatial domains. 
For PDE systems describing the flow of a physical continuum, 
such as a gas, a fluid, or a plasma, 
the most important kind of conservation laws and constants of motion 
are ones that hold on moving domains transported by the flow. 
Of particular interest are material conservation laws that have vanishing fluxes. 
These conservation laws will be shown to be closely connected to ``frozen-in'' quantities which are very useful for understanding the physical and analytical properties of solutions. 

We further plan to study the different types of conservation laws on static and moving domains for PDE systems in two spatial dimensions.

\section*{Acknowledgements}

The authors are each supported by the Discovery grant program of NSERC.

\begin{appendix}

\section{Relationships among the different types of local conservation laws}\label{sec:interrelations}

Local volumetric, surface-flux, and circulatory conservation laws 
\eqref{3Dvolconslaw}, \eqref{3Dsurfconslaw}, \eqref{3Dcurvconslaw}
are related to each other in several ways.
These relationships can be expressed succinctly in terms of 
a time-independent vector function $\vec{\xi}(x)$ in $\Rnum^3$. 
Hereafter we will use subscripts $\V,\S,\C$ to distinguish the three respective types of 
conserved densities and spatial fluxes. 

Firstly, we take the dot product of $\vec{\xi}(x)$ 
with any local circulatory conservation law \eqref{3Dcurvconslaw}. 
The time-derivative term yields 
$\vec{\xi}\cdot D_t\curvT_\C= D_t(\curvT_\C\cdot\vec{\xi})$,
while the spatial gradient term yields 
$\vec{\xi}\cdot(\Grad\,\curvX_\C)= \Div(\curvX_\C\vec{\xi}) + \curvX_\C\div\vec{\xi}$. 
Therefore, if we take $\vec{\xi}(x)$ to be divergence-free, 
then we obtain a set of local volumetric conservation laws \eqref{3Dvolconslaw}
given by 
\begin{equation}\label{CL:curvtovol}
\volT_\V = \curvT_\C\cdot\vec{\xi},
\quad
\volX_\V = \curvX_\C\vec{\xi} ,
\quad
\div\vec{\xi} =0. 
\end{equation} 
In a similar way, 
the dot product of $\vec{\xi}$ with any local surface-flux conservation law \eqref{3Dsurfconslaw}
yields the terms $D_t(\surfT_\S\cdot\vec{\xi})$ 
and $\Div(\surfX_\S\times\vec{\xi}) -\surfX_\S\cdot\curl\vec{\xi}$. 
By taking $\vec{\xi}(x)$ to be curl-free, 
we obtain a set of local volumetric conservation laws \eqref{3Dvolconslaw} 
given by 
\begin{equation}\label{CL:surftovol}
\volT_\V = \surfT_\S\cdot\vec{\xi},
\quad
\volX_\V = \surfX_\S\times\vec{\xi} ,
\quad
\curl\vec{\xi} =0 . 
\end{equation}

Secondly, 
we take the cross product of $\vec{\xi}(x)$ 
with any local circulatory conservation law \eqref{3Dcurvconslaw}. 
The time-derivative term yields 
$\vec{\xi}\times D_t\curvT_\C= -D_t(\curvT_\C\times\vec{\xi})$,
while the spatial gradient term yields 
$\vec{\xi}\times(\Grad\,\curvX_\C)= -\Curl(\curvX_\C\vec{\xi}) -\curvX_\C\curl\vec{\xi}$. 
Therefore we obtain a set of local surface-flux conservation laws \eqref{3Dsurfconslaw}
given by 
\begin{equation}\label{CL:curvtosurf}
\surfT_\S = \curvT_\C\times\vec{\xi},
\quad
\surfX_\S = \curvX_\C\vec{\xi} , 
\quad
\curl\vec{\xi} =0 . 
\end{equation} 

The cross product of $\vec{\xi}$ 
with any local surface-flux conservation law \eqref{3Dsurfconslaw}
does not yield any type of local conservation law. 
In particular, if $\vec{\xi}$ is taken to be a constant vector for simplicity, 
the spatial curl term gives 
$\vec{\xi}\times(\Curl\,\surfX_\S)= \Grad(\vec{\xi}\cdot\surfX_\S) + \Curl(\vec{\xi}\times\surfX_\S) -\vec{\xi}\Div\surfX_\S$ 
which is a linear combination of all three types of spatial terms 
appearing in local conservation laws. 

The three relationships \eqref{CL:curvtovol}, \eqref{CL:surftovol}, \eqref{CL:curvtosurf} 
have direct counterparts for local spatial divergence, curl, and gradient conservation laws
\eqref{spatial-div-3Dconslaw},  \eqref{spatial-curl-3Dconslaw}, \eqref{spatial-grad-3Dconslaw}.
Specifically, 
we have
\begin{align}
& \volX_\S = \curvX \vec{\xi} ,
\quad
\div\vec{\xi} =0 , 
\label{spatialCL:gradtodiv}
\\
& 
\volX_\S = \surfX_\C\times\vec{\xi} ,
\quad
\curl\vec{\xi} =0 ,
\label{spatialCL:curltodiv}
\\
& 
\surfX_\C = \curvX\vec{\xi} ,
\quad
\curl\vec{\xi} =0 ,
\label{spatialCL:gradtocurl}
\end{align} 
where subscripts $\S$ and $\C$ denote (the domains of) spatial divergence and spatial curl conservation laws; and an empty subscript denotes a spatial gradient conservation law. 

A special case of all of these relationships is 
when $\vec{\xi}$ is a constant vector. 
In particular, we can work in Cartesian coordinates $\vec{x}=(x,y,z)$
and take $\vec{\xi}$ to be each of three corresponding unit vectors $\mathbf{i,j,k}$.
Then the three components of local surface-flux and circulatory conservation laws 
produce three corresponding local volumetric conservation laws, 
and likewise the three components of local spatial curl and gradient conservation laws 
produce three corresponding local spatial divergence conservation laws.

The general form of the relationships
\eqref{CL:curvtovol}--\eqref{CL:curvtosurf}
and \eqref{spatialCL:gradtodiv}--\eqref{spatialCL:gradtocurl}
provides a mapping from a local conservation law into a set of local conservation laws
parameterized by a vector function $\vec{\xi}(x)$. 
To understand the properties of these sets of local conservation laws, 
we will use the result (Poincar\'e's lemma) that, in $\Rnum^3$, 
$\curl\vec{\xi} =0$ is equivalent to $\vec{\xi}=\grad\zeta$ 
for some scalar function $\zeta(x)$, 
and similarly $\div\vec{\xi} =0$ is equivalent to $\vec{\xi}=\curl\vec{\zeta}$ 
for some vector function $\vec{\zeta}(x)$. 

Therefore, 
we can write the sets of dynamical conserved currents \eqref{mapCL:curvtovol}, \eqref{mapCL:surftovol}, \eqref{mapCL:curvtosurf}
in the respective forms
\begin{align}
& \volT_\V = \Div(\vec{\zeta}\times\curvT_\C) + \vec{\zeta}\cdot(\Curl\,\curvT_\C) , 
\quad
\volX_\V|_\Esp = -D_t(\vec{\zeta}\times\curvT_\C) + \Curl(\curvX_\C\vec{\zeta}) ,
\label{mapCL:curvtovol}
\\
& \volT_\V = \Div(\zeta\surfT_\S) -\zeta\Div\,\surfT_\S , 
\quad
\volX_\V|_\Esp = -D_t(\zeta\surfT_\S) - \Curl(\zeta\surfX_\S) ,
\label{mapCL:surftovol}
\\
& \volT_\S = -\Curl(\zeta\curvT_\C) + \zeta\Curl\,\curvT_\C , 
\quad
\volX_\S|_\Esp = D_t(\zeta\curvT_\C) + \Grad(\zeta\curvX_\C) ,
\label{mapCL:curvtosurf}
\end{align}
which will be useful for studying the corresponding sets of global conservation laws. 

For an arbitrary regular volume $\V\in\Rnum^3$, 
the two sets of volumetric conserved currents \eqref{mapCL:curvtovol} and \eqref{mapCL:surftovol}
respectively yield the global conservation laws
\begin{equation}\label{mapCL:curvtovol:global}
\frac{d}{dt} \int_\V \vec{\zeta}\cdot(\Curl\,\curvT_\C)\, \dV\big|_\Esp = 0
\end{equation}
and 
\begin{equation}\label{mapCL:surftovol:global}
\frac{d}{dt} \int_\V \zeta\Div\,\curvT_\S\, \dV\big|_\Esp = 0
\end{equation}
after the divergence terms have been converted into flux terms by Gauss' theorem.
In these conservation laws, 
the net flux integrals on $\p\V$ are zero by Stokes' theorem 
since this boundary surface $\p\V$ is closed. 
Thus, these two sets of global volumetric conservation laws 
comprise constants of motion. 

Similarly, for an arbitrary regular surface $\S\in\Rnum^3$, 
the set of surface-flux conserved currents \eqref{mapCL:curvtosurf}
yields 
\begin{equation}\label{mapCL:curvtosurf:global}
\frac{d}{dt} \int_\S \zeta(\Curl\,\curvT_\C)\cdot \dS\big|_\Esp = 0
\end{equation}
after the curl term has been converted into a circulation flux term by Stokes' theorem. 
Here the net circulation line integral on $\p\S$ is zero 
by the fundamental theorem of line integrals 
since the boundary curve $\p\S$ is closed. 
This set of global surface-flux conservation laws thus comprises constants of motion. 

In an analogous way, 
the sets of time-independent conserved densities \eqref{spatialCL:gradtodiv}, \eqref{spatialCL:curltodiv}, \eqref{spatialCL:gradtocurl}
can be expressed in the respective forms
\begin{align}
& \volX_\S = \vec{\zeta}\times\Grad\,\curvX + \Curl(\curvX\vec{\zeta}) ,
\label{mapspatialCL:gradtodiv}
\\
& \volX_\S = \zeta\Curl\,\surfX_\C - \Curl(\zeta\surfX_\C) ,
\label{mapspatialCL:curltodiv}
\\
& \volX_\C = -\zeta\Grad\,\curvX + \Grad(\zeta\curvX) . 
\label{mapspatialCL:gradtocurl}
\end{align}
But since we have $\Grad\,\curvX|_\Esp=0$ and $\Curl\,\surfX_\C|_\Esp=0$,
the spatial divergence densities \eqref{mapspatialCL:gradtodiv} and \eqref{mapspatialCL:curltodiv}
reduce to curl expressions, 
while the spatial curl density \eqref{mapspatialCL:gradtodiv} 
reduces to a gradient expression. 
As a consequence, 
for an arbitrary closed regular surface $\S\in\Rnum^3$, 
the global form of the spatial divergence conservation laws 
arising from the mapping formulas \eqref{mapspatialCL:gradtodiv} and \eqref{mapspatialCL:curltodiv}
is just an identity:
\begin{equation}\label{mapspatialCL:div:global}
\int_\S (\Curl(\curvX\big|_\Esp \vec{\zeta})\cdot\dS \equiv 0,
\quad
\int_\S \Curl(\zeta\surfX_\C\big|_\Esp )\cdot\dS \equiv 0 
\end{equation}
by Stokes' theorem. 
Likewise, for an arbitrary closed regular curve $\C\in\Rnum^3$, 
the global form of the spatial curl conservation laws 
arising from the mapping formula \eqref{mapspatialCL:gradtocurl}
is an identity 
\begin{equation}\label{mapspatialCL:curl:global}
\int_\C \Grad(\zeta\curvX\big|_\Esp )\cdot\ds \equiv 0 . 
\end{equation}

\subsection{Triviality relationships}\label{sec:interrelations-triv}

The various relationships \eqref{mapCL:curvtovol}, \eqref{mapCL:surftovol}, \eqref{mapCL:curvtosurf} 
among local volumetric, surface-flux, and circulatory conservation laws 
can be shown to preserve local triviality. 
Specifically, we have the following results. 

\begin{theorem}\label{thm:mapCL-nontriv}
For any PDE system \eqref{3Dpde} in $\pdedom$:
(i) If the conserved density in a local circulatory conservation law \eqref{3Dcurvconslaw} 
is curl-free \eqref{3Dcurvdensity-curlfree}, 
then all of local surface-flux conservation laws 
and all of the local volumetric conservation laws 
in the respective sets produced by the mappings \eqref{mapCL:curvtosurf} and \eqref{mapCL:curvtovol}
are locally trivial. 
Conversely, 
if the conserved density in a local circulatory conservation law \eqref{3Dcurvconslaw} is 
not curl-free, 
then at least one local surface-flux conservation law in the set produced by the mapping \eqref{mapCL:curvtosurf} 
is locally non-trivial,
and at least one volumetric conservation law in the set produced by the mapping \eqref{mapCL:curvtovol} 
is locally non-trivial. 
(ii) If the conserved density in a local surface-flux conservation law \eqref{3Dsurfconslaw} 
is divergence-free \eqref{3Dsurfdensity-divfree}, 
then all of the local volumetric conservation laws 
in the set produced by the mapping \eqref{mapCL:surftovol}
are locally trivial. 
Conversely, 
if the conserved density in a local surface-flux conservation law \eqref{3Dsurfconslaw} is not divergence-free, 
then at least one local volumetric conservation law in the set produced by the mapping \eqref{mapCL:surftovol}
is locally non-trivial. 
\end{theorem}

The proof of parts (i) and (ii) are analogous, 
so we will give only the proof of part (i). 
For the first mapping \eqref{mapCL:curvtosurf} in part (i),
we use the observation that 
the surface-flux conserved current \eqref{mapCL:curvtosurf} 
has the locally-trivial form \eqref{3Dsurfconslaw-triv} 
apart from the term $\zeta\Curl\,\curvT_\C|_\Esp$ in the conserved density. 
This term vanishes on the solution space of the given PDE system 
if $\Curl\,\curvT_\C|_\Esp=0$ holds,
whereby the conserved current is locally trivial. 
Conversely, 
if the conserved current is locally trivial, 
then the term $\zeta\Curl\,\curvT_\C|_\Esp$ must be a total curl,
whereby each of its Cartesian components (with respect to $\mathbf{i,j,k}$) 
must be a total divergence. 
This will hold for all functions $\zeta(x)$ iff $\zeta\Curl\,\curvT_\C|_\Esp$ 
belongs to the kernel of the spatial Euler operator (cf Lemma~\ref{Eulerop-props}), 
which yields $0=\hat\Eop_{\zeta}(\zeta\Curl\,\curvT_\C|_\Esp)=\Curl\,\curvT_\C|_\Esp$. 
For the second mapping \eqref{mapCL:curvtovol} in part (i), 
the proof uses the same steps and will be omitted. 
This completes the proof of Theorem~\ref{thm:mapCL-nontriv}. 

In the analogous relationships 
\eqref{mapspatialCL:gradtodiv}, \eqref{mapspatialCL:curltodiv}, \eqref{mapspatialCL:gradtocurl}
that hold among local spatial divergence, curl, and gradient conservation laws,
we see that all of the time-independent conservation laws produced by these mappings
are both locally and globally trivial.

\section{Formulation of locally and globally (non) trivial conservation laws by differential forms}\label{sec:diffforms}

All of the types of local conservation laws, 
on spatial domains consisting of volumes, surfaces, and curves in three spatial dimensions, 
have an alternative formulation using differential forms. 
The general mathematical setting for this formulation is given by 
the variational bi-complex, as explained in Ref~\cite{Olv-book}. 

Here we will first show how to express local volumetric, surface-flux, and circulatory conservation laws 
\eqref{3Dvolconslaw}, \eqref{3Dsurfconslaw}, \eqref{3Dcurvconslaw}
by using differential forms. 
Only a few basic aspects of the variational bi-complex will be needed. 
In particular, 
volumetric conservation laws are shown to coincide with 3-form conservation laws, 
whereas surface-flux and circulatory conservation laws are respectively shown to be 
a strict generalization of 2-form and 1-form conservation laws. 

Then we will state in terms of differential forms
the definition and properties of local triviality and global triviality 
for local volumetric, surface-flux, and circulatory conservation laws, 
which are given in Propositions~\ref{prop:3Dvolconslaw-nontrivflux}, \ref{prop:3Dsurfconslaw-nontrivcirc}, \ref{prop:3Dcurvconslaw-nontrivflow},
respectively. 
In this language we will explain the conditions under which 
a locally trivial conservation law on a spatial domain can yield 
a constant of motion given by a non-trivial global conservation law 
on the boundary of the domain.
We will also show how this extends to 3-form, 2-form, and 1-form conservation laws. 

This will provide a transcription of our main results from vector calculus into differential forms. 

Finally, the example of Maxwell's equations from Section~\ref{sec:ex:temporalCLs}
will be re-worked using differential forms to illustrative how 
non-trivial boundary conservation laws arise directly for dynamical PDE systems 
that contain differential identities enforcing the compatibility between 
spatial constraint equations and evolution equations in the system.

\subsection{Differential forms and local conservation laws}

For dynamical PDE systems in three-dimensional space, 
the independent variables $t$ and $\x=(x^1,x^2,x^3)$ can be viewed as coordinates 
for a four-dimensional space-time manifold $\Rnum\times\Rnum^3$. 
A corresponding index notation will be useful. 
In particular, 
let $x^\mu=(t,x^1,x^2,x^3)$ denote the space-time coordinates, 
where $\mu=0,1,2,3$, with $x^0=t$. 
Similarly, let $D_\mu = (D_t,D_{1},D_{2},D_{3})$ denote the space-time total derivative, 
given in terms of the total derivatives \eqref{eq:totD_t}--\eqref{eq:totD_i}
with respect to $t$ and $(x^1,x^2,x^3)$. 
Hereafter, summation is assumed for any repeated space-time index. 

The differential forms $\d x^\mu$ constitute a basis of 1-forms on the four-dimensional space-time manifold, 
where $\d$ is the standard exterior derivative, satisfying $\d^2=0$. 
The corresponding total exterior derivative will be denoted $\D$,
which likewise satisfies $\D^2=0$. 

A basis of 2-forms is given by $\d x^\mu\wedge\d x^\nu$,
which spans a $\binom{4}{2}=6$ dimensional space; 
a basis of 3-forms is given by $\d x^\mu\wedge\d x^\nu\wedge\d x^\sigma$,
which spans a $\binom{4}{3}=4$ dimensional space; 
and a basis of 4-forms is given by $\d x^\mu\wedge\d x^\nu\wedge\d x^\sigma\wedge\d x^\tau$,
which spans a $\binom{4}{4}=1$ dimensional space. 
Here $\wedge$ is the standard wedge product of differential forms. 

The coordinate 4-form 
\begin{equation}
\volform = \d x^1\wedge\d x^2\wedge\d x^3\wedge\d t
\end{equation}
has components $\vol_{\mu\nu\sigma\tau}$ given by the Levi-Civita symbol,
with $\vol_{1230}=1$. 
Similarly, 
the spatial coordinate 3-form 
\begin{equation}
\volformV = \d x^1\wedge\d x^2\wedge\d x^3
\end{equation}
has components $\volV_{\mu\nu\sigma} = \vol_{\mu\nu\sigma0}$,
with $\volV_{123}=1$. 
Some useful relations are:
\begin{gather}
\d x^\mu\wedge\d x^\nu\wedge\d x^\sigma\wedge\d x^\tau
=\vol^{\mu\nu\sigma\tau}\volform,
\quad
\vol_{\mu\nu\sigma\tau}\d x^\mu\wedge\d x^\nu\wedge\d x^\sigma\wedge\d x^\tau
=4! \volform,
\label{volform}\\
\begin{aligned}
& \vol^{\mu\nu\sigma\alpha}\vol_{\mu\nu\sigma\beta} =3!\,\delta_{\beta}^{\alpha} , 
\quad
\vol^{\mu\nu\sigma\alpha}\vol_{\mu\nu\tau\beta} =(2!)^2\,\delta_{[\tau}^{\sigma}\delta_{\beta]}^{\alpha}, 
\\
& 
\vol^{\mu\nu\sigma\alpha}\vol_{\mu\gamma\tau\beta} =3!\,\delta_{[\gamma}^{\nu}\delta_{\tau}^{\sigma}\delta_{\beta]}^{\alpha},
\quad
\vol^{\mu\nu\sigma\alpha}\vol_{\kappa\gamma\tau\beta} =4!\,\delta_{[\kappa}^{\mu}\delta_{\gamma}^{\nu}\delta_{\tau}^{\sigma}\delta_{\beta]}^{\alpha},
\label{volform_contract}
\end{aligned}
\end{gather}
where indices are raised by using the Kronecker symbol $\delta_{\beta}^{\alpha}$,
and where square brackets denote antisymmetrization of indices. 
Subscript indices refer to differential forms,
while superscript indices refer to vectors and contravariant tensors. 
Contraction of a pair of space-time indices represents 
the interior product between space-time tensors and differential forms. 

This 4-form $\volform$ together with the interior product $\hook$ 
play a central role in converting 
local conservation laws into expressions in terms of differential forms. 

To begin,
we consider local volumetric conservation laws \eqref{3Dvolconslaw}:
the conserved density $\volT[\depvars]$ and the spatial flux $\volX[\depvars]$
can be viewed as time and space components of a space-time vector 
\begin{equation}\label{3Dvol-cov-current}
\Phi^\mu = (T,\Psi^1,\Psi^2,\Psi^3) = \mbs{\Phi}_\text{vol}
\end{equation}
namely the conserved current \eqref{3Dvolcurrent}. 
A conservation law \eqref{3Dvolconslaw} then takes the form of a space-time divergence 
\begin{equation}\label{3Dvol-cov-conslaw}
D_\mu \Phi^\mu|_\Esp = 0
\end{equation}
where $\Esp$ denotes the solution space of the given PDE system. 
The conserved current corresponds to a 3-form 
\begin{equation}\label{vol-3form}
\volform\hook \mbs{\Phi}_\text{vol}
= \tfrac{1}{3!} \vol_{\mu\nu\sigma\tau}\Phi^\tau \d x^\mu\wedge\d x^\nu\wedge\d x^\sigma
\end{equation}
Its exterior derivative is given by the 4-form 
\begin{equation}
\D(\volform\hook\mbs{\Phi}_\text{vol}) 
= \tfrac{1}{4!} \vol_{\mu\nu\sigma\tau}D_\alpha\Phi^\alpha \d x^\tau\wedge\d x^\mu\wedge\d x^\nu\wedge\d x^\sigma
\end{equation}
using the chain rule $\D\Phi^\tau= D_\alpha\Phi^\tau \d x^\alpha$ 
and the relation \eqref{volform}, along with the first identity \eqref{volform_contract}.
Hence, a local volumetric conservation law has the equivalent formulation 
\begin{equation}
\D(\volform\hook \mbs{\Phi}_\text{vol})|_\Esp = 0
\end{equation}
which states $\volform\hook \mbs{\Phi}_\text{vol}$ is a closed 3-form 
for all solutions of the PDE system.

Hence, 
a local volumetric conservation law \eqref{3Dvolconslaw} in three spatial dimensions 
is the same as a 3-form conservation law in the space-time manifold $\Rnum\times\Rnum^3$. 

Continuing, 
we consider local surface-flux conservation laws \eqref{3Dsurfconslaw}: 
the conserved flux density $\surfT[\depvars]$ and the spatial circulation flux $\surfX[\depvars]$
can be viewed as components of a space-time skew tensor 
\begin{equation}\label{3Dsurf-cov-current}
\Phi^{\mu\nu} = 
\begin{pmatrix} 
0 & T^1 & T^2 & T^3 \\ 
-T^1& 0& -\Psi^3 & \Psi^2 \\
-T^2 & \Psi^3 &  0 & -\Psi^1 \\
-T^3 & -\Psi^2 & \Psi^1 & 0
\end{pmatrix} 
= \mbs{\Phi}_\text{surf}
\end{equation}
namely the conserved flux current \eqref{3Dsurfcurrent}
expressed in a matrix form. 
The conserved flux current corresponds to a 2-form 
\begin{equation}\label{surf-2form}
\volform\hook\mbs{\Phi}_\text{surf} 
= \tfrac{1}{2!}\vol_{\mu\nu\sigma\tau}\Phi^{\sigma\tau} \d x^\mu\wedge\d x^\nu
\end{equation}
whose exterior derivative is given by the 3-form 
\begin{equation}\label{surf-3form}
\D(\volform\hook\mbs{\Phi}_\text{surf}) 
= \tfrac{1}{3!}\vol_{\mu\nu\sigma\tau}D_\alpha\Phi^{\alpha\tau} \d x^\mu\wedge\d x^\nu\wedge\d x^\sigma
\end{equation}
This follows from the chain rule $\D\Phi^{\sigma\tau}= D_\alpha\Phi^{\sigma\tau} \d x^\alpha$ 
and the relation \eqref{volform}, combined with the second identity \eqref{volform_contract}.

A conservation law \eqref{3Dsurfconslaw} thereby takes the form of 
a set of three space-time divergences
\begin{equation}\label{3Dsurf-cov-conslaw}
D_\mu \Phi^{\mu i}|_\Esp = 0,
\quad
i=1,2,3
\end{equation}
This has the equivalent formulation that the space-time part of the 3-form 
$\D(\volform\hook\mbs{\Phi}_\text{surf})$ vanishes on the solution space of the given PDE system:
$\vol_{ij0\tau}D_\alpha\Phi^{\alpha\tau}|_\Esp \d x^i\wedge\d x^j\wedge\d t
= \volV_{ijk}D_\alpha\Phi^{\alpha k}|_\Esp\d x^i\wedge\d x^j\wedge\d t =0$, 
with $i,j=1,2,3$. 
The restriction to considering only this space-time part has the geometrical and physical meaning 
that the conservation law is associated to a spatial surface $\S$ in $\Rnum^3$ for all $t$. 
In particular, 
$\Rnum\times\S$ is a submanifold, analogous to a three-dimensional cylinder, 
in the space-time manifold $\Rnum\times\Rnum^3$,
and hence the projection of 
$(\D(\volform\hook\mbs{\Phi}_\text{surf}))|_\Esp$ into this submanifold 
yields 
$D_\alpha\Phi^{\alpha k}|_\Esp \nor_k\,dA\,dt$,
where $\nor\,dA$ is the surface element corresponding to the area 2-form
$\volformV\hook\nor$ of $\S$ in $\Rnum^3$. 
Thus, if  
\begin{equation}\label{surf-3form-projection}
\big((\D(\volform\hook\mbs{\Phi}_\text{surf}))|_\Esp\big)_{\Rnum\times\S}  = 0 
\end{equation}
holds for all surfaces $\S$, 
then this provides an equivalent formulation of a local surface-flux conservation law \eqref{3Dsurfconslaw}. 
Note that the 2-form 
$\tfrac{1}{2!}\volform\hook\mbs{\Phi}_\text{surf} = \tilde{\mbs{\Phi}}_\text{surf}$
is the space-time dual of the skew tensor \eqref{3Dsurf-cov-current}, 
with components 
\begin{equation}\label{surf-dual2form}
\tilde{\Phi}_{\alpha\beta}
= \tfrac{1}{2!}\vol_{\alpha\beta\mu\nu}\Phi^{\mu\nu}
= \begin{pmatrix} 
0 & \Psi^1 & \Psi^2 & \Psi^3 \\ 
-\Psi^1& 0& T^3 & -T^2 \\
-\Psi^2 & -T^3 &  0 & T^1 \\
-\Psi^3 & T^2 & -T^1 & 0
\end{pmatrix} 
\end{equation}

The surface-flux continuity equation \eqref{surf-3form-projection} for this 2-form 
$\tilde{\mbs{\Phi}}_\text{surf}$
is strictly weaker than requiring that $\tilde{\mbs{\Phi}}_\text{surf}$ 
is a closed 2-form for all solutions of the PDE system.
If we impose the equation $(\D\tilde{\mbs{\Phi}}_\text{surf})|_\Esp = 0$,
whereby the 2-form is closed in the whole space-time manifold, 
then we obtain the local surface-flux conservation law \eqref{3Dsurf-cov-conslaw}
plus an additional local spatial divergence conservation law 
$D_\mu \Phi^{\mu 0}|_\Esp = \Div \surfT|_\Esp =0$ 
holding for the conserved flux density $\surfT$. 
This pair of local conservation laws constitute a 2-form conservation law
in four space-time dimensions \cite{BCA-book,CheBlu2010},
specifically
$D_\mu \Phi^{\mu\nu}|_\Esp =\tfrac{1}{2!}\vol^{\mu\sigma\tau\nu}D_\mu \tilde\Phi_{\sigma\tau}|_\Esp = 0$. 
In comparison, 
a local surface-flux conservation law \eqref{3Dsurf-cov-conslaw} 
has the differential consequence 
\begin{equation}
D_i D_\mu \Phi^{\mu i}|_\Esp = D_t D_i \Phi^{0i}|_\Esp = D_t \Div \surfT|_\Esp =0
\end{equation}
which is a local temporal conservation law. 
This additional conservation law is equivalent to 
$(\D\big(\D\tilde{\mbs{\Phi}}_\text{surf}\big)_{\Rnum\times\S})|_\Esp = 0 $. 

Hence, 
a local surface-flux conservation law \eqref{3Dsurfconslaw} in three spatial dimensions 
is a strict generalization of a 2-form conservation law in $\Rnum\times\Rnum^3$.

Last we consider local circulatory conservation laws \eqref{3Dcurvconslaw}:
the conserved circulation density $\curvT[\depvars]$ and the spatial endpoint flow $\curvX[\depvars]$
can be viewed directly as components of a 1-form
\begin{equation}\label{curv-1form}
\tilde{\mbs{\Phi}}_\text{curv} = \tilde\Phi_{\alpha} \d x^\alpha
\end{equation}
given by 
\begin{equation}
\tilde\Phi_{\alpha} 
=\begin{pmatrix} 
\Psi \\ 
-T^1 \\
-T^2 \\
-T^3 
\end{pmatrix} 
\end{equation}
namely the conserved circulation current expressed as a dual vector. 
Its exterior derivative is a 2-form 
\begin{equation}
\D\tilde{\mbs{\Phi}}_\text{curv}
= \tfrac{1}{2!} \vol_{\mu\nu\sigma\tau}D_\alpha\Phi^{\alpha\sigma\tau} \d x^\mu\wedge\d x^\nu
\end{equation}
where $\Phi^{\mu\nu\sigma}= \vol^{\mu\nu\sigma\alpha}\tilde\Phi_{\alpha}$ 
are components of a totally-antisymmetric tensor of rank 3, $\mbs{\Phi}_\text{curv}$, 
which is the space-time dual of $\tilde{\mbs{\Phi}}_\text{curv}$, 
with 
\begin{equation}\label{3Dcurv-cov-current}
\Phi^{\mu\nu 0}
= \begin{pmatrix} 
0 & 0 & 0 & 0 \\ 
0 & 0& T^3 & -T^2 \\
0 & -T^3 &  0 & T^1 \\
0 & -T^2 & -T^1 & 0
\end{pmatrix},
\quad
\Phi^{ijk} = \volV^{ijk}\Psi
\end{equation}
This 2-form is obtained from 
the chain rule $\D\Phi^{\nu\sigma\tau}= D_\alpha\Phi^{\nu\sigma\tau} \d x^\alpha$ 
and the relation \eqref{volform}, plus the third identity \eqref{volform_contract}. 

A conservation law \eqref{3Dcurvconslaw} takes the form of a set of three space-time divergences
\begin{equation}\label{3Dcurv-cov-conslaw}
D_\mu \Phi^{\mu ij}|_\Esp = 0,
\quad
i,j=1,2,3
\ (i\neq j)
\end{equation}
An equivalent formulation is that the space-time part of the 2-form 
$\D\tilde{\mbs{\Phi}}_\text{curv}$ 
vanishes on the solution space of the given PDE system:
$\vol_{i0\sigma\tau}D_\alpha\Phi^{\alpha\sigma\tau}|_\Esp \d x^i\wedge\d t
= \volV_{ijk}D_\alpha\Phi^{\alpha jk}|_\Esp \d x^i\wedge\d t$, 
with $i,j,k=1,2,3$. 
Similarly to the formulation of surface-flux conservation laws, 
here the restriction to considering the space-time part of the 2-form 
has the geometrical and physical meaning 
that the conservation law is associated to a spatial curve $\C$ in $\Rnum^3$ for all $t$.
In particular, 
the projection of $(\D\tilde{\mbs{\Phi}}_\text{curv})|_\Esp$ into 
the two-dimensional submanifold $\Rnum\times\C$ in the space-time manifold $\Rnum\times\Rnum^3$
yields $D_\alpha\Phi^{\alpha ij}|_\Esp \nor_{ij}\,d\ell\,dt$,
where $\mbs{\nor}$ is the normal bi-vector of $\C$ 
(namely, the exterior product of any orthogonal pair of normal vectors),
and where $d\ell$ is the arclength of $\C$, 
corresponding to the arclength 1-form 
$\volformV\hook\mbs{\nor}$ of $\C$ in $\Rnum^3$. 
Thus, if  
\begin{equation}\label{curv-2form-projection}
\big((\D\tilde{\mbs{\Phi}}_\text{curv})|_\Esp\big)_{\Rnum\times\C}  = 0 
\end{equation}
holds for all curves $\C$, 
then this provides an equivalent formulation of a local circulatory conservation law \eqref{3Dcurvconslaw}. 

The circulation continuity equation \eqref{curv-2form-projection} 
for the 1-form $\tilde{\mbs{\Phi}}_\text{curv}$ 
is strictly weaker than requiring that $\tilde{\mbs{\Phi}}_\text{curv}$ 
is a closed 1-form for all solutions of the PDE system.
If we impose the equation $(\D\tilde{\mbs{\Phi}}_\text{curv})|_\Esp = 0$
whereby the 1-form is closed in the whole space-time manifold, 
then we obtain the local circulatory conservation law \eqref{3Dcurv-cov-conslaw}
plus an additional local spatial curl conservation law 
$D_\mu \Phi^{\mu i0}|_\Esp = -\Curl \curvT|_\Esp =0$ 
holding for the conserved circulation density $\curvT$. 
This pair of local conservation laws constitute a 1-form conservation law
in four space-time dimensions \cite{BCA-book,CheBlu2010},
specifically
$D_\mu \Phi^{\mu\nu\sigma}|_\Esp =\vol^{\nu\sigma\mu\tau}D_\mu \tilde\Phi_{\tau}|_\Esp = 0$. 
In comparison, 
a local circulatory conservation law \eqref{3Dcurv-cov-conslaw} 
has the differential consequence 
\begin{equation}
D_i D_\mu \Phi^{\mu ij}|_\Esp = D_t D_i \Phi^{0ij}|_\Esp = -D_t \Curl \curvT|_\Esp =0
\end{equation}
which is a local temporal conservation law. 
This additional conservation law is equivalent to 
$(\D\big(\D\tilde{\mbs{\Phi}}_\text{curv}\big)_{\Rnum\times\C})|_\Esp = 0 $. 

Hence, 
a local circulatory conservation law \eqref{3Dcurvconslaw} in three spatial dimensions 
is a strict generalization of a 1-form conservation law in $\Rnum\times\Rnum^3$.

\subsection{Locally trivial conservation laws and exact differential forms}

If a volumetric conservation law \eqref{3Dvol-cov-current}--\eqref{3Dvol-cov-conslaw} 
is locally trivial \eqref{3Dvolconslaw-triv}, 
then it can be expressed in the space-time curl form 
\begin{equation}\label{3Dvol-triv-cov-conslaw}
\Phi^\mu|_\Esp = D_\nu \Theta^{\nu\mu}|_\Esp
\end{equation}
with 
\begin{equation}\label{trivvol-skewtens}
\Theta^{\nu\mu} = 
\begin{pmatrix} 
0 & -\Theta^1 & -\Theta^2 & -\Theta^3 \\ 
\Theta^1& 0& -\Lambda^3 & \Lambda^2 \\
\Theta^2 & \Lambda^3 &  0 & -\Lambda^1 \\
\Theta^3 & -\Lambda^2 & \Lambda^1 & 0
\end{pmatrix} 
= \mbs{\Theta}_\text{vol}
\end{equation}
where 
$\vec{\Theta}=(\Theta^1[\depvars],\Theta^2[\depvars],\Theta^3[\depvars])$
and $\vec{\Lambda}=(\Lambda^1[\depvars],\Lambda^2[\depvars],\Lambda^3[\depvars])$
are arbitrary vector functions. 
Equivalently, 
the space-time conserved current \eqref{3Dvol-cov-current} 
whose components are given by $\Phi^\mu$ 
is locally trivial iff the corresponding 3-form \eqref{vol-3form} is given by 
\begin{equation}\label{3form-triv-conslaw}
\volform\hook \mbs{\Phi}_\text{vol}|_\Esp 
= (\D (\tfrac{1}{2}\volform\hook\mbs{\Theta}_\text{vol}))|_\Esp
\end{equation}
This states that
$\volform\hook \mbs{\Phi}_\text{vol}=\tilde{\mbs{\Phi}}_{(3)}$ 
is an exact 3-form 
for all solutions $\depvars(t,x)$ of the PDE system, 
where $\tfrac{1}{2}\volform\hook\mbs{\Theta}_\text{vol}=\tilde{\mbs{\Theta}}_{(2)}$ 
is the 2-form given by the space-time dual of the skew tensor $\mbs{\Theta}_\text{vol}$. 
In particular, 
local triviality amounts to the identity 
$(\D\tilde{\mbs{\Phi}}_{(3)})|_\Esp = (\D^2\tilde{\mbs{\Theta}}_{(2)})|_\Esp =0$,
due to the basic property $\D^2=0$ of the total exterior derivative. 
We note that local triviality of volumetric conservation laws coincides with 
the standard notion of triviality for 3-form conservation laws \cite{HT-book} in four space-time dimensions. 

Local triviality for surface-flux conservation laws and circulatory conservation laws 
is similar to local triviality for volumetric conservation laws. 
We will first look at the standard notion of triviality 
for 2-form and 1-form conservation laws \cite{HT-book} in four space-time dimensions. 

A 1-form conservation law $(\D\tilde{\mbs{\Phi}}_{(1)})|_\Esp = 0$
is called trivial iff 
$\tilde{\mbs{\Phi}}_{(1)}|_\Esp 
= (\D \tilde{\Theta}_{(0)})|_\Esp$
is an exact 1-form, 
where $\tilde{\Theta}_{(0)}=\Theta[\depvars]$
is an arbitrary scalar function. 
Note this notion is strictly local. 
Likewise,
a 2-form conservation law $(\D\tilde{\mbs{\Phi}}_{(2)})|_\Esp = 0$
is called trivial iff 
$\tilde{\mbs{\Phi}}_{(2)}|_\Esp 
= (\D \tilde{\mbs{\Theta}}_{(1)})|_\Esp$
is an exact 2-form, 
where $\tilde{\mbs{\Theta}}_{(1)}=\tilde{\Theta}_\alpha[\depvars]\d x^\alpha$
is an arbitrary 1-form function. 
This notion again is strictly local. 
For later comparison with local triviality of surface-flux conservation laws, 
it is helpful to use a space-time dual formulation
by writing 
$\tilde{\mbs{\Phi}}_{(2)} = \tfrac{1}{2!}\volform\hook\mbs{\Phi}_\text{surf}$,
where $\mbs{\Phi}_\text{surf}$ is a skew tensor,
and $\tilde{\mbs{\Theta}}_{(1)} = \tfrac{1}{3!}\volform\hook\mbs{\Theta}_\text{surf}$,
where $\mbs{\Theta}_\text{surf}$ is a totally-antisymmetric tensor of rank 3. 
Triviality of a 2-form conservation law then states 
$\volform\hook\mbs{\Phi}_\text{surf}|_\Esp 
= (\D (\tfrac{1}{3}\volform\hook\mbs{\Theta}_\text{surf}))|_\Esp$, 
namely $\volform\hook\mbs{\Phi}_\text{surf}$ is an exact 2-form
for all solutions $\depvars(t,x)$ of the PDE system. 
In components, 
the dual formulation of a 2-form conservation law is given by 
$D_\mu \Phi^{\mu\nu}|_\Esp = 0$
with $\Phi^{\mu\nu}=\tfrac{1}{2!}\vol^{\mu\nu\sigma\tau}\tilde\Phi_{\sigma\tau}$,
and triviality is expressed as 
\begin{equation}\label{2form-triv-conslaw}
\Phi^{\mu\nu}|_\Esp = D_\sigma \Theta^{\sigma\mu\nu}|_\Esp
\end{equation}
with $\Theta^{\sigma\mu\nu}=\vol^{\sigma\mu\nu\tau}\tilde\Theta_{\tau}$.

Now we turn to locally triviality \eqref{3Dsurfconslaw-triv} 
for surface-flux conservation laws. 
In space-time form, 
if a surface-flux conservation law \eqref{3Dsurf-cov-current} and \eqref{3Dsurf-cov-conslaw} 
is locally trivial, 
then we have
$\Phi^{0i}|_\Esp = D_j \Theta^{0ij}|_\Esp$
and 
$\Phi^{ji}|_\Esp = D_t \Theta^{0ji}|_\Esp + D_k \Theta^{kji}|_\Esp$, 
with
\begin{equation}\label{3Dsurf-triv-antisymmtens}
\Theta^{0ij}
= \begin{pmatrix} 
0& \Theta^3 & -\Theta^2 \\
-\Theta^3 &  0 & \Theta^1 \\
-\Theta^2 & -\Theta^1 & 0
\end{pmatrix},
\quad
\Theta^{kji} = -\volV^{kji}\Lambda, 
\end{equation}
where 
$\Lambda[\depvars]$ is an arbitrary scalar function
and 
$\vec{\Theta}=(\Theta^1[\depvars],\Theta^2[\depvars],\Theta^3[\depvars])$
is an arbitrary vector function. 
By cyclically defining $\Theta^{ij0}=\Theta^{j0i}=\Theta^{0ij}$,
we obtain components $\Theta^{\mu\nu\sigma}$ of a totally-antisymmetric tensor of rank 3,
$\mbs{\Theta}_\text{surf}$,
in space-time. 
Locally triviality then can be expressed in the equivalent form 
\begin{equation}\label{3Dsurf-triv-cov-conslaw}
\Phi^{\mu i}|_\Esp = D_\sigma \Theta^{\sigma\mu i}|_\Esp,
\quad
i=1,2,3
\end{equation}
which represents a set of three space-time curls.
Note that $\Phi^{\mu i}$ contains all of the non-zero components $\Phi^{\mu\nu}$, 
by antisymmetry 
$\Phi^{0i}=-\Phi^{i0}$ and $\Phi^{ji}=-\Phi^{ij}$ with $j\neq i$; 
similarly $\Theta^{\sigma\mu i}$ contains all of the non-zero components $\Theta^{\mu\nu\sigma}$, 
by antisymmetry $\Theta^{0ji}= -\Theta^{0ij}$
combined with cyclic symmetry $\Theta^{ij0}=\Theta^{j0i}=\Theta^{0ij}$. 
Hence, local triviality \eqref{3Dsurf-triv-cov-conslaw}
implies that $\Phi^{\mu\nu}$
has the form \eqref{2form-triv-conslaw}. 

Therefore, 
a surface-flux conservation law \eqref{3Dsurf-cov-current} and \eqref{3Dsurf-cov-conslaw} 
is locally trivial \eqref{3Dsurfconslaw-triv} iff 
the corresponding 2-form \eqref{surf-2form} is exact, 
\begin{equation}
\volform\hook \mbs{\Phi}_\text{surf}|_\Esp
= (\D (\tfrac{1}{3}\volform\hook\mbs{\Theta}_\text{surf}))|_\Esp 
\end{equation}
for all solutions $\depvars(t,x)$ of the PDE system, 
where $\tfrac{1}{3}\volform\hook\mbs{\Theta}_\text{surf}=\tilde{\mbs{\Theta}}_\text{surf}$
is the 1-form given by the space-time dual of the totally-antisymmetric tensor $\mbs{\Theta}_\text{surf}$.
In particular,
the components of $\tilde{\mbs{\Theta}}_\text{surf}$ are given by 
\begin{equation}\label{3Dsurf-triv-1form}
\tilde\Theta_{\alpha} 
= \begin{pmatrix} 
\Lambda \\ 
-\Theta^1 \\
-\Theta^2 \\
-\Theta^3 
\end{pmatrix} 
\end{equation}

Thus, local triviality is the same for surface-flux conservation laws as for 2-form conservation laws. 

Last, we consider locally triviality \eqref{3Dcurvconslaw-triv} 
for circulatory conservation laws. 
In space-time form, 
if a circulatory conservation law \eqref{3Dcurv-cov-current} and \eqref{3Dcurv-cov-conslaw} 
is locally trivial,
then we have
$\Phi^{0ij}|_\Esp = \volV^{ijk}D_k \Theta|_\Esp$
and 
$\Phi^{kji}|_\Esp =-\volV^{kji}D_t \Theta|_\Esp$,
where 
$\Theta[\depvars]$
is an arbitrary scalar function.
Hence
\begin{equation}
\Phi^{\mu ij}|_\Esp = \vol^{\sigma\mu ij} D_\sigma\Theta|_\Esp,
\quad
i,j=1,2,3
\ (i\neq j)
\end{equation}
represents a set of three space-time curls.
Note that $\Phi^{\mu ij}$ contains all of the non-zero components $\Phi^{\mu\nu\sigma}$, 
due to cyclic symmetry 
$\Phi^{0ij}=\Phi^{j0i}=\Phi^{ij0}$. 
Locally triviality then can be expressed in the equivalent form 
\begin{equation}\label{3Dcurv-triv-cov-conslaw}
\Phi^{\mu\nu\sigma}|_\Esp = \vol^{\tau\mu\nu\sigma}D_\tau \Theta|_\Esp 
\end{equation}

Therefore, 
a circulatory conservation law \eqref{3Dcurv-cov-current} and \eqref{3Dcurv-cov-conslaw} 
is locally trivial \eqref{3Dcurvconslaw-triv} iff 
the corresponding 1-form \eqref{curv-1form} is exact, 
\begin{equation}
\tilde{\mbs{\Phi}}_\text{curv}|_\Esp
= (\D\tilde\Theta_\text{curv})|_\Esp 
\end{equation}
for all solutions $\depvars(t,x)$ of the PDE system, 
where $\tilde{\Theta}_\text{curv}=\Theta$. 

Thus, local triviality is the same for circulatory conservation laws as for 1-form conservation laws.

\subsection{Global (non-)triviality of conservation laws}

As shown by Propositions~\ref{prop:3Dcurvconslaw-nontrivflow}, \ref{prop:3Dsurfconslaw-nontrivcirc}, \ref{prop:3Dvolconslaw-nontrivflux}, 
if a locally trivial dynamical conservation law on a spatial domain
has vanishing flux,
then it reduces to a temporal conservation law on the domain boundary.
This temporal conservation law can, under certain conditions,
yield a non-trivial global conservation law representing a constant of motion
on the domain boundary.

This main result extends to 3-form, 2-form, and 1-form conservation laws,
since they share the same notion of local triviality as 
dynamical conservation laws of volumetric, surface-flux, and circulatory type,
respectively.
Thus, there is distinction between local and global triviality
for differential-form conservation laws in the context of space-time manifolds.
This result, which we will now explain in detail, 
has not been widely recognized in the literature \cite{HT-book}.

We start by considering 3-form conservation laws in $\Rnum\times\Rnum^3$:
$(\D\tilde{\mbs{\Phi}}_{(3)})|_\Esp = 0$, 
with
\begin{equation}
\tilde{\mbs{\Phi}}_{(3)}
= \tilde{\Phi}_{\mu\nu\sigma}\d x^\mu\wedge\d x^\nu\wedge\d x^\sigma
= T\volV_{ijk} \d x^i\wedge\d x^j\wedge\d x^k -3\Psi^k\volV_{kij}\d x^i\wedge\d x^j\wedge\d t
\end{equation}
On a spatial volume $\V$ in $\Rnum^3$,
the global form of this conservation law is given by
\begin{equation}\label{3Dglobal-3form-conslaw}
\begin{aligned}
& \frac{d}{dt} \int_{\V} \tilde{\mbs{\Phi}}_{(3)}\Big|_\Esp
= 3! \frac{d}{dt} \int_{\V} T \volformV \Big|_\Esp
\\&\qquad
= \int_{\p\V} \tilde{\mbs{\Phi}}_{(3)}\hook\vec{t}\,\Big|_\Esp
= -3!\int_{\p\V} \Psi^k \nor_k \volformV\hook\nor \Big|_\Esp
\end{aligned}
\end{equation}
where $\nor$ is the outward normal vector of the boundary surface $\p\V$ of $\V$,
and where $\vec{t}$ is the temporal vector that is dual to the 1-form $\d t$
(namely, $\d t\hook \vec{t}=1$ and $\d x^i\hook \vec{t}=0$, $i=1,2,3$).
This integral equation \eqref{3Dglobal-3form-conslaw} is the same as
a global volumetric conservation law \eqref{3Dglobalvolconslaw}.

Suppose that a 3-form conservation law is locally trivial,
$\tilde{\mbs{\Phi}}_{(3)}|_\Esp = (\D\tilde{\mbs{\Theta}}_{(2)})|_\Esp$,
such that the spatial flux 2-form vanishes, 
$\tilde{\mbs{\Phi}}_{(3)}\hook\vec{t}\,|_\Esp=(\D\tilde{\mbs{\Theta}}_{(2)})\hook\vec{t}\,|_\Esp =0$,
with
\begin{equation}
\tilde{\mbs{\Theta}}_{(2)} 
= \tilde{\Theta}_{\mu\nu}\d x^\mu\wedge\d x^\nu
= 3(\Theta^k\volV_{ijk} \d x^i\wedge\d x^j -2\Lambda_k \d x^k\wedge\d t)
\end{equation}
where $(\Theta^1[\depvars],\Theta^2[\depvars],\Theta^3[\depvars])$
and $(\Lambda_1[\depvars],\Lambda_2[\depvars],\Lambda_3[\depvars])$
are arbitrary vector functions. 
Then, in the integral equation \eqref{3Dglobal-3form-conslaw},
the flux integral vanishes
while the volume integral reduces to a surface integral
\begin{equation}
\begin{aligned}
& \int_{\V} \tilde{\mbs{\Phi}}_{(3)}\Big|_\Esp 
= 3! \int_{\V} T \volformV \Big|_\Esp
\\&\qquad
= \int_{\p\V} \tilde{\mbs{\Theta}}_{(2)} \Big|_\Esp
= 3!\int_{\p\V} \Theta^k \nor_k \volformV\hook\nor \Big|_\Esp
\end{aligned}
\end{equation}
by Gauss' theorem.
Hence the integral equation \eqref{3Dglobal-3form-conslaw} becomes
\begin{equation}\label{3Dboundary2form-conslaw}
\frac{d}{dt} \int_{\p\V} \tilde{\mbs{\Theta}}_{(2)}\Big|_\Esp
= 3! \frac{d}{dt} \int_{\p\V} \Theta^k \nor_k \volformV\hook\nor \Big|_\Esp
=0
\end{equation}
which is a temporal 2-form conservation law on the boundary surface $\p\V$ of the volume.
This global boundary conservation law will be non-trivial whenever
the spatial part of $\tilde{\mbs{\Theta}}_{(2)}$ is locally non-trivial,
namely $\big(\tilde{\mbs{\Theta}}_{(2)}|_\Esp\big)_{\p\V}$
is not an exact spatial 2-form,
or equivalently,
$\vec{\Theta}|_\Esp =(\Theta^1,\Theta^2,\Theta^3)|_\Esp$
is not equal to a curl.

Next we consider 2-form conservation laws in $\Rnum\times\Rnum^3$:
$(\D\tilde{\mbs{\Phi}}_{(2)})|_\Esp = 0$, 
with
\begin{equation}
\tilde{\mbs{\Phi}}_{(2)}
= \tilde{\Phi}_{\mu\nu}\d x^\mu\wedge\d x^\nu
= T^k\volV_{kij} \d x^i\wedge\d x^j +2\Psi_i\d x^i\wedge\d t
\end{equation}
On a spatial surface $\S$ with a boundary curve $\p\S$ in $\Rnum^3$,
the global form of this conservation law is given by
\begin{equation}\label{3Dglobal-2form-conslaw}
\begin{aligned}
& \frac{d}{dt} \int_{\S} \tilde{\mbs{\Phi}}_{(2)}\Big|_\Esp
= 2!\frac{d}{dt} \int_{\S} T^k\nor_k \volformV\hook\nor \Big|_\Esp
\\&\qquad
= -\int_{\p\S} \tilde{\mbs{\Phi}}_{(2)}\hook\vec{t}\,\Big|_\Esp
= -2!\int_{\p\S} \Psi^k \hat\ell_k \volformV\hook\mbs{\nor} \Big|_\Esp
\end{aligned}
\end{equation}
where $\nor$ is the normal vector of the surface $\S$
and $\hat\ell$ is the unit tangent vector of the boundary curve $\p\S$
such that $\hat\ell\times\nor$ is pointing outward,
and where $\mbs{\nor}$ is the normal bi-vector dual to the 2-form $\volformV\hook\hat\ell$,
while $\vec{t}$ is the temporal vector dual to the 1-form $\d t$.
This integral equation \eqref{3Dglobal-2form-conslaw} looks the same as
a global surface-flux conservation law \eqref{3Dglobalsurfconslaw},
but here $\vec{T}=(T^1,T^2,T^3)$ is divergence-free $\Div\vec{T}|_\Esp=0$
due to the spatial part of the 2-form conservation law,
$\big(\D\tilde{\mbs{\Phi}}_{(2)}\big)_{\Rnum^3}
=\tfrac{1}{3} D_l T^l\volV_{ijk} \d x^i\wedge\d x^j\wedge\d x^k
=2!\Div\vec{T}\,\volformV$.

Suppose that a 2-form conservation law is locally trivial,
$\tilde{\mbs{\Phi}}_{(2)}|_\Esp = (\D\tilde{\mbs{\Theta}}_{(1)})|_\Esp$,
such that the spatial circulation 1-form vanishes, 
$\tilde{\mbs{\Phi}}_{(2)}\hook\vec{t}\,|_\Esp=(\D\tilde{\mbs{\Theta}}_{(1)})\hook\vec{t}\,|_\Esp =0$,
with
\begin{equation}
\tilde{\mbs{\Theta}}_{(1)} 
= \tilde{\Theta}_{\mu}\d x^\mu 
= 2(\Theta_k\d x^k +\Lambda \d t)
\end{equation}
where $(\Theta_1[\depvars],\Theta_2[\depvars],\Theta_3[\depvars])$
is an arbitrary vector function,
and $\Lambda[\depvars]$ is an arbitrary scalar function. 
Then, in the integral equation \eqref{3Dglobal-2form-conslaw},
the circulation integral vanishes
while the surface integral reduces to a line integral
\begin{equation}
\begin{aligned}
& \int_{\S} \tilde{\mbs{\Phi}}_{(2)}\Big|_\Esp 
= 2! \int_{\S} T^i\nor_i \volformV\hook\nor \Big|_\Esp
\\&\qquad
= \int_{\p\S} \tilde{\mbs{\Theta}}_{(1)} \Big|_\Esp
= 2!\int_{\p\S} \Theta^k \hat\ell_k \volformV\hook\mbs{\nor} \Big|_\Esp
\end{aligned}
\end{equation}
by Stokes' theorem.
Hence the integral equation \eqref{3Dglobal-2form-conslaw} becomes
\begin{equation}\label{3Dboundary1form-conslaw}
\frac{d}{dt} \int_{\p\S} \tilde{\mbs{\Theta}}_{(1)}\Big|_\Esp
= 2! \frac{d}{dt} \int_{\p\S} \Theta^k\hat\ell_k \volformV\hook\mbs{\nor} \Big|_\Esp
=0
\end{equation}
which is a temporal 1-form conservation law on the boundary curve $\p\S$ of the surface.
This global boundary conservation law will be non-trivial whenever
the spatial part of $\tilde{\mbs{\Theta}}_{(1)}$ is locally non-trivial,
namely $\big(\tilde{\mbs{\Theta}}_{(1)}|_\Esp\big)_{\p\S}$
is not an exact spatial 1-form,
or equivalently,
$\vec{\Theta}|_\Esp =(\Theta^1,\Theta^2,\Theta^3)|_\Esp$
is not equal to a gradient.

Finally, we consider 1-form conservation laws in $\Rnum\times\Rnum^3$:
$(\D\tilde{\mbs{\Phi}}_{(1)})|_\Esp = 0$, 
with
\begin{equation}
\tilde{\mbs{\Phi}}_{(1)}
= \tilde{\Phi}_{\mu}\d x^\mu
= T_k\d x^k -\Psi\d t
\end{equation}
On a spatial curve $\C$ with endpoints $\p\C$ in $\Rnum^3$,
the global form of this conservation law is given by
\begin{equation}\label{3Dglobal-1form-conslaw}
\begin{aligned}
& \frac{d}{dt} \int_{\C} \tilde{\mbs{\Phi}}_{(1)}\Big|_\Esp
= \frac{d}{dt} \int_{\C} T^k\hat\ell_k \volformV\hook\mbs{\nor} \Big|_\Esp
\\&\qquad
= \Big(\tilde{\mbs{\Phi}}_{(1)}\hook\vec{t}\,\big|_{\p\C}\Big)\Big|_\Esp
= -\Big( \Psi\big|_{\p\C}\Big)\Big|_\Esp
\end{aligned}
\end{equation}
where $\hat\ell$ is the unit tangent vector of the curve $\C$
and $\mbs{\nor}$ is the normal bi-vector dual to the 2-form $\volformV\hook\hat\ell$,
while $\vec{t}$ is the temporal vector dual to the 1-form $\d t$.
This integral equation \eqref{3Dglobal-1form-conslaw} looks the same as
a global circulatory conservation law \eqref{3Dglobalcurvconslaw},
but here $\vec{T}=(T_1,T_2,T_3)$ is curl-free $\Curl\vec{T}|_\Esp=0$
due to the spatial part of the 1-form conservation law,
$\big(\D\tilde{\mbs{\Phi}}_{(1)}\big)_{\Rnum^2}
=D_j T_k\wedge\d x^j\wedge\d x^k
=\volformV\hook\Curl\vec{T}$,
where $\Rnum^2$ is an arbitrary plane in $\Rnum^3$. 

Suppose that a 1-form conservation law is locally trivial,
$\tilde{\mbs{\Phi}}_{(1)}|_\Esp = (\D\tilde{\Theta})|_\Esp$,
such that the spatial circulation 1-form vanishes, 
$\tilde{\mbs{\Phi}}_{(1)}\hook\vec{t}\,|_\Esp=(\D\tilde{\Theta})\hook\vec{t}\,|_\Esp =0$,
where $\tilde{\Theta}[\depvars]$
is an arbitrary scalar function. 
Then, in the integral equation \eqref{3Dglobal-1form-conslaw},
the boundary terms vanish
while the line integral reduces to endpoint terms 
\begin{equation}
\int_{\C} \tilde{\mbs{\Phi}}_{(1)}\Big|_\Esp 
= \int_{\C} T^i\hat\ell_i \volformV\hook\mbs{\nor} \Big|_\Esp
= \Big( \tilde{\Theta}\big|_{\p\C}\Big)\Big|_\Esp
\end{equation}
by the line integral theorem.
Hence the integral equation \eqref{3Dglobal-1form-conslaw} becomes
\begin{equation}\label{3Dboundary0form-conslaw}
\frac{d}{dt} \Big( \tilde{\Theta}\big|_{\p\C}\Big)\Big|_\Esp
=0
\end{equation}
which is a temporal 0-form conservation law at the boundary endpoints $\p\C$ of the curve.
This global boundary conservation law will be non-trivial whenever
the function $\tilde{\Theta}|_\Esp$ is not equal to a constant.

\subsection{Example of Maxwell's equations}

Maxwell's equations (in vacuum) have a well-known formulation
in which the electromagnetic field is represented by the 2-forms
$\mbs{F} = F_{\mu\nu}\d x^\mu\wedge \d x^\nu$
and
$*\mbs{F} = {*F}_{\mu\nu}\d x^\mu\wedge \d x^\nu$
with components 
\begin{equation}\label{ME:F}
F_{\mu\nu}
= \begin{pmatrix} 
0 & E^1 & E^2 & E^3 \\ 
-E^1& 0& B^3 & -B^2 \\
-E^2 & -B^3 &  0 & B^1 \\
-E^3 & -B^2 & -B^1 & 0
\end{pmatrix} ,
\quad
{*F}_{\mu\nu}
= \begin{pmatrix} 
0 & -B^1 & -B^2 & -B^3 \\ 
B^1& 0& -E^3 & E^2 \\
B^2 & E^3 &  0 & -E^1 \\
B^3 & E^2 & E^1 & 0
\end{pmatrix} 
\end{equation}
Here $*$ denotes the Hodge dual, which satisfies $*^2=-1$.
(More specifically,
${*F}_{\mu\nu}= \tfrac{1}{2}\vol_{\mu\nu}{}^{\alpha\beta}F_{\alpha\beta}$,
where $\vol_{\mu\nu}{}^{\alpha\beta} = \vol_{\mu\nu\sigma\tau}\eta^{\alpha\sigma}\eta^{\beta\tau}$ is given in terms of the Minkowski metric $\eta^{\alpha\beta}=\text{diag}(-1,1,1,1)$.)
In this representation, the electric and magnetic field equations \eqref{eq:EM:vac}
are given by
\begin{equation}
\d\mbs{F}=0,
\quad
\d{*\mbs{F}}=0
\end{equation}
These equations constitute 2-form local conservation laws.
On any spatial closed surface $\S$,
their global form is given by
\begin{equation}
\frac{d}{dt} \int_{\S} \mbs{F} \Big|_\Esp =0,
\quad
\frac{d}{dt} \int_{\S} {*\mbs{F}} \Big|_\Esp =0
\end{equation}
without a boundary circulation integral because $\S$ is boundaryless.
Both of these conservation laws are non-trivial.
They have the physical meaning that there is
no net non-zero electric and magnetic flux through closed surfaces,
since
\begin{equation}\label{ME:global-2form-conslaws}
\begin{aligned}
& \int_{\S} \mbs{F}
= \int_{\S} \tfrac{1}{2!} B^k\vol_{ijk} \d x^i\wedge \d x^j
= \int_{\S} B^k\nor_k dA,
\\
& \int_{\S} {*\mbs{F}}
= -\int_{\S} \tfrac{1}{2!} E^k\vol_{ijk} \d x^i\wedge \d x^j
= -\int_{\S} E^k\nor_k dA,
\end{aligned}
\end{equation}
where $\nor$ is the outward normal vector of $\S$ in $\Rnum^3$. 

Using the property $\d^2=0$, we can obtain two 3-form local conservation laws
\begin{subequations}\label{ME:3form-localconslaws}
\begin{align}
& \mbs{\Phi}_{(3)}= \d\mbs{F},
\quad
\d\mbs{\Phi}_{(3)}|_\Esp=0
\label{ME:3form-F-conslaw}
\\
& \mbs{\Phi}_{(3)}= \d{*\mbs{F}},
\quad
\d\mbs{\Phi}_{(3)}|_\Esp=0
\label{ME:3form-*F-conslaw}
\end{align}
\end{subequations}
Clearly, both conservation laws are locally trivial,
and they describe differential identities holding on the electric and magnetic field equations.

The global form of the first conservation law \eqref{ME:3form-F-conslaw}
on a spatial volume $\V$ consists of
\begin{equation}
\frac{d}{dt} \int_{\V} \d\mbs{F} \Big|_\Esp 
= \int_{\p\V} (\d\mbs{F})\hook\vec{t}\,\Big|_\Esp
=0
\end{equation}
where, by Stokes' theorem for differential forms, 
\begin{equation}
\int_{\V} \d\mbs{F} = \int_{\p\V} \mbs{F},
\quad
\int_{\p\V} (\d\mbs{F})\hook\vec{t}
= \int_{\p\V} (\p_t\mbs{F} + \d(\mbs{F}\hook\vec{t})),
\quad
\int_{\p\V} \d(\mbs{F}\hook\vec{t}) =0
\end{equation}
because the boundary surface $\p\V$ is closed.
As a consequence,
the 3-form conservation law \eqref{ME:3form-F-conslaw} yields
a temporal boundary conservation law 
coinciding with the first of the two global 2-form conservation laws \eqref{ME:global-2form-conslaws}. 
Similarly, the second global 2-form conservation law coincides
with the global form of the second conservation law \eqref{ME:3form-*F-conslaw}.

Thus, the two differential identities \eqref{ME:3form-localconslaws}
representing locally trivial 3-form conservation laws 
give rise to non-trivial global 2-form conservation laws
describing constants of motion on spatial boundary surfaces.

\section{Jet space formalism}\label{sec:jetspace}

The \emph{jet space} associated to a set of 
independent variables $z^i$, $i=1,\ldots,n$,
and dependent variables $w^\alpha$, $\alpha=1,\ldots,m$,
is the coordinate space $J= (z^i,w^\alpha,w^\alpha_{z^i},w^\alpha_{z^iz^j}, \ldots)$.

A \emph{differential function} $f[w]$, on a domain in $J$, is
a smooth function of finitely many variables in $J$.
Total derivatives of $f[w]$ with respect to $z^i$ are given by the derivative operator
$D_i = \parderop{z^i} + w^\alpha_{z^i}\parderop{w^\alpha} + w^\alpha_{z^iz^j}\parderop{w^\alpha_{z^j}} + \cdots$.
$D=(D_1,\ldots,D_n)$ will denote the set of total derivatives
with respect to all independent variables $z=(z^1,\ldots,z^n)$.

For a PDE system $G[w]=0$,
the solution space $\Esp$ is the set of all functions $w(z)$ that satisfy
all of the PDEs in the system,  $G^a[w(z)]=0$, $a=1,\ldots,M$.
This space $\Esp$ can be identified with a subspace $\Esp_J$ in $J$ as follows.
If $w(z)$ is any solution of the PDE system,
then for each $z=z_0$ the values
$w^\alpha(z_0)=w^\alpha_0$,
$\p_{z^i}w^\alpha(z_0)=w^\alpha_{0z^i}$,
$\p_{z^i}\p_{z^j} w^\alpha(z_0)= w^\alpha_{0z^iz^j}$, 
and so on, 
yield a point in $J$ given by the coordinate values
$(z_0^i,w^\alpha_0,w^\alpha_{0z^i},w^\alpha_{0z^iz^j},\ldots)$.
The set of these points for all $z=z_0$ in the domain common to all solutions $w(z)$
defines a subspace $\Esp_J\subset J$.
This subspace can be defined equivalently as the set of points in $J$ defined by
the equations $G^a[w]=0$, $D G^a[w]=0$, $D^2 G^a[w]=0$, $\ldots$, $a=1,\ldots,N$,
provided that this prolonged PDE system is locally solvable.
The equivalence consists of showing that for each point
$(z_0^i,w^\alpha_0,w^\alpha_{0z^i},w^\alpha_{0z^iz^j},\ldots)$ in $E\subset J$
there exists a solution $w(z)$ of $G^a[w(z)]=0$ having the values
$w^\alpha(z_0)=w^\alpha_0$,
$\p_{z^i}w^\alpha(z_0)=w^\alpha_{0z^i}$,
$\p_{z^i}\p_{z^j} w^\alpha(z_0)= w^\alpha_{0z^iz^j}$, $\ldots$
at $z=z_0$.
Local solvability is precisely the existence of such solutions $w(z)$
in a neighbourhood of $z=z_0$ for any given values
$w^\alpha_0=w^\alpha(z_0)$,
$w^\alpha_{0z^i}=\p_{z^i}w^\alpha(z_0)$,
$w^\alpha_{0z^iz^j}= \p_{z^i}\p_{z^j} w^\alpha(z_0)$, $\ldots$
to an arbitrary finite order.

For locally solvable PDE systems,
$f|_\Esp$ has two equivalent meanings.
First, $f|_\Esp = f[w(z)]$ for an arbitrary function $w(z)$ in $\Esp$.
Second, $f|_\Esp = f[w]|_{\Esp_J}$ for an arbitrary point in $\Esp_J$.

There is another way to view this equivalence.
The evaluation of $w(z)$ and its derivatives $\p^k w(z)$ at a point $z=z_0$
provides an embedding of $\Esp$ into $J$.
Given any function $f$ defined on all of $j$,
the embedding map can be used to pull back $f$ from $J$ to $\Esp$,
which defines $f|_\Esp$.

It is important to have a characterization of differential functions
$f[w]$ that vanish on $\Esp$.
This requires some preliminaries about the algebraic form of the equations
$G^a[w]=0$, $D G^a[w]=0$, $D^2 G^a[w]=0$, $\ldots$, $a=1,\ldots,M$
in a prolonged PDE system.
A partial derivative variable in $J$ is a \emph{leading derivative} of
a given prolonged PDE system,
with respect to some chosen ordering of partial derivatives,
if no differential consequences of it appear in the system.
The prolonged system is called {\em regular}
if the following conditions hold \cite{Anc-review}. 
First, each equation in the system can be expressed in a solved form
in terms of a set of leading derivatives
such that the right-hand side of each equation does not contain any of these derivatives.
Second, the system is closed in the sense that it has no integrability conditions 
and all of its differential consequences produce PDEs that have a solved form 
in terms of differential consequences of the set of leading derivatives.
Note that if a PDE system is not closed then it can always be enlarged to obtain a closed system
by appending any integrability conditions that involve the introduction of more leading derivatives.
Also note that a regular PDE system may possess differential identities.

The following result, known as \emph{Hadamard's lemma}, holds:
For any regular PDE system $G[w]=0$,
if a differential function $f[w]$ vanishes on the solution space $\Esp_J\subset J$,
then $f[w]= C^{(0)}{}_a[w]G^a[w] + C^{(1)}{}_a^i[w] D_i G^a[w] + C^{(2)}{}_a^{ij}[w] D_iD_j G^a[w] + \cdots$ (which terminates at some finite order)
holds identically in $J$,
where the coefficients $C^{(0)}{}_a[w]$, $C^{(1)}{}_a^i[w]$, $C^{(2)}{}_a^{ij}[w]$, $\ldots$
are differential functions that are non-singular on $\Esp$.
(see Refs.\cite{Nes-book,Anc-review} for a proof).

If a regular PDE system possesses one or more differential identities,
then the coefficient functions $C^{(0)}{}_a[w]$, $C^{(1)}{}_a^i[w]$, $C^{(2)}{}_a^{ij}[w]$, $\ldots$
in Hadamard's lemma will be non-unique,
since an arbitrary multiple of a differential identity
can be added to the righthand side.
A sufficient condition under which these coefficient functions will be unique
is if the set of leading derivatives in a regular PDE system
consists of  pure derivative of all dependent variables
with respect to some single independent variable.
Such PDE systems are a generalization of Cauchy-Kovalevskaya systems
and are usually called \emph{normal} 
\cite{Olv-book,Mar-Alo1979}.

\end{appendix}

\end{document}